\documentclass[rmp,aps]{revtex4-1}
\usepackage{multirow,graphics,epsfig}

\begin{document}
\title{Pre-galactic metal enrichment -- \\
The chemical signatures of the first stars}

\author{\bf Torgny Karlsson}
\email{torgny.karlsson@physics.uu.se, torgny@physics.usyd.edu.au}
\affiliation{Sydney Institute for Astronomy, School of Physics, The University of Sydney, NSW 2006, Australia\\} 
\affiliation{Visiting Research Fellow, University of Oxford, Oxford, OX1 3RH, UK}
\author{\bf Volker Bromm}
\affiliation{Department of Astronomy and Texas Cosmology Center, University of Texas at Austin, Austin, TX 78712, USA}
\author{\bf Joss Bland-Hawthorn}
\affiliation{Sydney Institute for Astronomy, School of Physics, The University of Sydney, NSW 2006, Australia\\} 
\affiliation{Leverhulme Visiting Professor, University of Oxford, Oxford, OX1 3RH, UK}

\begin{abstract}
\noindent
The emergence of the first sources of light at redshifts of $z\sim 10-30$ signaled the transition from the simple initial state of the Universe to one of increasing complexity. We review recent progress in our understanding of the formation of the first stars and galaxies, starting with cosmological initial conditions, primordial gas cooling, and subsequent collapse and fragmentation. We emphasize the important open question of how the pristine gas was enriched with heavy chemical elements in the wake of the first supernovae. We conclude by discussing how the chemical abundance patterns conceivably allow us to probe the properties of the first stars, and allow us to test models of early metal enrichment.
\end{abstract}

\date{September 2012}
\maketitle
\tableofcontents

\section{Introduction}\label{intro}
One of the key goals in modern cosmology is to understand the formation of the first generations of stars and the assembly process of the first galaxies. With the advent of the first stars -- referred to historically as Population~III (Pop~III) -- the Universe was rapidly transformed into an increasingly complex system, due to the energy and heavy element input from stellar sources and accreting black holes \citep{bl01,miralda03,bl04,cf05}.  Anisotropies in the cosmic microwave background (CMB) allow us to probe the state of the Universe $370,000$ years after the Big Bang and provide us with some details of early structure formation. With the best available ground- and space-based telescopes, we can probe cosmic history all the way from the present-day Universe to roughly a billion years after the Big Bang. In between lies the remaining frontier, and the first stars and galaxies are the signposts of this early, formative epoch.

To simulate the build-up of the first stellar systems, we have to address the feedback from the very first stars on the surrounding intergalactic medium (IGM), and the formation of the second generation of stars out of material that was enriched by the the first stellar generation. There are a number of reasons why addressing the feedback from the first stars and understanding second-generation star formation is crucial:
\begin{itemize}
\item Over the past fifty years, there has been extraordinary progress in our understanding of the origin and evolution
of the chemical elements over cosmic time. But there remains considerable uncertainty on the physical processes that
seeded many elements in the first stars. \\
\item The initial burst of Pop~III star formation may have been rather brief due to the strong negative feedback effects that likely acted to self-limit this formation mode \citep{yoshida04,gb06}.  Second-generation star formation, therefore, may well have been cosmologically dominant compared to Pop~III stars. A subset of the second-generation stars with masses $\lesssim 0.8~\mathcal{M_{\odot}}$ must have survived to the present day. These stars provide an indirect window into the Pop~III era, as they reflect the enrichment from a single, or at most a small multiple of supernova (SN) events \citep{bc05,k05,kg05,k06,tumlinson06,frebel07,salvadori07,karlsson08}.  By scrutinizing their chemical abundance patterns, we can extract empirical constraints which we apply to supercomputer simulations, which in turn allow us to derive theoretical abundance yields to be compared with the data. \\
\item The first steps in the hierarchical build-up of structure provide us with a simplified laboratory for studying galaxy formation.
Did the reionization epoch influence the subsequent evolution of galaxies? What was the role of galactic winds in the first galaxies? Did star formation occur in bursts, or in a steady, self-regulated mode? How were the first nuclear black holes seeded? We can probe these formative times by reconstructing the conditions in the first galaxies from the chemical signatures\footnote{In this review, we make a clear distinction between a chemical {\it fingerprint} and a chemical {\it signature} (see Sec.~\ref{fingerprint}). The term chemical fingerprint is used when the elemental abundances provide direct evidence of a specific reaction mechanism: the r-process or triple-$\alpha$ process, for example. This review is concerned with identifying the chemical signatures of the first stars in the surface abundances of the oldest stellar populations. A chemical signature is inherently more complex because the elemental yields expelled from a dying star are likely to depend on more than one physical process. The signature then reflects multiple parameters, such as stellar mass, rotation, explosion energy, and the amount of fallback onto the remnant.} in the most ancient stars.
\end{itemize}

Existing and planned observatories, such as the Hubble Space Telescope (HST), the 8-10m class telescopes and the James Webb Space Telescope (JWST), planned for launch in 2018, yield data on stars and quasars less than a billion years after the Big Bang. The ongoing {\it Swift} gamma-ray burst (GRB) mission provides us with a window into massive star formation at the highest observable redshifts \citep{lr00,bl02,bl06}.  Measurements of the near-IR cosmic background radiation, both in terms of the spectral energy distribution and the angular fluctuations provide additional constraints on the overall energy production due to the first stars \citep{santos02,mag03,dwek05,kashlinsky05,fk06}.  Focusing on the nearby Universe, the {\it Gaia} mission\footnote{http://www.rssd.esa.int/gaia} will provide six-dimensional phase-space information on no less than one billion stars in the Milky Way and projects like HERMES\footnote{http://www.aao.gov.au/hermes}, APOGEE\footnote{http://www.sdss3.org/surveys/apogee.php}, LAMOST\footnote{http://www.lamost.org}, and the Southern Sky Survey\footnote{http://www.mso.anu.edu.au/skymapper}, will measure detailed chemical abundances for millions of Galactic stars, out of which some fraction will likely probe the very first metal enrichment phase. Understanding the formation of the first stars and galaxies is thus of great interest to observational studies conducted both at high redshifts and in our local Galactic neighborhood.

In this review, we will focus on the interplay between primordial star formation and pre-galactic metal enrichment, 
since it provides one of the main feedback mechanisms that shaped the early IGM \citep{madau01}.  
The chemical feedback from the first SNe had a far-reaching impact on early cosmic history
\citep{cf05}. Our present understanding is that the character of star formation
changed from the early, high-mass dominated Pop~III mode to the more normal, lower-mass
Pop~II mode once a critical level of enrichment had been reached, the so-called critical
metallicity, $Z_{\rm crit} \sim 10^{-4}~\mathcal{Z_{\odot}}$ (Omukai, 2000\nocite{omukai00}; Bromm {\it et al.}, 2001\nocite{bromm01}; 
Bromm and Loeb, 2003\nocite{bl03}; Schneider {\it et al.}, 2003\nocite{schneider03}). It is then crucially
important to understand the topology of early metal enrichment, and when a certain
region in the Universe becomes supercritical. In general, metals produced by Pop III SNe
will initially be dispersed into the IGM, and a fraction of them will later be re-incorporated
into more massive systems during bottom-up structure formation. The metal-enriched protogalaxies
then drive further heavy element synthesis and ejection into the intergalactic medium on a more
massive scale. They are also candidate drivers of the early hydrogen reionization that has
been inferred from the relatively large optical depth in the CMB polarization measured with
the Wilkinson Microwave Anisotropy Probe (WMAP). This second wave of metal injection pollutes a 
larger fraction of the Universe, and
sets the stage for pervasive gravitational fragmentation in those dark halos that have avoided
complete photoevaporation in the emerging cosmic ionizing background.

We now introduce our working definition of what we mean by a ``first star''. A Pop~III star
contains no heavy elements, such that its metallicity is $Z=0$. 
A Pop~II star, on the other hand, has $Z>Z_{\rm crit}$. In principle, there could have been stars with
$0<Z<Z_{\rm crit}$, but none of them, by the definition of critical metallicity, would
have survived until today. More importantly, such hypothetical stars were extremely rare,
if current cosmological simulations are on the right track. Specifically, those
simulations are indicating that even a single Pop~III SN can enrich its surroundings
to values above $Z_{\rm crit}$, or that gas with $Z<Z_{\rm crit}$ resides in low-density
regions of the IGM, where star formation does not occur. Once we learn more about
the topology of early metal-mixing, we may wish to revisit the definition for Pop~III,
but for now, our terminology should work well. For the record, no metal-free
star has been observed in any of the major surveys that seek to identify the most metal-poor stars at the present
epoch, i.e. a spectrum exhibiting H absorption lines with no trace metals (T. Beers, personal communication, 2012).
The lack of metal-free stars fits within a developing paradigm in which the first stars
were very massive and short-lived (see Sec. \ref{form}), although if such an object is eventually found, there will be
no shortage of theories to support its existence!

A prime objective for current astrophysics is to predict the properties of the first galaxies with the goal
of detecting them with the JWST or an Extremely Large Telescope (ELT).
What is the required minimum stellar and, by extension,
virial mass in order for them to be detectable? This will determine the number of sources
accessible through deep field campaigns \citep{pawlik11}. In turn, it will depend on the stellar population
mix within the first galaxies. Is the fine-grain mixing of the metals that fall into the center of
the galactic potential well incomplete, so that Pop~III stars would form simultaneously with
Pop~I/II stars? Or is the turbulent mixing so efficient as to wipe out all remaining pockets of
primordial gas? Thus, arguably, the most important ingredient that defines the physical
state of the first galaxies is their metal content. This is clearly a very complex
problem. Recently, it has become possible to attack it with numerical simulations, utilizing the latest
breakthroughs in algorithm development and supercomputing power.

Here, we raise a number of issues that are highly controversial in the general literature, and appear in 
need of clarification. Because of recent developments in galactic dynamics, we argue that it is not obvious 
where astronomers should look for the 
most ancient stars. Published numerical simulations are already unclear on whether the oldest 
stars are solely the preserve of the inner bulge \citep{ws00,bhp06} or spread 
over the entire Galaxy \citep{scannapieco06,brook07}. Given that galaxies appear to grow 
inside-out \citep{zolotov09,cooper10}, it is often assumed that the majority of these stars must be confined
to the inner bulge, consistent with its age \citep{zoccali06}. But regardless of the galaxy formation model, 
the long-term dynamical evolution of the galaxy must also be considered, and this has important
consequences for our understanding of the early Milky Way (Sec. \ref{churning}).

The uncertainty in the birth site, and therefore the birth process, leads us to question another tenet of 
stellar astrophysics. Does it necessarily follow that targeting the most metal-poor stars is an efficient route to
learning about the yields of the first stars? This needs some clarification. A low value\footnote{$[\mathrm{A}/\mathrm{B}]=\log_{10}(n_{\mathrm{A}}/n_{\mathrm{B}})_{\star}-\log_{10}(n_{\mathrm{A}}/n_{\mathrm{B}})_{\odot}$, where $n_{\mathrm{X}}$ is the number density of element $\mathrm{X}$.} of [Fe/H]
is no guarantee that a star is ancient since it may reflect environmental conditions (e.g. 
low star formation efficiency, shallow potential well).
But radioactivity in some extremely metal poor stars provides clear independent 
evidence that many are indeed ancient \citep{sneden08}.

Conversely, a high value of [Fe/H] does not indicate that a star is of young or intermediate
age. Super-solar abundances are detected in sources out to the highest detectable redshifts because the dynamical
times are very short in the cores of galaxies \citep{hamann99,freeman02,savaglio12}.
It is now well established that the bulge, the halo and all dwarf galaxies show a spread
in [Fe/H] and comprise stellar populations that are 10 Gyr or older, equivalent to a formation redshift prior to 
$z\gtrsim 2$. Globular clusters, which are some of the most ancient systems with well determined ages, 
exhibit a bimodal distribution in metallicity in the range $-2.5\lesssim [\mathrm{Fe}/\mathrm{H}] \lesssim 0$ \citep{brodie06}. 
These are unlikely to retain a clear chemical
signature of the first generation of stars \citep{larsen12}.

Given the large errors in the ages of old stars, the relative fractions of these stars that formed 
before, during or after the reionization epoch is an open question.
In other words, there are no 
obvious stellar age-metallicity relations that we can appeal to at the present time to help us unravel the 
sequence of events in the early universe. We believe that this complex situation can
only be sorted out with a far greater understanding of the origin and evolution of the chemical elements.

Our working definition of what constitutes a ``first star'' needs further clarification regarding the timing of star 
formation. A relatively small number of widely dispersed stars (and conceivably
star clusters) triggered the reionization epoch in the redshift interval $z\simeq 15-20$.
Evidently, these stars belong to the first stellar population, i.e. to Pop~III. But we can extend the temporal 
definition of a ``first star'' to include those that were the first to enrich their immediate 
environment regardless of epoch. Most of the gas confined by collapsing dark matter will not have
experienced any form of chemical enrichment. The stars that formed in these regions during or after the
reionization epoch are the first to enrich the {\it local} gas. This stellar population is identified by location rather
than by epoch but since it specifically is formed out of virtually metal-free gas, it too is ascribed to
Pop~III. As we shall see, however, due to the all-pervasive ionizing field, the two sub-populations 
are thought to have been physically distinct. 

It is not known at what stage in cosmic 
time the last vestiges of pristine gas finally succumbed to stellar enrichment, but it
is not inconceivable that some Pop III stars formed a billion years 
after the reionization epoch. This raises the tantalizing prospect that we can identify such regions in direct
observations of the intermediate and high-redshift Universe \citep{scannapieco03,johnson10}.
We anticipate that Pop III stars will have distinct chemical signatures that can be identified in large stellar surveys. 

We direct the reader to excellent reviews on early star formation from which we have drawn inspiration. \textcite{bromm09} discuss the state-of-the-art numerical simulations of primordial star formation and the assembly of the first galaxies. \textcite{bc05} discuss the search for, and detailed observations of the most-metal poor stars in the Galaxy, while \textcite{tolstoy09} focus on the kinematics and chemical abundances of stars in dwarf galaxies, and \textcite{helmi08} describes the formation of the Galactic stellar halo and its connection to dwarf galaxies. In this review, we focus on
the chemical signatures of the first generations of SNe. These signatures provide us with an important empirical probe to test our theories of high-redshift star formation with observations of metal-poor stars in our cosmic neighborhood \citep{tumlinson06,karlsson08,frebel09}, an approach often called Stellar Archaeology or Near-Field Cosmology \citep{bhf00}.

\section{Formation of the first stars}\label{form}
\begin{figure}[t]
\begin{center}
 \includegraphics[width=5.4in]{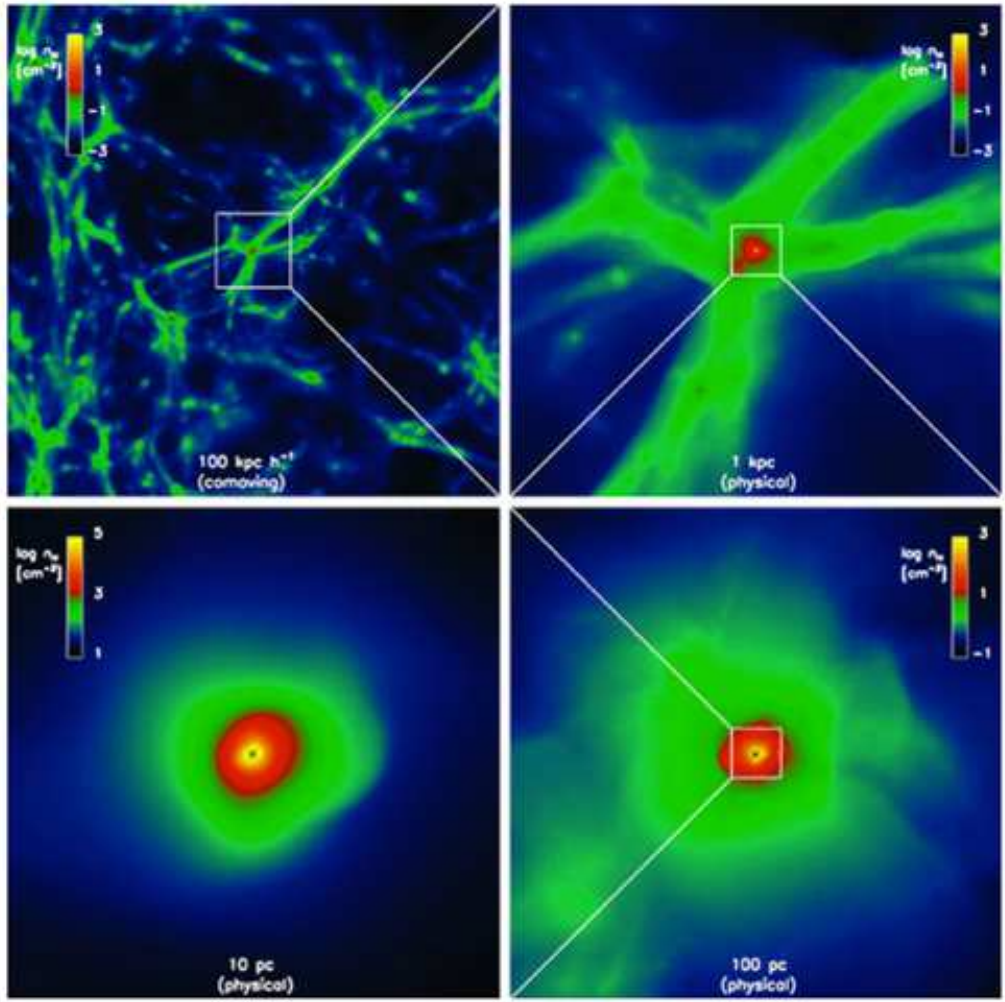} 
 \caption{The first stars form in minihalos.
Shown is the projected density on progressively smaller scales, reaching from the entire simulation box ({\it top left}) down to the center of the newly virialized minihalo on scales of 10~pc ({\it bottom left}). In the bottom two panels, the asterisk denotes the location of the first sink formed. The numerical technique employed here uses sink particles as proxies for growing protostars. Notice how the morphology approaches an increasingly smooth distribution, corresponding to the characteristic loitering state just prior to gravitational runaway collapse. From Stacy {\it et al.} (2010).}
   \label{Fig1_VB}
\end{center}
\end{figure}

The first stars in the Universe likely formed roughly 200 Myr after the Big Bang \citep{bromm99,bromm02,abel02}, when the primordial gas was first able to cool and collapse into dark matter minihalos with masses of the order of 10$^6$ $\mathcal{M_{\odot}}$ (see Fig.~\ref{Fig1_VB}). These stars are believed to have been predominantly very massive, with masses of $\sim 30-100 \mathcal{M_{\odot}}$, owing to the limited cooling ability of primordial gas, which in minihalos could only cool through the radiation from H$_2$ molecules. While the initial conditions for the formation of these stars are, in principle, known from precision measurements of cosmological parameters \citep{komatsu11}, Pop~III star formation may have occurred in different environments which may have allowed for different modes of star formation.  Indeed, it has become evident that Pop~III star formation may actually consist of two distinct modes: one where the primordial gas collapses into a dark matter (DM) minihalo, and one where the metal-free gas becomes significantly ionized prior to the onset of gravitational runaway collapse \citep{jb06}.  This latter mode of primordial star formation was originally termed `Pop~II.5' \citep{gb06,jb06,mackey03}. To more clearly indicate that both modes pertain to {\it metal-free} star formation, we here follow the new classification scheme \citep{mt08,johnson08}, in which the minihalo Pop~III mode is termed Pop~III.1, whereas the second mode (formerly `Pop~II.5') is now called Pop~III.2.

\subsection{Population III: The first mode}
The formation of the very first stars, Pop~III.1 in the new terminology, can be understood with two basic ingredients: $\Lambda$CDM structure formation and the atomic and molecular physics of the primordial hydrogen and helium. Here, $\Lambda$CDM refers to a Universe composed of cold dark matter (CDM), but dominated by dark energy, possibly Einstein's cosmological constant ($\Lambda$). It has been recognized for some time that within variants of CDM, minihalos provide the first star forming sites, where cooling relies on molecular hydrogen \citep{cr86,haiman96,tegmark97}. The properties of the dark matter must be drastically modified before one sees significant deviations from this robust basic prediction, such as assuming warm dark matter (WDM) models, or self-annihilating dark matter \citep{yoshida03,gt07,mapelli07,spolyar08}. The question then is: What kind of stars emerge during the H$_2$-facilitated collapse of the pure H-He gas into the minihalo DM potential wells? This problem, although still beyond our observational horizon, is much simpler than the corresponding star formation process in the local, well observed Universe where the environment of giant molecular clouds is extremely complex. Indeed, prior to the formation of the first stars, the Universe had no heavy elements, and therefore no dust to complicate the physics of cooling and opacity. It was also likely that magnetic fields did not yet play a dynamically significant role (but see Schleicher {\it et al.}, 2009, 2010\nocite{schleicher09,schleicher10}). Finally, the early post-recombination Universe was devoid of external ionizing radiation fields, and strong drivers of turbulence, such that the collapse into the minihalos proceeded in a rather quiescent fashion.
The formation of the first stars inside minihalos, where we know the initial conditions as given by $\Lambda$CDM cosmology, and where we have a complete understanding of the relevant physical processes, thus provides us with a well-posed problem, amenable to rigorous numerical studies.  Since the late 1990s, a number of groups \citep{bromm99,bromm02,abel02,abel00,nu01,yoshida06,oshea07,oshea08} have simulated the formation of the first stars with sophisticated numerical algorithms, utilizing either smoothed-particle hydrodynamics (SPH), or adaptive-mesh refinement (AMR) techniques. These calculations have converged on a number of main results, leading to the current `standard model' of first star formation, although important open questions remain (see below).

The most important result, where there is general agreement, is that Pop~III.1 stars were predominantly massive. To first order, this can be understood as the consequence of a large Jeans mass in gas that cools only via H$_2$. Prior to undergoing runaway collapse, the primordial gas settles into what is sometimes termed a quasi-hydrostatic `loitering' state \citep{bl04}. This state is characterized by typical values for the temperature and number density: $T_{\rm char}\simeq 200$\,K and $n_{\rm char}\simeq 10^4$\,cm$^{-3}$. Recent numerical simulations indicate that Pop~III.1 stars formed with characteristic masses of $M_{\star}\gtrsim 30~\mathcal{M_{\odot}}$. It is likely that these stars formed with a range of masses, described by the so-called initial mass function (IMF). The IMF gives the number of stars formed per unit mass, where present-day star formation is often described with a power-law: $dN/dm_{\star}\propto m_{\star}^{-x}$, or a sequence of such power-laws (Kroupa 2001). For Pop~III, we do not know the complete functional form of the underlying IMF with any certainty. There is, however, a simpler, first-order way to characterize the outcome of the star formation process by focusing on the mass-average: $M_{\star}\propto \int m_{\star} dN$. This mass is what we mean by ``characteristic mass'' as it describes the typical result of star formation, with the understanding that some stars form with lower, and some with higher masses. It has, however, not yet been possible to self-consistently simulate the assembly of an entire Pop~III star, starting from realistic cosmological initial conditions. Recently, such ab initio calculations \citep{yoshida08} have traced the evolution up to the point where a small protostellar core has formed at the center of a minihalo.  This initial hydrostatic core has a mass, $m_{\star}\sim 10^{-2}~\mathcal{M_{\odot}}$, very similar to present-day, Pop~I, protostellar seeds. The subsequent growth of the protostar through accretion, however, is believed to proceed in a markedly different way \citep{bl04}.  In the early Universe, protostellar accretion rates are believed to have been much larger, due to the higher temperatures in the star forming clouds, which in turn is a consequence of the limited ability of the primordial gas to cool below the $\sim 100$~K accessible to H$_2$-cooling. The higher accretion rates, together with the absence of dust grains and the correspondingly reduced radiation pressure that could in principle shut off the accretion, conspire to yield heavier final stars in the early Universe. Estimates for the masses thus built up are somewhat uncertain, but a rough range is $m_{\star}\sim 30-100~\mathcal{M_{\odot}}$. It is possible that this range may extend to lower, as well as higher values, depending on the details of the still very uncertain primordial IMF. The current frontier in numerical simulations attempts to carry out fully self-consistent radiation-hydrodynamical calculations of the Pop~III accretion process, taking into account effects of negative protostellar feedback and of centrifugal support in the accretion disk around the first stars \citep{mt08}. First attempts have now been made (Hosokawa {\it et al.} 2011; Stacy {\it et al.} 2012\nocite{hosokawa11,stacy12}), suggesting the range of masses as given above.

\begin{figure}[t]
\begin{center}
 \includegraphics[width=2.7in]{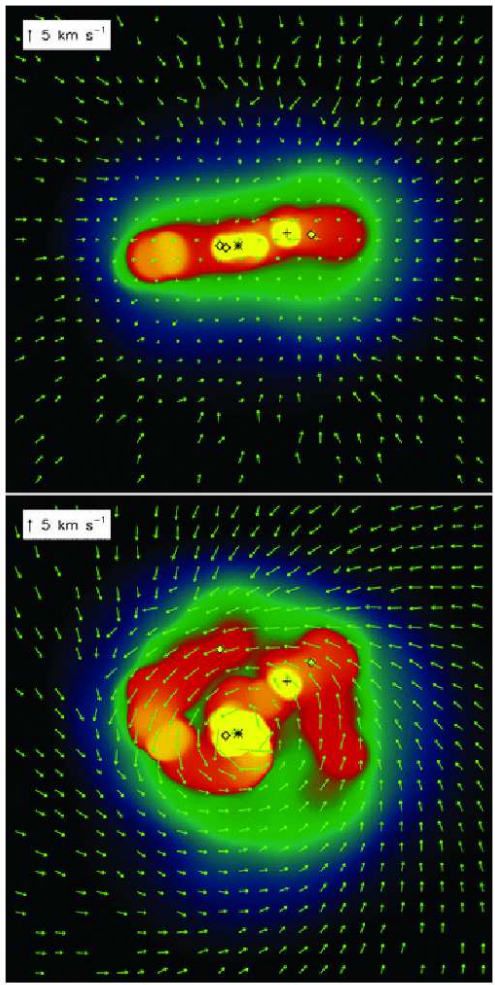}
\caption{Kinematics in the disk-like configuration in the center of a minihalo (cf. Fig.~\ref{Fig1_VB}). Asterisks denote the location of the most massive sink, crosses of the second most massive sink, and diamonds of the other sinks. {\it Top}: Density projection in the $x-z$ plane after 5000 yr, shown together with the velocity field. Velocities are measured with respect to the center of mass of the gas distribution. {\it Bottom}: Same as above, but in the orthogonal ($x-y$) plane. It can be seen how an ordered, nearly Keplerian, velocity structure has been established within the disk. From Stacy {\it et al.} (2010).}
   \label{Fig2_VB}
\end{center}
\end{figure}

Although such fully self-consistent simulations are not yet available, an
important new development is that Pop~III.1 stars may have typically formed
as members of a binary or small multiple system \citep{turk09,stacy10}.
In simulations with sink particles, where the
evolution can be followed beyond the initial collapse of the first
high density peak, one sees the emergence of a compact disk around the
first protostar \citep{clark11b}. This disk is gravitationally unstable and fragments into
a dominant binary, possibly with a few more lower-mass companions (see
Fig.~\ref{Fig2_VB}). These simulations have not yet been able to
reach the asymptotic end state, where all available mass is either
accreted or permanently expelled. The reason is again that radiative
feedback processes in partially optically thick material cannot be
neglected once a protostar has grown to $10 \mathcal{M_{\odot}}$, when full
radiation-hydro simulations are required \citep{hosokawa11,stacy12}. Currently, the binary statistics
of Pop~III, regarding quantities like mass ratios, orbital parameters,
and overall binary fraction are not yet known with any certainty. Rapid
progress is to be expected, however.

\subsection{Population III: The second mode}

\begin{figure}[t]
\begin{center}
 \includegraphics[width=7.2in]{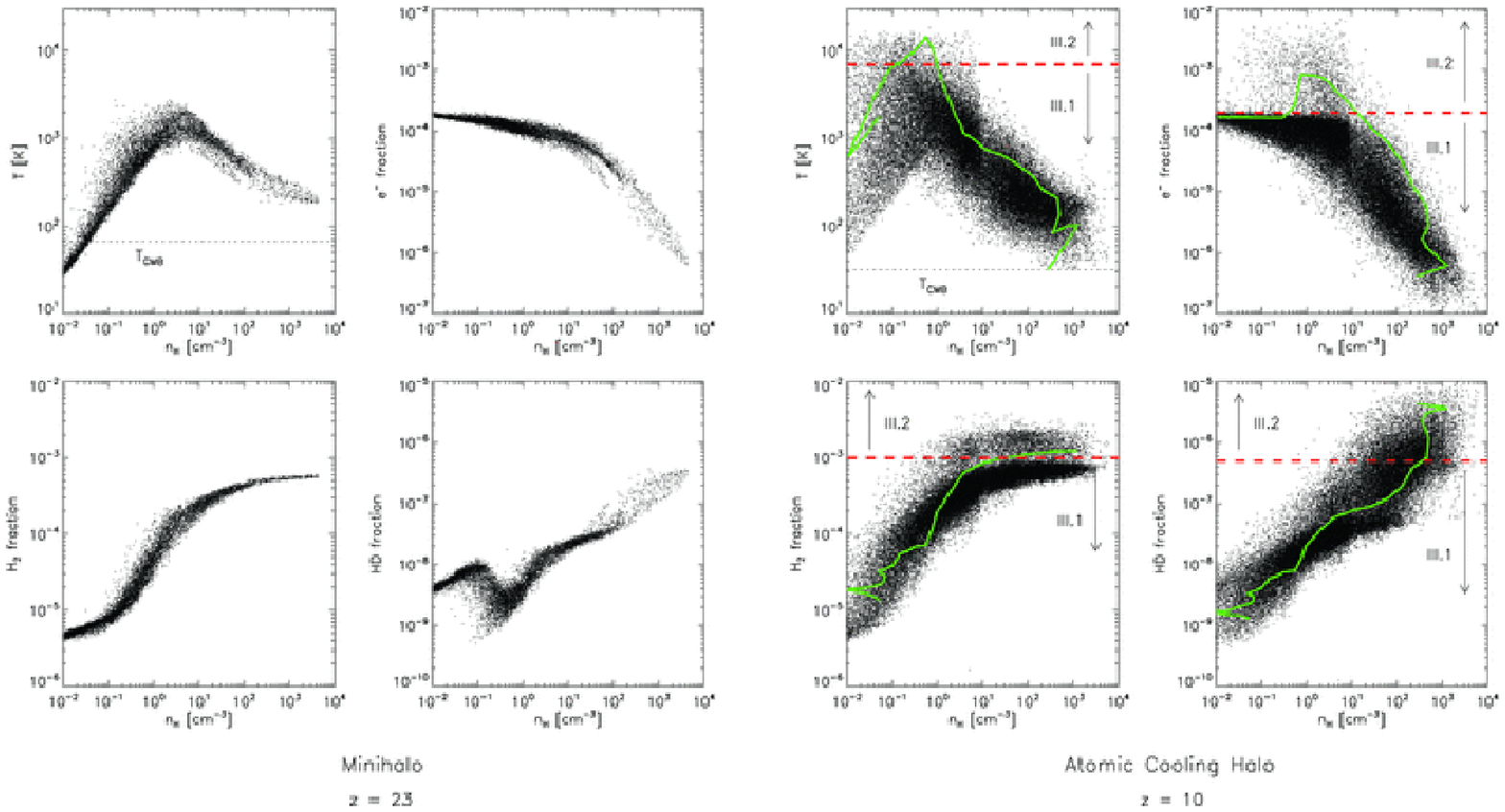}
 \caption{Cooling channels in primordial gas.
Gas properties inside a minihalo ({\it left-hand panel}) and an atomic cooling
halo ({\it right-hand panel}). Shown are the temperature, electron fraction, 
HD fraction and H$_{2}$ fraction as a function of hydrogen number density,
clockwise from top left to bottom left. 
{\it Left-hand panel}: In the minihalo, adiabatic collapse drives the
temperature to $\sim 10^3$\,K and the density to $n_{\rm{H}}\sim 1$\,cm$^{-3}$,
where molecule formation sets in allowing the gas to cool to 200 K. At this
stage, the central clump becomes gravitationally unstable and eventually
forms a Pop III.1 star. 
{\it Right-hand panel}: In the first galaxy, a second cooling channel has
emerged due to an elevated electron fraction at the virial shock. The 
molecule fraction is enhanced in turn, thus enabling the gas to cool to the
temperature of the CMB. Under these conditions, Pop III.2 stars are predicted
to form. The solid green lines denote the path of a representative
fluid element that follows the Pop III.2 channel.
From Greif {\it et al.} (2008).}
   \label{Fig3_VB}
\end{center}
\end{figure}

While the very first Pop~III stars (so-called Pop III.1), with masses of the order of $100~\mathcal{M_{\odot}}$, formed within DM minihalos in which primordial gas cools by H$_2$ molecules alone \citep{abel02,bromm02}, the HD molecule can play an important role in the cooling of primordial gas in situations where the gas experiences substantial ionization. The temperature can then drop well below $200~\rm{K}$. In turn, this efficient cooling may lead to the formation of primordial stars with characteristic masses of the order of $M_{\star}\sim 10~\mathcal{M_{\odot}}$ \citep{jb06}, so-called Pop III.2 stars. In general, the formation of HD, and the concomitant cooling that it provides, is found to occur efficiently in primordial gas which is strongly ionized. This is largely due to the high abundance of electrons which serve as catalyst for molecule formation in the early Universe \citep{sk87}.

Efficient cooling by HD can be triggered within the relic H~{\sc ii} regions that surround Pop~III.1 stars at the end of their brief lifetimes, owing to the high electron fraction that persists in the gas as it cools and recombines \citep{no05,johnson07,yoshida07}.  The efficient formation of HD can also take place when the primordial gas is collisionally ionized, such as behind the shocks driven by the first SNe or in the virialization of massive DM halos \citep{gb06,jb06,machida05,sv06}. There is a critical HD fraction, necessary to allow the primordial gas to cool to the temperature floor set by the CMB at high redshifts: $X_{\rm HD}=n_{\rm HD}/n\sim 10^{-8}$, where $n_{\rm HD}$ is the number density of HD molecules and $n$ that of all particles. Except for the gas collapsing into the virtually un-ionized minihalos, the fraction of HD typically increases quickly enough to play an important role in the cooling of the gas, allowing the formation of Pop~III.2 stars. 
An interesting environment for Pop~III.2 formation may be the so-called atomic cooling halos \citep{oh02}. They have been suggested as candidates for the first galaxies to form (see Bromm {\it et al.}, 2009\nocite{bromm09}). If these systems manage to remain metal-free \citep{johnson08}, the HD cooling channel will become important in tying temperatures to the CMB (see Fig.~\ref{Fig3_VB}).

There may thus be a progression in the characteristic masses of the various stellar populations that form in the early Universe. In the wake of Pop~III.1 stars formed in DM minihalos, where $M_{\star}\gtrsim 30 \mathcal{M_{\odot}}$, Pop~III.2 star formation, where $M_{\star}\sim 10 \mathcal{M_{\odot}} $, ensues in regions which have been previously ionized, typically associated with relic H~{\sc ii} regions left over from massive Pop~III.1 stars collapsing to black holes, while even later, when the primordial gas is locally enriched with metals, Pop~II stars begin to form \citep{bl03,gb06}.  Recent simulations confirm this picture, as Pop~III.2 star formation ensues in relic H~{\sc ii} regions in well under a Hubble time, while the formation of Pop~II stars after the first SN explosions is delayed by more than a Hubble time (Greif {\it et al.}, 2007\nocite{greif07}; Yoshida {\it et al.}, 2007a,b\nocite{yoshida07,yoshida07b}; but see Whalen {\it et al.}, 2008\nocite{whalen08}). Another key question is related to the role of turbulence in shaping the 
primordial mass function \citep{clark11a}. If sufficient turbulence were present,
a broad range of fragment masses could result, similar to the power-law
extension towards high masses, observed in the present-day IMF.

\section{Chemical feedback}\label{chemfeedback}
In this section, we discuss the transport and mixing of metals in the gaseous medium between the stars. The efficiency of this mixing is crucial to how, where, and when the transition from Pop III to Pop II star formation occurred. We begin by briefly reviewing the final fate of metal-free stars, stellar nucleosynthesis, and the current status of stellar yield calculations.

\subsection{Final fate of the first stars}\label{finalfate}
\begin{figure}[t]
\begin{center}
 \includegraphics[width=5.4in]{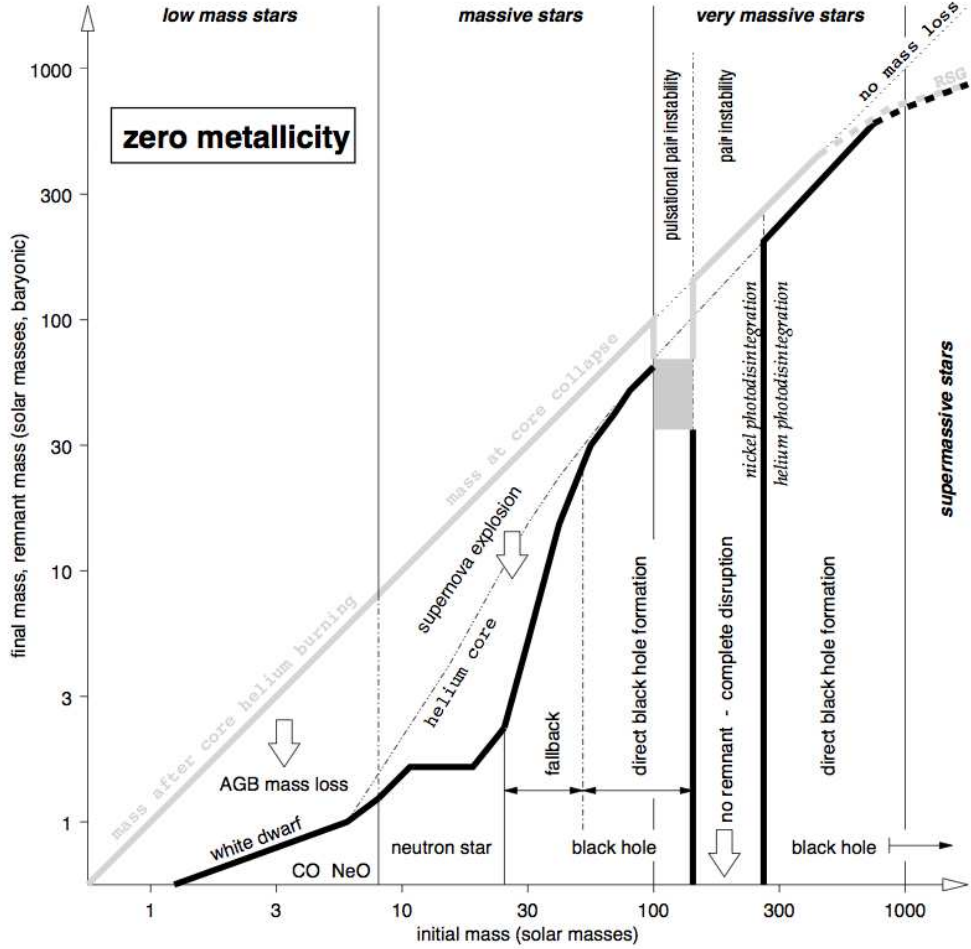} 
\caption{Initial-final mass function of non-rotating primordial stars. Notice that there are distinct mass regimes where enriched material is expected to be ejected into the surroundings, separated by ranges where no or very little enrichment would result. From \textcite{hw02}.}
   \label{fates_hw02}
\end{center}
\end{figure}

Unless a star loses a significant fraction of its mass in a stellar wind, the final fate of the star is predominantly determined by its initial mass (see Fig.~\ref{fates_hw02}).  Generally speaking, this holds true also for the nucleosynthesis in the star's interior.  Primordial massive stars should generally have weak stellar winds due to lower atmospheric opacities but this depends critically on whether the outer layers essentially remain metal-free during the lifetime of the star \citep{ekstrom08b}.  Likewise, mass loss driven by nuclear pulsations \citep{sh59} for metal-free, very massive stars is probably low in most cases \citep{baraffe01}.  

Table \ref{fates} shows, for non-rotating stellar models (however, see Ohkubo {\it et al.}, 2006\nocite{ohkubo06} for rotating $500$ and $1000~\mathcal{M_{\odot}}$ models), the predicted final fates of massive primordial stars and the remnants that are left behind, for different mass regimes. The various ranges of initial stellar masses should be taken as indicative as they may change if properties like rotation and magnetic fields are shown to be significant for the evolution of metal-free stars \citep{ekstrom08a,stacy11}.

At the end of their lives, stars below $\sim 9~\mathcal{M_{\odot}}$ commonly undergo a series of deep mixing events where He burning products, typically carbon, are dredged up to the surface. In particular, for metal-free stars below $\sim 3.5~\mathcal{M_{\odot}}$, this process is triggered by a mechanism \citep{fujimoto90} called helium flash-driven deep mixing (He-FDDM) rather than, or in addition to, the third dredge-up (TDU), operating in metal-rich, asymptotic giant branch (AGB) stars. Primordial stars $\gtrsim3.5~\mathcal{M_{\odot}}$, may experience TDU, but no He-FDDM, if convective overshoot is able to inject freshly synthesized carbon into the hydrogen-rich outer layer \citep{chieffi01,siess02}.  With carbon mixed into the atmosphere, metal-free intermediate-mass stars should be able to develop a dust-driven wind \citep{mattsson08} in a similar fashion to their metal-rich cousins.  Consequently, the entire envelope would eventually be blown away and leave behind a white dwarf. 

\begin{table}[t]
\caption{Final fates and remnants of massive primordial stars for different initial masses.}
{\begin{tabular}{@{}cclc@{}} \toprule
Mass range & Final fate &  \hphantom{o}Metal & Remnant \\
($\mathcal{M_{\odot}}$) & & \,ejection & \\ \colrule
$\!\hphantom{000000}m \lesssim \hphantom{00}9$ & Planetary nebula\footnote{\textcite{heger03}.} & \hphantom{ol.}Yes & White dwarf \\
$\hphantom{00}9\lesssim m < \hphantom{0}10$ & O/Ne/Mg core collapse SN$^{\mathrm{a}}$ &  \hphantom{ol.}Yes & Neutron star \\
$\hphantom{0}10\le m < \hphantom{0}25$ & Fe core collapse SN$^{\mathrm{a}}$ &  \hphantom{ol.}Yes & Neutron star \\
$\hphantom{0}25 \le m < \hphantom{0}40$ & Weak Fe core collapse SN$^{\mathrm{a}}$ (fallback) &  \hphantom{ol.}Yes & Black hole \\ 
$\hphantom{0}25 \le m < \hphantom{0}40$ & Hypernova\footnote{\textcite{nomoto06}.} (fallback) &  \hphantom{ol.}Yes & Black hole \\ 
$\hphantom{0}40\le m < 100$ & Direct collapse$^{\mathrm{a}}$ (no SN) &  \hphantom{ol.}No & Black hole \\
$100\le m < 140$ & Pulsational pair-instability SN$^{\mathrm{a}}$ (fallback) &  \hphantom{ol.}No? & Black hole \\
$140 \le m < 260$ & Pair-instability SN$^{\mathrm{a}}$ &  \hphantom{ol.}Yes & No remnant \\
$260\le m \lesssim10^5$ & Direct collapse\footnote{\textcite{fryer01}.}/core collapse$^{\mathrm{c,}}$\footnote{\textcite{ohkubo06}.} (no SN?) &  \hphantom{ol.}No? & Black hole \\
$10^5\lesssim m \! \hphantom{000000}$ & Direct collapse before reaching main sequence$^{\mathrm{d}}$ &  \hphantom{ol.}No & Black hole \\ \botrule
\end{tabular} \label{fates}}
\end{table}

Stars in the mass range $9\lesssim m/\mathcal{M_{\odot}} < 40$ are believed to explode as core collapse SNe. At the end of hydrostatic burning, stars above $\sim 10~\mathcal{M_{\odot}}$ contain an electron degenerate iron core. As the Fe core continues to grow by silicon shell burning, the electron pressure can eventually no longer counterbalance the increasing gravitational pull and the core collapses.  An explosive instability develops when the infalling outer layers bounce off the collapsed core (proto-neutron star) and are ejected into the interstellar medium (ISM). The details of the explosion mechanism is still a topic of debate \citep{burrows06,janka07}.  Stars around $9-10~\mathcal{M_{\odot}}$ instead form a degenerate O$+$Ne$+$Mg core which may collapse due to rapid electron captures on $^{20}$Ne and $^{24}$Mg, prior to the ignition of Ne \citep{barkat74,nomoto87}.  If so, an electron-capture, or ONeMg SN is formed.  Alternatively, the star loses its outer layers and forms a white dwarf \citep{gi94}.  At the high mass end of the core collapse SN regime, $25-40~\mathcal{M_{\odot}}$, the potential energy of the stellar envelope is comparable to the kinetic energy of the explosion and the innermost layers of the star fall back onto the central neutron star which eventually collapses to a black hole \citep{fryer99}.  A weak, or faint SN is formed if the black hole is non-rotating. For rotating black holes, however, a much stronger explosion may instead be expected \citep{fh00,nomoto06} in the form of a hypernova (HN).  SN1998bw \citep{galama98,iwamoto98} and SN2003lw \citep{mazzali06} are two examples of unusually energetic SNe commonly ascribed to the HN branch.  Stars with masses above $40~\mathcal{M_{\odot}}$ collapse directly to a black hole (see, however, Ekstr\"{o}m {\it et al.}, 2008a\nocite{ekstrom08b}).

For Pop. III stars with masses $\gtrsim 100~\mathcal{M_{\odot}}$, the pair instability kicks in after central carbon burning, driven by the creation of electron-positron pairs. Below $\sim 140~\mathcal{M_{\odot}}$, this instability causes the star to pulsate violently. While the outer layers are lost in SN-like explosions, the star settles down to form a massive Fe core.  Eventually, the star collapses quietly to form a black hole. On the other hand, if the star is in the mass range $140\le m/\mathcal{M_{\odot}} < 260$ it will face complete disruption as a result of the pair instability, and no remnant is formed \citep{hw02}.  In still more massive stars, the energy released from explosive oxygen and silicon burning, caused by the pair instability, is instead used to photodisintegrate the nuclei in the central core. The explosion is halted and a black hole is formed \citep{fryer01}.  As discussed by \textcite{fryer01}, rotation could, however, change this scenario (see also Ohkubo {\it et al.}, 2006\nocite{ohkubo06}) and stars above $\sim 300~\mathcal{M_{\odot}}$ may be able to explode as SNe, in which tremendous amounts of energy exceeding \mbox{$10^{54}$ erg} would be released.  In contrast, supermassive stars $\gtrsim 10^5~\mathcal{M_{\odot}}$, if these exist, will collapse even before settling down onto the main sequence, owing to a general relativistic instability \citep{shapiro02,ss02,ohkubo06}.

\subsection{Nucleosynthesis in the first stars}\label{nucleosynthesis}
Whether the primordial stars will eject any metals into the ISM at the end of their lives is of utmost importance for the early cosmic chemical enrichment, as well as for how subsequent (Pop. II) star formation proceeds.  In Sec. \ref{fingerprint}, we shall discuss in detail which chemical signatures are observed and what these may be telling us about the Pop. III stars and the primordial IMF.  In preparation for that section, we will here briefly review what possible chemical fingerprints and signatures to expect from primordial stars of various masses.  Table \ref{fates} shows in which mass regimes we expect the stars to enrich their surroundings in metals and in which regimes they do not.  In principle, stars below rougly $40~\mathcal{M_{\odot}}$ will be able to eject freshly synthesized material, through SNe or winds. The same goes for stars in the mass range $140\le m/\mathcal{M_{\odot}} \le 260$, which are completely disrupted as they explode as pair-instability SNe (PISNe).  Stars in the range  $40\lesssim m/\mathcal{M_{\odot}} \lesssim 100$ are expected to undergo direct collapse without an explosion, while stars in the pulsational pair-instability SN regime presumably do not eject any elements heavier than helium \citep{hw02}. These two groups are therefore omitted in the detailed discussion below. Similarly, stars in the mass range $260 \le m/\mathcal{M_{\odot}}\lesssim 10^5$ are expected not to eject any metals, at least not in non-rotating models, as everything would be swallowed directly by the central black hole before an explosion takes place.  However, in the case of rotation, bipolar outflows may develop and significant amounts of metals could possibly be ejected \citep{ohkubo06}.

It should be noted that current SN yield calculations are performed without knowledge of the proper explosion mechanism(s).  In the literature, various methods such as a piston, energy injection, or enhanced neutrino opacity, are used to drive the explosion (see Fryer {\it et al.}, 2008\nocite{fryer08} for a discussion).  Moreover, due to the unknown amount of angular momentum in the stellar core, the fraction of mass falling back onto the proto-neutron star is uncertain and, like the explosion energy and the electron fraction $Y_{\mathrm{e}}$, is treated as a model parameter.  Consequently, yields of elements synthesized in the layers closest to the center of the star, e.g. the Fe-peak, are marred by significant uncertainties.  Only when the explosion can be treated in a self-consistent way can these issues be resolved.  Most models also lack a description of rotation. \textcite{mm02} have shown that the surface abundance of $^{14}$N is greatly enhanced in metal-poor, massive stars as a result of rotationally induced mixing (see also Meynet {\it et al.}, 2006; Hirschi, 2007\nocite{meynet06,h07}).  Increased mass loss, due to the increased surface opacity, can further affect the post-main sequence evolution of the stars, and hence the final SN yields.  In contrast, it seems unlikely that rotation alone can significantly alter the evolution of Pop III massive stars, without invoking anisotropic winds and strong magnetic fields \citep{ekstrom08a}.

\begin{table}[t]
\caption{Nucleosynthetic signatures of massive primordial stars.}
{\begin{tabular}{@{}ccccc@{}} \toprule

Mass range & Expl. energy & Ejected Fe & Metal & Nucleosynthetic \\

($\mathcal{M_{\odot}}$) & ($\times 10^{51}~\mathrm{erg}$) & ($\mathcal{M_{\odot}}$) & enrichment & characteristics \\ \colrule

\multirow{2}{*}{$\!\hphantom{000000}m \lesssim \hphantom{00}9$} & \multirow{2}{*}{Wind} & \multirow{2}{*}{$\,0$\,} & $^7$Li(?), C, N, O, & $(\mathrm{C}+\mathrm{N})/\mathrm{O}>1$ \\
& & & Na, Mg, s-process & e.g., $[\mathrm{Pb}/\mathrm{Ba}]\simeq 1.2$ \\[3pt] \hline

\vline \multirow{4}{*}{$\hphantom{00}9\lesssim m < \hphantom{0}10$}  & \multirow{4}{*}{\,\,$\sim0.1$} & \multirow{4}{*}{$\sim 0.002-0.004$} & \multirow{3}{*}{Carbon to Fe-peak\footnote{Based on a solar metallicity $8.8~\mathcal{M_{\odot}}$ 1D-model \citep{wanajo09}.}} &  \hfill $[\mathrm{C},\alpha/\mathrm{Fe}]\ll 0$ \hfill \vline \\
\vline \hfill \hphantom{0}& & & \multirow{3}{*}{r-process?} & \hfill $[\mathrm{Mg}/\mathrm{Ca}]\ll 0$ \hfill \vline \\
\vline \hfill \hphantom{0}& & & & \hfill $[\mathrm{Ni,Zn}/\mathrm{Fe}]\gg 0$ \hfill \vline \\
\vline \hfill \hphantom{0}& & & & \hfill e.g., $[\mathrm{Ba}/\mathrm{Eu}]\simeq-0.6$\footnote{Deduced from observations of metal-poor stars \citep{barklem05}.} \hfill \vline \\[3pt]
\vline $\hphantom{0}10\le m < \hphantom{0}25$ & $\sim1\,\,\,$ & $\sim0.07$ & Carbon to Fe-peak & \hfill $[\alpha/\mathrm{Fe}]>0$ \hfill \vline \\[3pt]

\vline \multirow{2}{*}{$\hphantom{0}25 \le m < \hphantom{0}40$} & \multirow{2}{*}{$<1\hphantom{0}$} & \multirow{2}{*}{$\lesssim0.01$} & Carbon to Fe-peak &  \hfill $[\mathrm{C,O}/\mathrm{Fe}]\gg 0$ \hfill \vline \\ 
\vline \hfill \hphantom{0} & & & r-process? & \hfill e.g., $[\mathrm{Ba}/\mathrm{Eu}]\simeq -0.6^{\mathrm{b}}$ \hfill \vline \\[3pt]

\vline \multirow{3}{*}{$\hphantom{0}25 \le m < \hphantom{0}40$} & \multirow{3}{*}{$\gtrsim10$} & \multirow{3}{*}{$\sim 0.08-0.3$} & \multirow{3}{*}{Carbon to Fe-peak} & \hfill larger [Si,S$/$C,O] \hfill \vline \\
\vline \hfill \hphantom{0} & & & & \hfill larger [V,Co,Cu,Zn$/$Fe] \hfill \vline \\
\vline \hfill \hphantom{0} & & & & \hfill smaller [Mn,Cr$/$Fe] \hfill \vline \\[3pt]

\vline $\hphantom{0}40\le m < 100$ & No expl. & 0 & $-$ & \hfill $-$ \hfill \vline \\[3pt]

\vline $100\le m < 140$ & $\sim 1\,\,\,$ & $0$ & Only H, He(?) & \hfill $-$ \hfill \vline \\[3pt]

\vline \multirow{3}{*}{$140 \le m < 260$} & \multirow{3}{*}{$\sim10-10^2$} & \multirow{3}{*}{$\sim0.01-\hphantom{.}40$} & \multirow{3}{*}{Carbon to Fe-peak} & \hfill $[\mathrm{Mg,Si}/\mathrm{Na,Al}]\gg 0$ \hfill \vline\\
\vline \hfill \hphantom{0}& & & & \hfill $[\mathrm{Si,S}/\mathrm{C}]\sim 1-1.5$ \hfill \vline\\
\vline \hfill \hphantom{0}& & & & \hfill $[\mathrm{Zn}/\mathrm{Fe}]\ll 0$, no r-proc. \hfill \vline\\[3pt]

\vline \multirow{3}{*}{$260\le m \lesssim10^5$} & \multirow{3}{*}{$\sim10^3-10^4$} & \multirow{3}{*}{$\sim 5-20$} & \multirow{3}{*}{Carbon to Fe-peak\footnote{Ejection of metals may only occur if bipolar jets are generated \citep{ohkubo06}.}} & \hfill $[\mathrm{Mg,Si}/\mathrm{Na,Al}] \gg 0$ \hfill \vline \\
\vline \hfill \hphantom{0}& & & & \hfill $[\mathrm{C,}\alpha/\mathrm{Fe}]\ll 0$ \hfill \vline \\
\vline \hfill \hphantom{0}& & & & \hfill $[\mathrm{Zn}/\mathrm{Fe}] \gg 0$ \hfill \vline \\ \hline
\\[-8pt]

$10^5\lesssim m \! \hphantom{000000}$ & No expl. & $0$ & $-$ & $-$ \\ \botrule
\end{tabular} \label{nucl}}
\end{table}

Crucial for the nucleosynthetic identification of the metal-free stellar population is the uniqueness of its chemical signature. A clear dependence of the Pop. III signature on stellar mass, or ranges of masses, provides the means to probe the primordial IMF. Table \ref{nucl} shows, for each of the mass regimes given in Table \ref{fates}, the predicted explosion energy (if an explosion occurs), the ejected iron mass, which of the elements are expected to be synthesized and ejected, and the associated characteristic nucleosynthetic signatures, if any.  The box roughly indicates the range of primordial stellar masses as predicted by theory/simulations.  It is encouraging that the chemical signatures in col. 5 show significant variation with stellar mass. In principle, these signatures can tell us a great deal about the first stars, assuming the models are realistic. 

Our working definition of Pop~III demands zero metallicity, but we here sometimes 
refer to work done at higher metallicities, if $Z=0$ models are not available,
or if it can further illustrate the physics involved. In addition, we here
do not discuss the important problem of how nucleosynthesis proceeds
in Pop~III binary stars, or in those with a significant degree of rotation, and
refer the reader to the specialized literature (Yoon {\it et al.}, 2008, 2012;
Meader and Meynet, 2012)\nocite{yoon08,yoon12,maeder12}.

\subsubsection{Asymptotic giant branch stars: $m/\mathcal{M_{\odot}} \lesssim 9$}
Detailed yields of intermediate-mass stars are in general very difficult to estimate due to several poorly constrained physical processes such as overshooting, mixing and dredge-up, and mass-loss \citep{cl08}.  Some things, however, we do know.  Intermediate-mass stars cannot be a source of e.g., iron-peak elements, as they never go beyond the helium burning stage. On the other hand, lighter elements, in particular $^{12}$C, $^{14}$N, and $^{16}$O, are produced efficiently by the CNO cycle and 3$\alpha$ reactions. These are brought up to the surface by deep mixing processes, specifically the TDU, and the He-FDDM (also named dual core flash, see Campbell and Lattanzio, 2008\nocite{cl08}) operating in lower mass stars. Generally, the $\mathrm{C}/\mathrm{O}$ ratio\footnote{Here, the ratio $\mathrm{C}/\mathrm{O} \equiv n_{\mathrm{C}}/n_{\mathrm{O}}$ denotes the absolute number density ratio of C and O, respectively. Hence, a ratio of $\mathrm{C}/\mathrm{O} = 1$ is equivalent to $[\mathrm{C}/\mathrm{O}] = 0.26$, given the new solar abundances by \textcite{asplund09}.} exceeds unity.  Although metal-free stars $\gtrsim 2~\mathcal{M_{\odot}}$ \citep{siess02} can efficiently convert $^{12}$C to $^{14}$N through the CN-cycle at the base of the convective envelope, which is deep enough to penetrate into the H-burning shell (so called hot bottom burning), the $(\mathrm{C}+\mathrm{N})/\mathrm{O}$ remains above unity.  As a consequence of  the presence of carbon (through so called ``carbon injections'') in the H-burning shell, the Pop. III low- and intermediate-mass stars continue to evolve as normal AGB stars with thermal pulses similar to their metal-rich counterparts (Siess {\it et al.}, 2002\nocite{siess02}, however see e.g., Fujimoto {\it et al.}, 1984; Chieffi and Tornamb\'{e}, 1984\nocite{ct84,fujimoto84}).  During these pulses, $^{23}$Na and $^{25,26}$Mg are produced, as well as neon and some aluminium.  Interestingly, s-process nucleosynthesis may occur also in primordial stars, despite the absence of heavy seed nuclei, i.e., $^{56}$Fe. Instead, the isotopes of C, N, O, F, and Ne act as seed nuclei.  The relatively large neutron production in these stars result in high neutron-to-seed nuclei ratios which ultimately favours the synthesis of the heaviest s-process elements,  like Pb and Bi \citep{gs01,sg03,busso01}.  If so, an overproduction of Pb relative to Ba, for example, is expected, contrary to what occurs in more metal-rich AGB stars. Furthermore, stars experiencing hot bottom burning can produce $^7$Li through the Cameron-Fowler mechanism \citep{cameron71}. This mechanism becomes active when $^7$Be, produced by $\alpha$ capture on $^3$He in the interior, occasionally is brought up to cooler regions by deep convection, captures an electron and forms $^7$Li. Whether this Li, however, survives long enough to be expelled, depends critically on the unknown amount of mass-loss at the time when the abundance of Li is high at the surface \citep{siess02}. This production branch may partly be responsible for the rise from the Spite plateau at $[\mathrm{Fe}/\mathrm{H}]\gtrsim -1$ \citep{travaglio01}. In contrast, \textcite{sbordone10} recently called attention to the apparent presence of a depression from the Spite plateau at the lowest metallicities, implying that Li have been destroyed (i.e., assuming that  the plateau indicates the original Li abundance). However, whether this destruction is due to internal stellar evolution within the observed sample or due to ejection and mixing of processed material from a primordial stellar population (see, e.g., Piau {\it et al.}, 2006\nocite{piau06}) is not yet known. Additional constraints and challenges are provided by the $^6$Li abundance in metal-poor stars; we refer the reader to Asplund {\it et al}., 2006\nocite{asplund06} for a comprehensive discussion.

\subsubsection{Electron capture supernovae: $9 \lesssim m/\mathcal{M_{\odot}} < 10$}\label{ecapsne}
The electron capture, or ONeMg core collapse SNe, have been largely neglected in models of Galactic chemical evolution. Few yield calculations are found in the literature, presumably because  these stars synthesize and eject only small amounts of metals, in particular Fe (i.e., $^{56}$Ni).  Hence, the chemical evolution for the majority of the main elements is, in general, unaffected by the presence of electron capture SNe.  In a metal-free or extremely metal-poor environment, however, the impact of the electron capture SNe may be significantly higher, at least in a fraction of the ISM volume.  This fraction depends on the ratio of electron capture SNe to Fe core-collapse SNe.  Various estimates of this ratio \citep{wanajo09,poelarends08}, based on solar metallicity models and a Salpeter IMF, quote a number between $\sim 0.04$ and $0.4$.  Despite this relatively low ratio, it may turn out to be of importance to account for electron capture SNe in low-metallicity environments, in particular for certain elements (see Sec. \ref{umps}).  

The amount of $^{56}$Ni ejected by electron capture SNe is very small, less than $0.01~\mathcal{M_{\odot}}$.  \textcite{wanajo09} estimated a final $^{56}$Fe yield (after decay) of $0.002-0.004~\mathcal{M_{\odot}}$ based on a $8.8~\mathcal{M_{\odot}}$, solar metallicity model. Even smaller amounts of carbon and $\alpha$-group elements are produced, which results in $[\mathrm{C,}~\alpha/\mathrm{Fe}]\ll 0$.  This goes along with small light-to-heavy $\alpha$-group element ratios.  Interestingly, the model by \textcite{wanajo09} predicts a large production of heavy iron-peak elements, such as Ni and Zn.  It is argued that the electron capture SNe could be the dominant source of Zn in the Galaxy.  Furthermore, the elements $^{70}$Ge and $^{90}$Zr are found to be produced in abundance.  Indeed, the electron capture SNe could be a potential site for the nucleosynthesis of the rapid neutron capture (i.e., the r-process) elements. In particular, the heaviest r-nuclei (with mass numbers $A>130$) have been proposed to form in electron capture SNe via the prompt explosion channel (Wheeler {\it et al.}, 1998; Wanajo {\it et al.}, 2003\nocite{wheeler98,wanajo03}, however, see Kitaura {\it et al.}, 2006\nocite{kitaura06}) or in the shocked surface layers of the ONeMg core (Ning {\it et al.}, 2007\nocite{ning07}, but see Janka {\it et al.}, 2008\nocite{janka08}).  The exact site of the r-process remains unknown. It also remains unclear whether there is more than one r-process site (Wasserburg and Qian, 2000\nocite{wq00}; Qian, 2003\nocite{qian03}; Sneden {\it et al.}, 2008\nocite{sneden08}; see also Sec. \ref{corecol}), as has been argued for the Solar System meteoritic abundances \citep{wasserburg96}.

\subsubsection{Core collapse supernovae: $10 \le m/\mathcal{M_{\odot}} < 40$}\label{corecol}
\begin{figure}[t]
\begin{center}
 \includegraphics[width=7.2in]{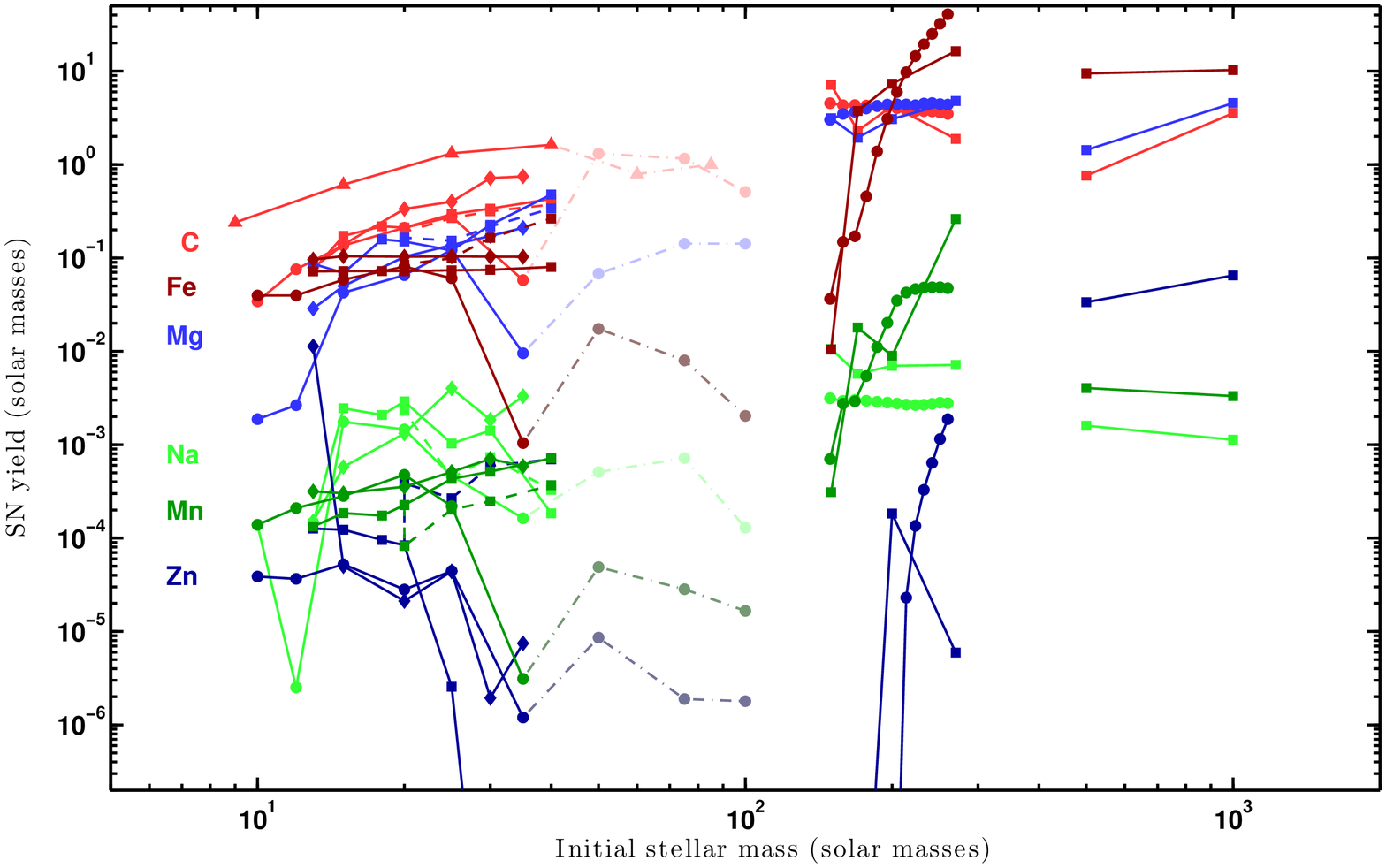}
 \caption{A selection of SN yields ($Z=0$) as a function of initial stellar mass.  Light colors denote light and intermediate mass elements (C, Na, and Mg) while dark colors denote heavy elements (Mn, Fe, and Zn). The specific color coding is displayed in the figure.  Yields by the Japanese group \citep{un02,nomoto06,ohkubo06} are denoted by squares, yields by Heger and Woosley (2002, 2008) are denoted by dots, yields by \textcite{cl04} by diamonds, and yields (only carbon) by \textcite{ekstrom08b} by triangles. Yields of core collapse SNe with masses $>40~\mathcal{M_{\odot}}$ are shaded and connected with dash-dotted lines. Yields of HNe are connected by dashed lines.  Note the sometimes large discrepancy between different models, which can be regarded as a measure of the uncertainties associated with these calculations.}
   \label{figyield}
\end{center}
\end{figure}

The classical core collapse SNe are found in the mass range $10\le m/\mathcal{M_{\odot}}< 25$.  They are believed to be responsible for the bulk of the Galactic inventory of intermediate-mass elements, like oxygen.  One of the most characteristic chemical signatures of core collapse SNe, including Pop. III, is the enhancement of $\alpha$-elements relative to Fe \citep{ww95,un02,cl04,kobayashi06,hw10}.  Due to the small neutron excess, the underproduction of odd-Z elements, like Na and Al, is pronounced as compared to the solar metallicity case.  Neutron-rich isotopes are also produced to a lesser extent. Furthermore, explosive nucleosynthesis can contribute more to the yield as a consequence of the lower mass-loss in Pop. III stars (cf. Ekstr\"{o}m {\it et al.}, 2008b\nocite{ekstrom08a}).  If so, the production of iron-peak elements is enhanced \citep{hw10} as compared to more metal rich core collapse SNe.  The nucleosynthesis of nitrogen is generally not realized in non-rotating stars (but see note by Heger and Woosley, 2010\nocite{hw10}).  As shown by \textcite{mm02}, rotation can radically boost the production of $^{14}$N in extremely metal-poor massive stars as a result of rotation-induced mixing. To some extent, excess $^{12}$C (see Fig.~\ref{figyield}) is also synthesized by this mechanism, as well as $^{13}$C, $^{17,18}$O, and $^{22}$Ne \citep{ekstrom08b,h07}.  

The more massive core collapse SNe ($25 \le m/\mathcal{M_{\odot}} < 40$) are increasingly affected by the deeper gravitational potential of the star as material falls back onto the forming black hole. The two identified branches have two distinctly different nucleosynthetic signatures. The weak SNe are generally characterized by very high C,O/Fe-peak ratios and slightly higher $\alpha$/Fe-peak ratios as compared to core collapse SNe in the $10-25~\mathcal{M_{\odot}}$ mass range. The deficiency of Fe-peak elements in the ejecta results from fallback of the innermost layers. Using the fallback/mixing algorithm developed by \citet{podsiadlowski02}, \textcite{un02,un05} proposed a mixing-fallback scenario in which the complete and incomplete Si-burning layers are presumed to be mixed during the explosion by Rayleigh-Taylor instabilities prior to the fallback, in order to obtain $\mathrm{Zn,Co}/\mathrm{Fe}$ ratios in accordance with observations.  In weak SNe, the complete Si-burning region does not reach high enough temperatures to produce Zn and Co to the required level of abundance.  In HNe, however, higher temperatures are realised and both the complete and incomplete Si-burning regions are shifted outwards in mass which promote the synthesis of V, Co, Cu, and Zn, as well as $^{44}$Ti and $^{48}$Cr which decay into $^{44}$Ca and $^{48}$Ti, respectively \citep{un02,tominaga07}.  On the other hand, Mn, Cr/Fe ratios are low in HNe, as compared to normal SNe (Fig.~\ref{figyield}). Moreover, as the oxygen-burning region increases with increasing explosion energy, oxygen-burning products like Si and S become more abundant relative to O and C \citep{umeda02}.  However, the observational need for core collapse SNe with high explosion energies is debated \citep{hw10}.  Core collapse SNe with fallback may also be a possible site for the r-process \citep{fryer06}.  When material falls back onto the neutron star, it gets strongly shock-heated. Some fraction of this material gains enough energy to escape the potential well of the neutron star and as it bubbles up through the accreting gas, it cools down quickly to a few$\times 10^{9}$ K by adiabatic expansion.  When the expansion slows down, the shock-heated ejecta enters a simmering phase, in which nucleon capture may occur.  \textcite{fryer06} obtain a ratio of [Ba/Eu ]$= - 0.2$ in the total ejecta, which is close to the r-process ratio of [Ba/Eu] $\simeq -0.6$ inferred from observations \citep{barklem05,francois07} of extremely metal-poor stars. Note that the solar r-process value coincides with that in the metal-poor stars \citep{arlandini99}.  There are several other potential r-process sites associated with core collapse SNe, such as neutrino-driven winds \citep{wh92,qw96} and neutron star-neutron star mergers \citep{freiburghaus99}.  It is, however, beyond the scope of this review to discuss the precise conditions for n-capture nucleosynthesis and to assess which astrophysical site(s) may predominantly be responsible for the inventory of r-process elements in the Galaxy.  This can be found elsewhere \citep{arnould07,sneden08}.   

We emphasise that there exist very different approaches to mixing and fallback exist in the literature. The Japanese school parameterise both the mixing and fallback, making specific mass cuts and assuming the mixing is homogenous, and the fallback is complete within a given mass coordinate. Thus one can essentially develop a model that is `made to order' which provides limited insight into the physics. These models do not specify what does the mixing or fixes the amount 
of fallback. In contrast, \textcite{joggerst09,joggerst10a,joggerst10b} take the 1D models of \textcite{hw10} and map these onto a higher dimensionality grid. They follow the Rayleigh-Taylor instability during the explosion (providing mixing) and trace the fall back of material, with the most recent models including the effects of stellar rotation.


\begin{figure}[t]
\begin{center}
 \includegraphics[width=6.8in]{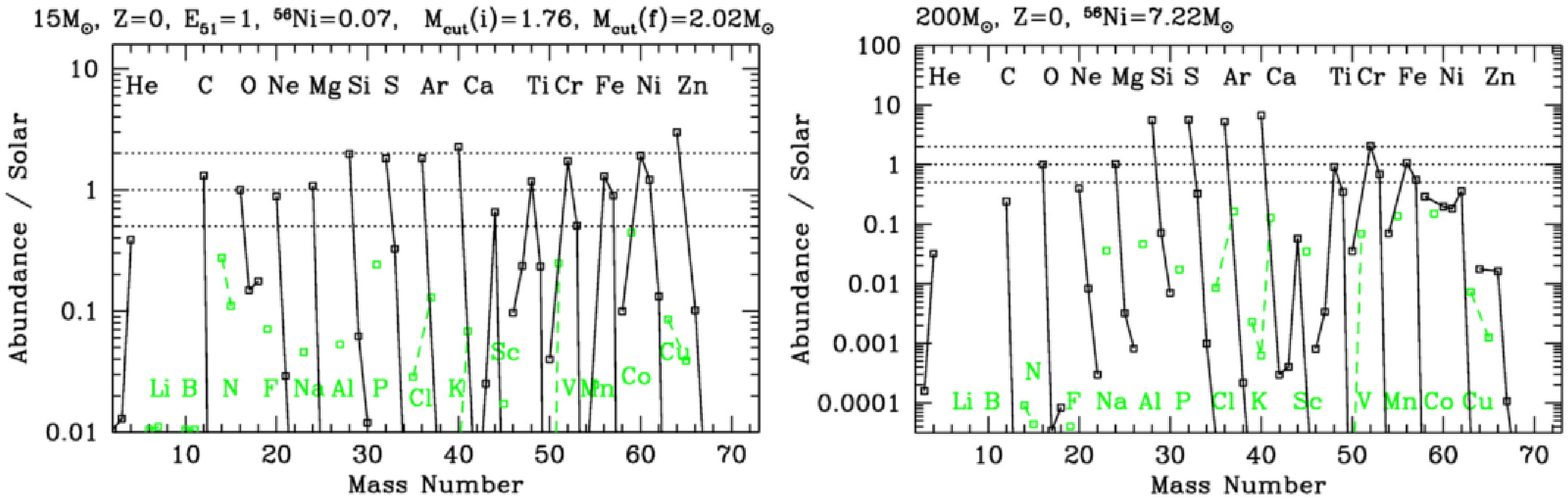}
 \caption{The abundance pattern in the ejecta (after radioactive decay) of a $15~\mathcal{M_{\odot}}$ (left) and a $200~\mathcal{M_{\odot}}$ (right) model star with initial zero metallicity, normalized to the solar $^{16}$O abundance.  Isotopes of even-Z (black) and odd-Z (green) elements are connected by solid and dashed lines, respectively.  Note that the two patterns are displayed on different scale.  Here, Si and S are clearly over-abundant with respect to C and O for the $200~\mathcal{M_{\odot}}$ star, while, e.g., the Zn/Fe ratio is significantly lower.  Adopted from \textcite{un02}.}
   \label{figccvspi}
\end{center}
\end{figure}

\subsubsection{Pair-instability supernovae: $ 140 \le m/\mathcal{M_{\odot}} < 260$}
In contrast to the core collapse SNe, the explosion mechanism for PISNe is well understood and there is no issue with fallback (Heger and Woosley, 2002\nocite{hw02}; Fowler and Hoyle, 1964\nocite{fowler64}).  As a consequence, following the remark above, PISN yields may therefore be considered more reliable than the yields of core collapse SNe, apart from the effects of rotation and magnetic fields.  Due to the very small neutron excess in their interiors \citep{hw02}, stars exploding as PISNe should exhibit a pronounced odd-even effect in which the ejecta show particularly low abundance ratios of odd- compared to even-$Z$ elements.  Interestingly, this effect is less distinct in the models by Umeda and Nomoto (2002\nocite{un02}, see Fig. \ref{figccvspi}). Furthermore, the lack of excess neutrons in addition to less-rapid expansion timescales during the explosion, inhibit the production of rapid n-capture elements \citep{hw02,un02}. Likewise, the production of s-process elements is inhibited in PISNe.  As a consequence of the extended oxygen-burning region, elements such as Si and S are significantly over-produced \citep{un02} leading to highly super-solar ratios of, e.g., $[\mathrm{Si}/\mathrm{O}]$  and $[\mathrm{S}/\mathrm{C}]$. The combination of these and other abundance characteristics, such as low  $\mathrm{Mg}/\mathrm{Ca}$ ratios (true over the majority of the PISN mass range) and very low ratios of $\mathrm{Co,Zn}/\mathrm{Fe}$ make up a chemical signature unique to the PISNe (Fig. \ref{figccvspi}).

\subsubsection{Supermassive core collapse supernovae: $260 \le m/\mathcal{M_{\odot}} \lesssim 10^5$}
Rotating stars in the mass range $260\lesssim m/\mathcal{M_{\odot}} \lesssim 10^{5}$ may escape direct black hole formation if an accretion disk is formed around the central remnant \citep{ss02}.  If so, freshly synthesized material is allowed to be ejected in jets towards the polar direction, after core collapse. In their $500-1000~\mathcal{M_{\odot}}$ model stars, \textcite{ohkubo06} showed that $>20\%$ of the mass undergoes explosive Si-burning with large amounts of Fe-peak elements being produced, in particular elements heavier than Fe, such as Co and Zn. On the other hand, relatively little intermediate-mass elements are synthesized and $[\mathrm{C},\alpha/\mathrm{Fe}]\ll0$ are realized in the ejecta.  In a sense, the yields of $260-10^{5}~\mathcal{M_{\odot}}$ core collapse SNe anticorrelates with those of normal core collapse SNe, especially the weak SNe. Overall, [X/Fe] decreases with heavier elements X for normal (weak) SNe while the opposite is true for $260-10^{5}~\mathcal{M_{\odot}}$ SNe.  Moreover, the odd-even effect in the ejecta of the $500-1000~\mathcal{M_{\odot}}$ SNe studied by \textcite{ohkubo06} is pronounced as compared to their lighter cousins.  We will further discuss the contribution from various SNe to the individual chemical abundance patterns of the most metal-poor Galactic halo stars in Sec. \ref{fingerprint}.

\subsection{Metal transport and mixing}\label{mixing}
The implications of ISM mixing on possible anomalies in stellar chemistry have been recognized for a long time \citep{searle77,norris78}.
So how did the first metals mix into the IGM, and into the gas clouds that would eventually give rise to the first Pop~II stars? This is an extremely complex problem, and we are only now beginning to get a handle on it. Again, progress relies on advances in theoretical modeling and numerical simulation techniques in combination with accurate and precise observations. The early metal dispersal problem can be separated into large-scale, coarse-grain mixing, and subsequent fine-grain mixing on small scales. We discuss these two aspects in turn.

\subsubsection{Early transport: The supernova explosion}
It is likely that Pop~III.1 stars will encounter different fates,
according to their mass (see Table~\ref{fates}), including both
core-collapse and pair-instability SNe. We will here illustrate
the feedback from such explosions by discussing the evolution of
a PISN remnant in greater detail. The overall hydrodynamics, however,
would be very similar for the case of a hypernova, where
metal yields are much lower, but explosion energies can be similarly
large (Umeda and Nomoto, 2002).

\begin{figure}
\begin{center}
\includegraphics[width=10.0cm,height=10.0cm]{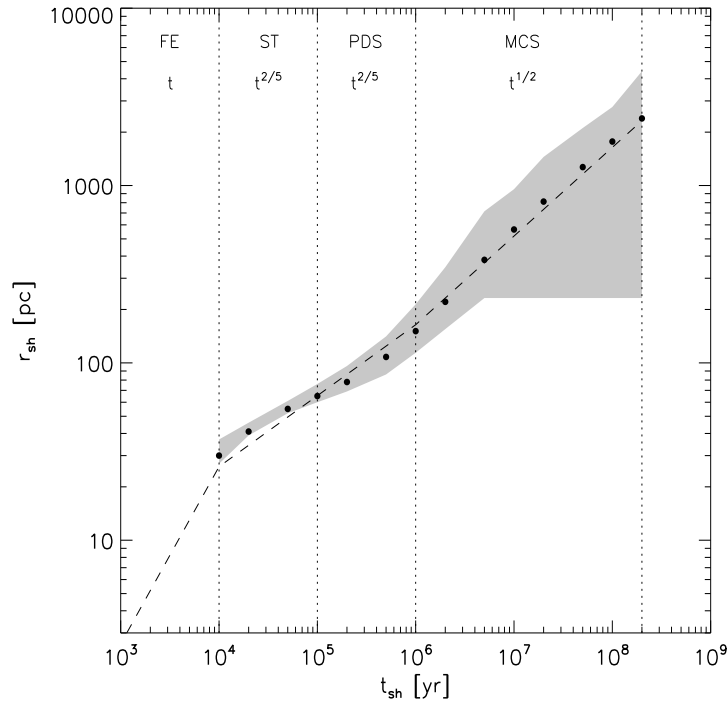}
\caption{Evolution of a Pop~III PISN remnant, from $z\simeq 20$ when the Pop~III.1 star explodes, to $z\simeq 12$ when the shock finally stalls (adopted from Greif {\it et al.}, 2007). The remnant lasts about $200~\rm{Myr}$. The black dots indicate the mass-weighted mean shock radius according to a simulation, while the dashed line shows the analytic solution. The final mass-weighted mean shock radius is $2.5~\rm{kpc}$. The shaded region shows the radial dispersion of the shock. Its increase towards late times indicates the growing deviations from 
spherical symmetry when the blast wave runs into the cosmic web, surrounding
the host minihalo.}
\label{phases}
\end{center}
\end{figure}

After the death of a Pop~III.1 star as a PISN, provided that the progenitor had the appropriate mass (see Table \ref{fates}), the resulting blast wave goes through four evolutionary stages \citep{greif07}.  At very early times, the remnant enters the free-expansion (FE) phase and propagates nearly unhindered into the surrounding medium. It expands with a constant velocity $v_{\rm{ej}}$ given by $v_{\rm{ej}}^{2}=2E_{\rm{sn}}/M_{\rm{ej}}$, where $M_{\rm{ej}}=M_{*}$, as is appropriate for the complete disruption encountered in a PISN. The duration of the FE phase is given by $t_{\rm{fe}}=r_{\rm{fe}}/v_{\rm{ej}}$, or:
\begin{equation}
t_{\rm{fe}}=r_{\rm{fe}}\sqrt{\frac{M_{\rm{ej}}}{2E_{\rm{sn}}}}\mbox{\ ,}
\label{tfe}
\end{equation}
where $r_{\rm{fe}}$ is the radius at which the swept-up mass equals the mass of the original ejecta, i.e:
\begin{equation}
r_{\rm{fe}}=\left(\frac{3XM_{\rm{ej}}}{4\pi m_{\rm{H}}n_{\rm{H}}}\right)^{1/3}\mbox{\ ,}
\label{rfe}
\end{equation}
where $X=0.76$ is the primordial mass fraction of hydrogen, $m_{\rm{H}}$ the proton mass, and $n_{\rm{H}}$ the number density of hydrogen nuclei.  After $t_{\rm{fe}}$, the inertia of the swept-up mass becomes important, and the shock undergoes a transition to the Sedov-Taylor (ST) phase \citep{sedov46,taylor50}. Typical values are $r_{\rm{fe}}\lesssim 20~\rm{pc}$ and $t_{\rm{fe}}\lesssim 10^{4}~\rm{yr}$ (see Figure \ref{phases}).

The remnant next enters its ST phase. The photoheating prior to the explosion leads to a much reduced circumstellar density, with an almost flat profile of $n_{\rm{H}}\simeq0.5~\rm{cm}^{-3}$ in the vicinity of the progenitor star \citep{greif07}.  The SN blast wave approaches $r_{\rm{vir}}/2$ after about $10^{5}~\rm{yr}$, catching up with the previously established photoheating shock, where the outlying density profile becomes isothermal. Subsequently, the ST phase ends and the remnant undergoes a second transition.
At the high temperatures behind the shock, cooling is due to H and He collisional ionization, excitation and recombination processes, bremsstrahlung and the inverse Compton (IC) effect. IC cooling and bremsstrahlung are important at high temperatures, whereas H/He line cooling becomes significant below $10^{6}~\rm{K}$.
Radiative losses affect the energetics of the SN remnant after about $10^{5}~\rm{yr}$. At this point the shocked gas separates into a hot, interior bubble with temperatures above $10^{6}~\rm{K}$, and a dense shell at $10^{4}~\rm{K}$, bounded by a high pressure gradient, thus leading to a multi-phase structure, which remains intact for $\lesssim 10~\rm{Myr}$ \citep{greif07}.  With energy conservation no longer valid, the ST phase ends.

After the ST phase ends, the pressurized, interior bubble drives an extremely dense shell, often called a pressure-driven snowplow (PDS). This new phase can be described with an equation of motion:
\begin{equation}
\frac{d\left(M_{\rm{sw}}v_{\rm{sh}}\right)}{dt}=4\pi r_{\rm{sh}}^{2}P_{\rm{b}}\mbox{\ ,}
\label{pdseq}
\end{equation}
where $P_{\rm{b}}$ is the pressure of the hot, interior bubble. Since $M_{\rm{sw}}\propto r_{\rm{sh}}$ in an $r^{-2}$ density profile, and $P_{\rm{b}}\propto r_{\rm{sh}}^{-5}$ in the adiabatically expanding interior, one can solve the above equation with a power-law of the form $r_{\rm{sh}}\propto t_{\rm{sh}}^{2/5}$, yielding the same scaling as the ST solution.
When does the PDS phase end? The pressure directly behind the shock after $10^{5}~\rm{yr}$ can be estimated to be $P_{\rm{b}}/k_{\rm{B}}\simeq 3\times 10^{6}~\rm{K}~\rm{cm}^{-3}$ \citep{greif07}.  With $P_{\rm{b}}\propto r_{\rm{sh}}^{-5}$ and $r_{\rm{sh}}\propto t_{\rm{sh}}^{2/5}$, one finds $P_{\rm{b}}\propto t_{\rm{sh}}^{-2}$, implying that after roughly $1~\rm{Myr}$ the interior pressure has dropped to $P_{\rm{b}}/k_{\rm{B}}\simeq 3\times 10^{4}~\rm{K}~\rm{cm}^{-3}$. At this point pressure equilibrium between the hot interior and the dense shell has been established, and the shock is driven solely by its accumulated inertia.
With the pressure gradient no longer dominant, the SN remnant is driven solely by the accumulated inertia of the dense shell, and becomes a momentum-conserving snowplow (MCS). The shock position as a function of time can be obtained by solving Eq. (\ref{pdseq}) in the absence of a pressure term. Since the shock has not yet propagated beyond the surrounding $r^{-2}$ density profile, this yields an initial scaling of $r_{\rm{sh}}\propto t_{\rm{sh}}^{1/2}$.
At later times, the shock finally leaves the host halo and encounters neighbouring minihalos and underdense voids. Figure~\ref{phases} summarizes the 4-stage remnant evolution.

\begin{figure}[t]
\begin{center}
 \includegraphics[width=3.6in]{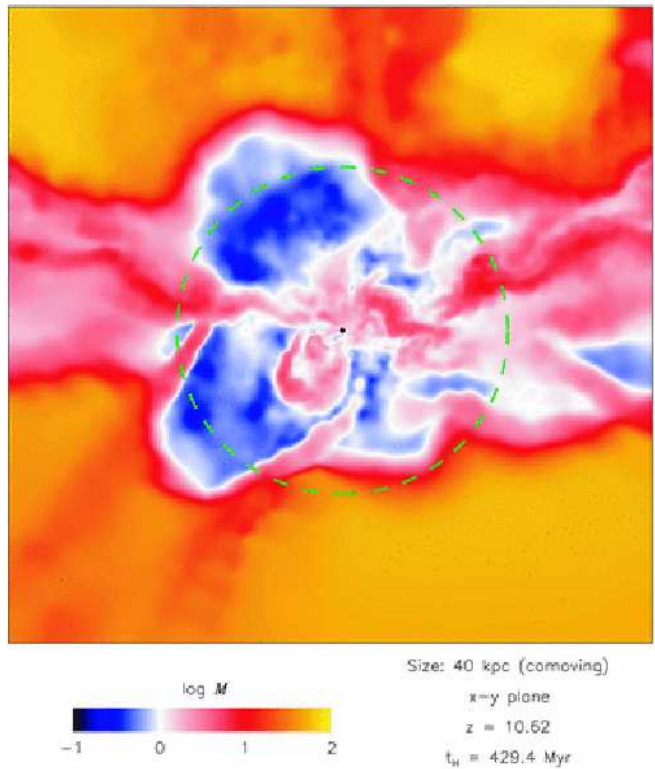} 
 \caption{Turbulent flows inside the first galaxies. Specifically,
the host halo here has a virial mass of $\sim 10^8 \mathcal{M_{\odot}}$, and
collapses at $z\simeq 10$.
Shown is the Mach number in the central few kpc of the galaxy.
The dashed line denotes the virial radius at $\simeq 1$\,kpc 
from the center. The Mach number approaches unity at the virial shock,
where gas accreted from the IGM is heated to the virial temperature.
Inflows of cold gas along filaments are supersonic and generate a high
level of turbulence at the center of the halo, where typical Mach numbers
are between 1 and 5.
From Greif {\it et al.} (2008).}
   \label{Fig4_VB}
\end{center}
\end{figure}

Mixing of enriched material with gas in existing star-forming minihalos is generally inefficient, indicating that the dispersal of metals can only occur via expulsion into the IGM. The bulk of the SN shock propagates into the voids surrounding the host halo, and chemical enrichment proceeds via the same channel \citep{pieri07}. 
During the first $10~\rm{Myr}$, the initial stellar ejecta expand adiabatically and apparently do not mix with the surrounding material. In reality, however, the high electron mean free path behind the shock leads to heat conduction and gas from the dense shell evaporates into the hot, interior bubble \citep{gull73}.  The onset of Rayleigh-Taylor (RT) and Kelvin-Helmholtz (KH) instabilities is predicted to mix the metals efficiently with primordial material evaporated from the dense shell, reducing the metallicity of the interior by a factor of a few to at most one order of magnitude \citep{madau01}.

Due to inefficient cooling, the evolution of the metal-enriched, interior bubble is governed by adiabatic expansion, and preferentially propagates into the cavities created by the shock. Once the shock leaves the host halo and becomes highly anisotropic, the interior adopts the same behavior and expands into the voids surrounding the host halo in the shape of an `hourglass', with a maximum extent similar to the final mass-weighted mean shock radius. The interior becomes substantially mixed with the initial stellar ejecta.
When the shock finally stalls, the interior bubble is in pressure equilibrium with its surroundings, but stays confined within the dense shell. To investigate the importance of RT instabilities in this configuration, one can estimate the mixing length $\lambda_{\rm{RT}}$ for large density contrasts between two media according to 

\begin{equation}
\lambda_{\rm{RT}}\simeq2\pi gt_{\rm{sh}}^{2}\mbox{\ ,}
\label{rt}
\end{equation}
where $g$ is the gravitational acceleration of the host halo. Mixing between the dense shell and the interior bubble takes place on scales $\lesssim 10~\rm{pc}$ in the course of a few $10~\rm{Myr}$. Such mixing, however, is generally inefficient, and that much larger potential wells must be assembled to recollect and mix all components of the shocked gas \citep{greif08}.  Turbulence arising in the virialization of the first galaxies could be an agent for this process \citep{greif08,wa08}. Recent simulations of first galaxy formation indicate that cold accretion streams deeply penetrate into the center of the corresponding potential well, giving rise to high Mach number, turbulent flows (see Fig.~\ref{Fig4_VB}).  

\begin{figure}[t]
\begin{center}
 \includegraphics[width=3.6in]{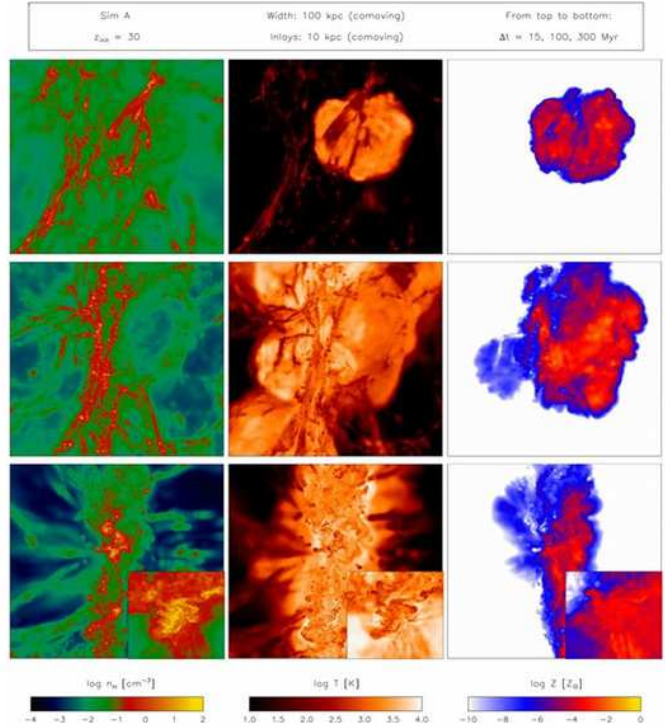}
 \caption{Chemical enrichment in the wake of the first supernovae.
Shown are the gas density, temperature, and metallicity in the central
100 kpc (comoving) of the simulation box ({\it left to right}).
{\it From top to bottom}: time series showing the simulation 15, 100, and
300~Myr after the SN explosion. The inlays show the central 10 kpc (comoving)
of the emerging galaxy. The metals are initially distributed by the bulk
motion of the SN remnant, and later by turbulent motions induced by 
photoheating from other stars and the virialization of the galaxy itself.
The gas within the newly formed galaxy is already substantially enriched.
From Greif {\it et al.} (2010).}
   \label{Fig5_VB}
\end{center}
\end{figure}

\begin{figure}[t]
\begin{center}
 \includegraphics[width=4.5in]{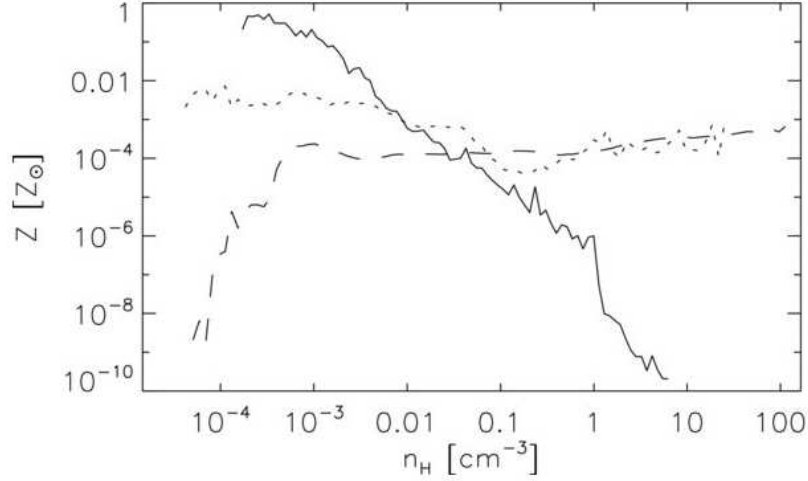} 
 \caption{Metallicity-density relation in the early Universe.
Mass-weighted, average metallicity as a function of gas
density, shown 15 ({\it solid line}), 100 ({\it dotted line}),
and 300 Myr ({\it dashed line}) after the SN explosion.
At early times, a distinct correlation is evident: underdense
region are highly enriched, while overdense region remain
largely pristine. Once the potential well of the galaxy assembles,
metal-rich gas becomes dense and the relation flattens.
From Greif {\it et al.} (2010).}
   \label{Fig6_VB}
\end{center}
\end{figure}

The minimum mass necessary to recollect the hot and underdense post-shock gas residing at $T\sim 10^{3}~\rm{K}$ and $n_{\rm{H}}\sim 10^{-2.5}~\rm{cm}^{-3}$ is at least $M_{\rm{vir}}\gtrsim 10^{8}~\mathcal{M_{\odot}}$. The final topology of metal enrichment could be highly inhomogeneous, with pockets of highly enriched material on the one hand, and regions with a largely primordial composition on the other hand. Recent simulations have reached a level of realism, in terms of taking
into account the radiative feedback from individual Pop~III stars, and modeling
the transport of metals from the first SNe, to give us a much better
understanding of early enrichment, at least on small scales (1~Mpc comoving).
In particular, they provide us with hints on the level and homogeneity
of metals inside the first galaxies, where second generation star
formation will occur \citep{wa08,greif10}. It
is interesting that even a single SN in one of the progenitor minihalos
already leads to a significant metallicity inside the first galaxies
(see Fig.~\ref{Fig5_VB}). Specifically, if such a minihalo hosted a PISN,
the resulting average enrichment would already be $\sim 10^{-3}~\mathcal{Z_{\odot}}$
(see Fig.~\ref{Fig6_VB}). If this is true, the {\it JWST} is unlikely to 
see any truly metal-free galaxies, even in very deep fields.

To further address the fine structure of metal enrichment, we now discuss the
physics of fine-grain mixing.

\subsubsection{Late transport: Turbulence}
Eventually, the characteristics of the SN blast wave as a dynamical system will be lost as it is overtaken by the random motions of the surrounding medium.  These gas motions are maintained by constant energy injections into the medium, e.g., by stellar winds, SN explosions, and by the conversion of potential to kinetic energy during the collapse of star-forming clouds and are generally turbulent in nature.  At early cosmic times in particular, gas motions were also generated by dark matter halo interaction.  

There is no existing general theory for hydrodynamic turbulence\footnote{See: http://www.claymath.org/millennium/Navier\_Stokes\_Equations/}, particularly if the turbulent flow is inhomogeneous and non-isotropic.  However, from a statistical viewpoint, turbulence may be seen as a random walk process and the transport of matter in a turbulent medium can be described as a diffusion process \citep{taylor21}, similar to molecular diffusion.  The derivation outlined below of the relation between time and the mean displacement of a fluid element in a turbulent medium follows that of \textcite{choudhuri98}.

Assume that the three dimensional displacement, $\mathbf{x}$, of a fluid element after time $t'=t$ is

\begin{equation}
\mathbf{x}(t) = \int\limits_0^t \mathbf{v}_{\mathrm{L}}(t')\mathrm{d}t',
\label{displ}
\end{equation}

\noindent
where $\mathbf{v}_{\mathrm{L}}(t')$ is the (Lagrangian) fluid velocity at the position of the element at time $t'$. The mean square displacement is thus given by 

\begin{equation}
\langle x^2(t)\rangle = \langle \mathbf{x}(t) \! \cdot \! \mathbf{x}(t) \rangle =  \int\limits_0^t \!\! \int\limits_0^t \langle \mathbf{v}_{\mathrm{L}}(t') \! \cdot \! \mathbf{v}_{\mathrm{L}}(t'') \rangle \mathrm{d}t'\mathrm{d}t''.
\label{displ2}
\end{equation} 

\noindent
This follows since, 

\begin{equation}
\langle \int\limits_0^t \mathbf{v}_{\mathrm{L}} \mathrm{d}t' \! \cdot \!  \int\limits_0^t \mathbf{v}_{\mathrm{L}} \mathrm{d}t'' \rangle = \langle \int\limits_0^t \!\! \int\limits_0^t \mathbf{v}_{\mathrm{L}} \! \cdot \! \mathbf{v}_{\mathrm{L}} \mathrm{d}t' \mathrm{d}t'' \rangle =  \int\limits_0^t \!\! \int\limits_0^t \langle \mathbf{v}_{\mathrm{L}} \! \cdot \! \mathbf{v}_{\mathrm{L}} \rangle \mathrm{d}t'\mathrm{d}t''.
\label{swap}
\end{equation} 

\noindent
The function $\langle \mathbf{v}_{\mathrm{L}}(t') \! \cdot \! \mathbf{v}_{\mathrm{L}}(t'') \rangle$ is the velocity correlation function of a fluid element measured at times $t'$ and $t''$.  Assuming that the turbulence does not change character with time (i.e., steady-state), this correlation function may be expressed as a symmetric function, $\mathcal{R}$, of $t''-t'$ alone,

\begin{equation}
\langle \mathbf{v}_{\mathrm{L}}(t') \! \cdot \! \mathbf{v}_{\mathrm{L}}(t'') \rangle = \langle v^2 \rangle \mathcal{R}(t''-t'),
\label{vcorr}
\end{equation}  

\noindent
where $\langle v^2 \rangle$ denotes the mean square velocity.  At $t''=t'$, we have maximal correlation and $\mathcal{R}=1$, while for $\mid \! t''-t' \! \mid > 0$, the velocities associated with the fluid element at times $t'$ and $t''$ become increasingly uncorrelated and $\mathcal{R}$ approaches zero as $\mid \! t''-t' \! \mid \to \infty$.  From this definition of $\mathcal{R}$, it is possible to define a turbulent correlation time, $\tau_{\mathrm{corr}}$, such that

\begin{equation}
\tau_{\mathrm{corr}} = \int\limits_0^{+\infty}\mathcal{R}(t') \mathrm{d}t'.
\label{tauc}
\end{equation}  

Now, for times $t \gg \tau_{\mathrm{corr}}$, the limits of integration of the inner integral in Eq. (\ref{displ2}) can be expanded to $]-\infty,+\infty[$.  Making use of Eq. (\ref{tauc}) and the fact that $\mathcal{R}$ is symmetric, the mean square displacement may be written as 

\begin{equation}
\langle x^2(t)\rangle = \int\limits_0^t \int\limits_{-\infty}^{+\infty} \langle v^2 \rangle \mathcal{R}(t''-t') \mathrm{d}t'\mathrm{d}t'' = 2 \tau_{\mathrm{corr}} \langle v^2 \rangle \int\limits_0^t \mathrm{d}t'' = 6D_{\mathrm{turb}}t,
\label{displ22}
\end{equation} 

\noindent
where $2\tau_{\mathrm{corr}}\langle v^2 \rangle \equiv 6 D_{\mathrm{turb}}$. Equation (\ref{displ22}) describes, in a statistical sense, the passive transport of matter in a turbulent medium. The coefficient $D_{\mathrm{turb}}$ is identified as the turbulent diffusion coefficient and is given by

\begin{equation}
D_{\mathrm{turb}} = \tau_{\mathrm{corr}} \langle v^2 \rangle/3 = \frac{1}{3}\langle v^2 \rangle \! \int\limits_0^{+\infty} \mathcal{R}(t')\mathrm{d}t' = l_{\mathrm{corr}}v_{\mathrm{rms}}/3,
\label{dturb}
\end{equation} 

\noindent
where $l_{\mathrm{corr}} = v_{\mathrm{rms}} \tau_{\mathrm{corr}}$ denotes the turbulent correlation length and $v_{\mathrm{rms}}=\sqrt{\langle v^2 \rangle}$ is the root mean square velocity. The above expressions for the mean square diplacement and $D_{\mathrm{turb}}$ corresponds exactly to those of molecular diffusion.

The continuous macroscopic ``stirring" of the gaseous medium is paramount to the transport and mixing of metals over large distances.  As compared to the microscopic, molecular diffusion coefficient, the turbulent diffusion coefficient is typically several orders of magnitude larger (Bateman and Larson, 1993; Oey, 2003\nocite{bl93,oey03}; cf. Pan and Scalo, 2007\nocite{pan07}). Hence, in the presence of turbulence, the mixing of gas is greatly amplified.  However, it should be kept in mind that turbulence still operates on a macroscopic level and only molecular diffusion is able to drive the mixing on the smallest scales. Turbulence merely increases the area-to-volume ratio so that molecular diffusion effectively becomes more efficient. The situation is similar to that of the human lung. Without the hundreds of millions of alveoli which drastically increase the contact area to the fine network of blood-vessels, the small volume of our lungs would make it very hard to breathe.

\subsection{Transition to Population II star formation}\label{pop2trans}
Chemical enrichment by the first SNe is among the most important processes in the formation of the first galaxies.  Efficient cooling by metal lines and thermal dust emission regulate the temperature of Pop~II star-forming regions in the first galaxies. The concept of a `critical metallicity' has been introduced\footnote{There may exist a critical metallicity which distinguishes the 
formation of individual stars from star clusters. This may relate to the critical column density that a collapsing cloud must reach
through effective cooling before the cluster can form, and may explain therefore why all globular clusters regardless of age have a metallicity above [Fe/H] $\sim -2.5.$}
to characterise the transition of the star-formation mode from predominantly high-mass, Pop~III, to low-mass Pop~II stars \citep{omukai00,bromm01,ss05}. Currently, two competing models for the \mbox{Pop\,III -- Pop\,II} transition are discussed: {\it (i)} atomic fine-structure line cooling \citep{bl03,sash06} and {\it (ii)} dust-induced fragmentation \citep{omukai05,schneider06,tsuribe06,clark08,dopcke11}.  Within the fine-structure model, C\,II and O\,I have been suggested as the main coolants \citep{bl03}, such that low-mass star formation can occur in gas that is enriched beyond critical abundances of $\mbox{[C/H]}_{\rm crit}\simeq -3.5\pm 0.1$ and $\mbox{[O/H]}_{\rm crit}\simeq -3 \pm 0.2$. For the atomic phase, one can robustly assume that carbon is singly ionized due to the strong UV background just below the Lyman limit predicted for the early stages of reionization.  The dust-cooling model \citep{omukai05}, on the other hand, predicts critical abundances that are typically smaller by a factor of $10-100$.  \textcite{sash06} have investigated the additional cooling provided by Si\,II and Fe\,II fine-structure lines, finding that their inclusion only leads to small corrections.  Dust cooling would only become important at higher densities, without being able to influence the evolution of primordial gas close to the characteristic, or `loitering', state \citep{bl04} with a temperature and density: $T_{\rm char}\simeq 200$\,K and $n_{\rm char}\simeq 10^4$\,cm$^{-3}$. Ultimately, this theoretical debate has to be decided empirically by probing the abundance pattern of extremely metal-poor, second-generation stars.

\begin{figure}[t]
\begin{center}
 \includegraphics[width=3.6in]{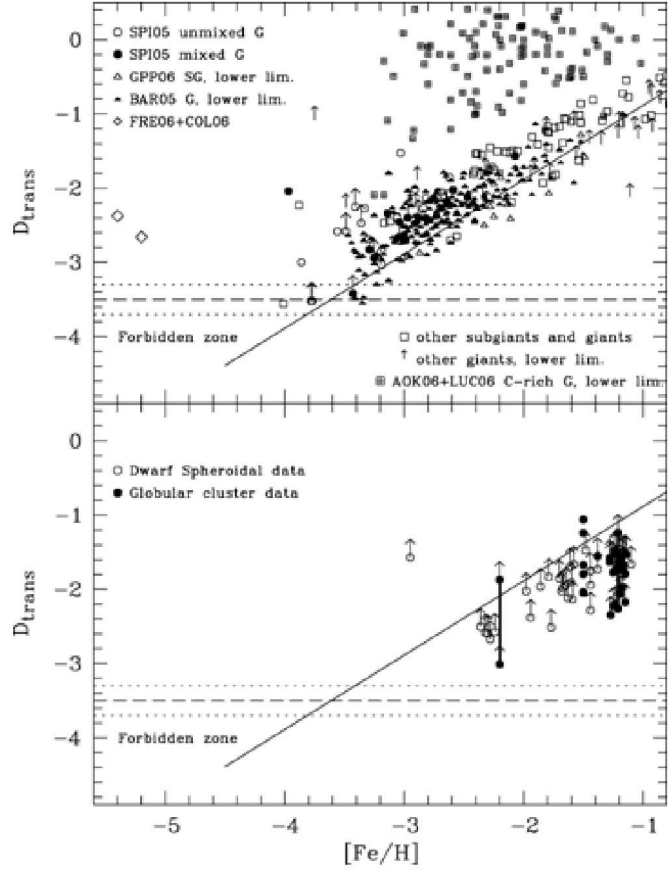} 
 \caption{Testing theories for the critical metallicity
with stellar archaeology. Shown is the transition discriminant,
$D_{\rm trans}$, for metal-poor stars
collected from the literature as a function of Fe abundance.
{\it Top panel}: Galactic halo stars.
{\it Bottom panel}: Stars in dSph galaxies and globular clusters.
G indicates giants, SG subgiants.
The critical limit is marked with a dashed line. The dotted
lines refer to the uncertainty. References for the data can be found
in the original journal paper.
From Frebel {\it et al.} (2007).}
   \label{Fig7_VB}
\end{center}
\end{figure}

To facilitate any such comparison with stellar abundance data, it is convenient to introduce a criterion \citep{frebel07} for low-mass star formation that combines cooling due to C\,II and O\,I.  An intuitive way to arrive at the so-called
`Frebel criterion' (following the
detailed calculation in Bromm and Loeb, 2003\nocite{bl03}) is to consider the balance between fine-structure line cooling and adiabatic compressional heating:
\begin{equation}
\Lambda_{\rm tot}=\Lambda_{\rm C\,II} + \Lambda_{\rm O\,I}\gtrsim \Gamma_{\rm ad}
\mbox{\ ,}
\end{equation}
where all terms have to be evaluated at $n_{\rm char}$, $T_{\rm char}$.
Heating due to adiabatic compression is given by
\begin{equation}
\Gamma_{\rm ad}\simeq 1.5 n_{\rm char}\frac{k_{\rm B}T_{\rm char}}{t_{\rm ff}}
\simeq 2\times 10^{-23}\mbox{ erg s$^{-1}$ cm$^{-3}$}
\mbox{\ ,}
\end{equation}
where $t_{\rm ff}\sim 5\times 10^5$\,yr is the free-fall time at the
characteristic state.  The cooling terms are \citep{sp05}:
\begin{equation}
\Lambda_{\rm O\,I}\simeq
2\times 10^{-20}\mbox{ erg s$^{-1}$ cm$^{-3}$}\left(\frac{n_{\rm O}}{n_{\rm H}}\right)/
\left(\frac{n_{\rm O}}{n_{\rm H}}\right)_{\odot}\mbox{\ ,}
\end{equation}
and
\begin{equation}
\Lambda_{\rm C\,II}\simeq
6\times 10^{-20}\mbox{ erg s$^{-1}$ cm$^{-3}$}\left(\frac{n_{\rm C}}{n_{\rm H}}\right)/
\left(\frac{n_{\rm C}}{n_{\rm H}}\right)_{\odot}\mbox{\ .}
\end{equation}
At the loitering state, C\,II is already in local thermodynamic equilibrium (LTE), leading to an approximate
scaling \citep{sp05} of the cooling rate: $\Lambda_{\rm C\,II}\propto
n_{\rm char}\exp{\left(-92{\rm \, K}/T_{\rm char}\right)}$, 
whereas O\,I is still in 
non-LTE, leading to:
$\Lambda_{\rm O\,I}\propto
n_{\rm char}^2\exp{\left(-230{\rm \, K}/T_{\rm char}\right)}$.
Here, LTE refers to the high-density case, where collisions dominate the
atomic level populations, whereas non-LTE denotes the opposite situation,
where radiative transitions dominate.
One can combine these expressions as follows:
\begin{equation}
10^{\mbox{[C/H]}} + 0.3\times 10^{\mbox{[O/H]}}\gtrsim 10^{-3.5}
\mbox{\ ,}
\end{equation}
leading to the `transition discriminant':
\begin{displaymath}
D_{\rm trans}\equiv \log_{10}\left(
10^{\mbox{[C/H]}} + 0.3\times 10^{\mbox{[O/H]}}\right)\mbox{\ ,}
\end{displaymath}
such that low-mass star formation requires $D_{\rm trans}>D_{\rm
trans, crit} \simeq -3.5\pm 0.2$.
Using this new quantity has the important
advantage that one can now accommodate an inhomogeneous data set of
stellar abundances, including stars where both carbon and oxygen
abundances are known, or even only one of them.

However, this critical gas metallicity is still poorly determined, possibly depending on environmental parameters, such as the presence of a soft UV (Lyman-Werner) radiation background \citep{safranek10}. It is not even clear if there exists such a sharp transition.  Since the enrichment from even a single PISN by a very massive Pop~III star likely leads to metallicities of $Z > 10^{-3}~\mathcal{Z_{\odot}}$ (see, e.g., Greif {\it et al.}, 2007, 2010\nocite{greif07,greif10}; Wise and Abel, 2008\nocite{wa08}, see also Karlsson {\it et al.}, 2008\nocite{karlsson08}), in excess of {\it any} predicted value for the critical metallicity, these arguments may be somewhat academic. The characteristic mass of prestellar gas clumps is likely determined by a number of physical processes (e.g., turbulence and, possibly, dynamo-amplified primordial magnetic fields) other than radiative cooling. The overall effect of gas metallicity on star formation may well be limited \citep{jappsen09}.

Star formation in the first galaxies may be influenced by yet another factor: the CMB likely played an important role in setting the minimum gas temperature in the high redshift Universe \citep{larson98}.  The radiation temperature of the CMB is $T_{\rm CMB} = 2.728\mbox{\ K} (1+z)$  such that $T_{\rm CMB}$ is greater than about 30~K at $z > 10$.  Hence, early molecular gas clouds cannot cool to the characteristic temperatures of present-day star-forming regions ($\sim 10$ K) even if molecules such as CO and H$_2$O act as efficient coolants.  If the minimum gas temperature is a critical physical quantity that determines the characteristic masses of stars, then the transition from high-mass Pop~III to the lower-mass Pop~II mode may be gradual and regulated by the CMB temperature \citep{schneider10}.

Anyhow, It is intriguing to see how well the fine-structure model, based
on C and O, does in terms of accounting for all existing metal-poor stellar
data (see Fig.~\ref{Fig7_VB}). While the criterion has been criticised on theoretical
grounds \citep{omukai05,jappsen07}, the simple model matches the observed
lower threshold very well, also in the regime below $[\mathrm{Fe}/\mathrm{H}]\sim -4$. Only one probable exception is currently known. With an Fe abundance of $[\mathrm{Fe}/\mathrm{H}]=-4.7$ (1D, LTE analysis), the star $\mathrm{SDSS}~\mathrm{J}102915+172927$ is among the most iron-poor stars observed to date \citep{caffau11}. Interestingly, it appears not to be particularly enhanced in carbon, with a measured upper limit on the C abundance of $[\mathrm{C}/\mathrm{H}] \le -3.8$ \citep{caffau11}. Together with a presumed O-to-Fe enhancement of $[\mathrm{O}/\mathrm{Fe}]=+0.6$ (typical for ``C-normal'', metal-poor stars), $\mathrm{SDSS}~\mathrm{J}102915+172927$ falls into the `forbidden zone' in Fig. \ref{Fig7_VB}. This implies the presence of a dust-cooling channel in $Z\sim10^{-5}~\mathcal{Z_{\odot}}$ star-forming gas \citep{klessen12,schneider12}. Since the majority of stars below $[\mathrm{Fe}/\mathrm{H}]=-4$ are enhanced in the CNO elements and fulfill the $D_{\rm trans}$ criterion, both cooling channels may in fact be viable. If so, the fraction of stars below the line-cooling threshold would then be a measure (although complex) of the relative importance of dust for the transition to Pop II star formation. Future observations should aim to constrain this fraction. Recently, evidence for carbon enhancement toward low [Fe/H] has also been discovered in an extremely metal-poor damped Lyman-$\alpha$ (DLA) system at $z\simeq 2.3$ \citep{cooke11}, opening up a new window into the conditions for early star formation. With a view to upcoming {\it JWST} observations, it is important to formulate similar empirical tests for alternative models for the Pop~III -- Pop~II transition that exploit dust cooling, say.

\section{Tracers of pre-galactic metal enrichment -- Concepts}\label{fingerprint} 
This review seeks to identify the chemical {\it signatures} of Pop III stars in the surface abundances of the oldest stellar populations. As already noted, we make a clear distinction between a chemical {\it fingerprint} and a chemical {\it signature} to 
readily distinguish between microscopic and (a series of) macroscopic processes respectively. We now discuss these in turn.

\subsection{Chemical fingerprints (microscopic processes)}
The many elements that can be measured in metal-poor stars provide evidence of physical processes that we are only now beginning to read (e.g., Sneden {\it et al.}, 2008\nocite{sneden08}). This approach has a long tradition. The celebrated papers of \textcite{b2fh57} and \textcite{cameron57} demonstrated how to infer the processes involved in stellar chemistry from detailed abundance measurements in meteorites and stars. This tradition can be traced back even further in the early interpretation of the Solar System abundances (see Suess and Urey, 1956\nocite{suess56}). In a seminal paper, \textcite{cameron73} showed how the meteoritic record can be separated into distinct neutron and proton capture processes. Remarkably, as we mention below, the fingerprints of these processes are observed directly in a substantial fraction of the most metal poor stars. As a general rule, we have a better understanding of fingerprints when compared to chemical
signatures. For example, while there is still no {\it in situ} evidence of r-process nucleosynthesis in supernovae, the main reaction networks that give rise to different heavy elements for a given neutron flux and number density are fairly well understood. How the s-process elements are synthesized in quiescently evolving stars is also well understood \citep{busso99}. After \textcite{sneden08}, we emphasize that the stellar record is not complete at the present time. There are no elements measured from Ag to Ba in any n-capture rich stars, which will be needed to rule out alternative n-capture models in the most metal poor stars (see Arnould {\it et al.}, 2007\nocite{arnould07}). Thus, chemical fingerprints are concerned with distinct hallmarks of a microscopic process that is broadly understood in terms of a limited set of physical parameters (e.g. flux, density).

\subsection{Chemical signatures (macroscopic processes)}
Most of this review is concerned with identifying and interpreting chemical signatures of the first stars and subsequent generations.
This is inherently more difficult. In discussing these signatures, we are largely dependent on the yield calculations from 
complex hydrodynamical simulations (Sec. \ref{nucleosynthesis}). The calculated yields are dependent on numerous factors, including stellar mass, metallicity, rotation, black-hole infall, and so forth. The numerical models make very detailed predictions up to 
atomic masses of $Z\approx 30$, which falls short of the neutron-capture elements (cf. Pignatari {\it et al.}, 2008\nocite{pignatari08}), so the predictions are limited at the present time. But these models predict distinct chemical signatures. For example, in intrinsically faint supernovae, \textcite{un03} find strong CNO enhancements compared to Fe due to mixing and fallback onto the black hole. In contrast, hypernova models predict strong Zn and Co as compared to Cr and Mn (e.g., Tominaga {\it et al.}, 2007\nocite{tominaga07}).

Apart from the underlying signatures of the preceding stellar generation(s), there are likely to be many macroscopic processes that influence the observed abundance pattern of metal-poor stars.  Stars that are born into binary pairs can be enriched by mass transfer from the primary donor, either through slow winds or Roche-lobe overflow. A significant fraction of the most metal-poor Galactic halo stars show evidence of mass transfer from a companion that is no longer directly visible, as we discuss below (e.g., Fujimoto {\it et al.}, 2000\nocite{fujimoto00}; Suda {\it et al.}, 2004\nocite{suda04}). In principle, this can complicate our interpretation of element groups within the spectrum of a metal-poor star as the original chemical signature is diluted with a secondary one. Another complicating factor is mixing within the interstellar gas. The efficiency of the macroscopic (turbulent) mixing of the ISM can transform the underlying signature and affect the overall trend as a function of metallicity (i.e., [Fe/H]). 

Some signatures only reveal themselves by comparing the abundances of an ensemble of stars. One example is 
clustered data points in the abundance plane as a result of clustered star formation (which produces
chemically uniform star clusters).  As discussed in Sec. \ref{clustering}, this is an extremely important phenomenon
because it allows us to reconstruct star clusters that have dissolved in the distant past. Of no less importance,
it allows us to identify stars that are not subject to chemical anomalies due to mixing or binary mass transfer, thereby
providing us with a cleaner signal of the preceding generations of stars.

\subsection{First versus second generation stars}\label{nfc} 
In this review, we have identified the Pop III stars (see Sec. \ref{intro} for definition and Sec. \ref{form} for discussion on the formation of Pop III.1 and Pop III.2 stars) as being metal-free, which in the context of early star formation implies a metallicity $Z<Z_{\mathrm{crit}}$ (see Sec. \ref{pop2trans}). If nothing else is stated, we will continue to use the terms `primordial stars' and `first stars', as synonyms for the Pop III. Chronologically, some of these designations, in particular `first stars', may be slightly ambiguous since the formation of Pop III.2 stars likely occurred as a result of feedback, e.g.,  from Pop III.1 stars (Sec. \ref{form}). The Pop III.2 stars were therefore not, in a literal sense, the very first stars in the Universe, although they were the first stars to form locally.  From a chemical point of view, however, both Pop III.1 and Pop III.2 stars can readily be defined as `first stars'.

Second generation stars are defined as stars formed out of gas only enriched by the primordial stellar generation. As such, they may also be called the first generation Pop II stars. As primordial star formation occurred in different places at different times \citep{brook07,tornatore07,maio10}, second generation stars may not necessarily belong to the oldest stellar population in the present Universe, although a fraction of the oldest stars should belong to the second generation. They will, however, generally be very old and probe a vast variety of star formation sites and initial conditions, from the Galactic bulge to dwarf galaxies. 

The identification of second generation stars is a difficult task as there is no way of labeling the heavy element atoms as `synthesized in a metal-free star' or `synthesized in a metal-enriched star' at the present time. The hope is, however, that the more deficient in metals a star is, the more likely it is that it belongs to a generation immediately following that of the first.  While this should be true in a general sense, despite a significant spread in the predicted age-metallicity relation (e.g., Argast {\it et al.}, 2000\nocite{argast00}; Tumlinson, 2010\nocite{tumlinson10}), it may not be true in particular cases, as we shall see (see, e.g., Sec. \ref{constrimf} and \ref{haloclstr}).  In any event, the identification of extremely metal-poor (EMP) stars (defined as stars with an iron abundance of $[\mathrm{Fe}/\mathrm{H}]<-3$, Beers and Christlieb, 2005\nocite{bc05}) in the Milky Way and in other galaxies, is essential for probing the Pop III era.  

\begin{figure}[t]
\begin{center}
 \includegraphics[width=6.0in]{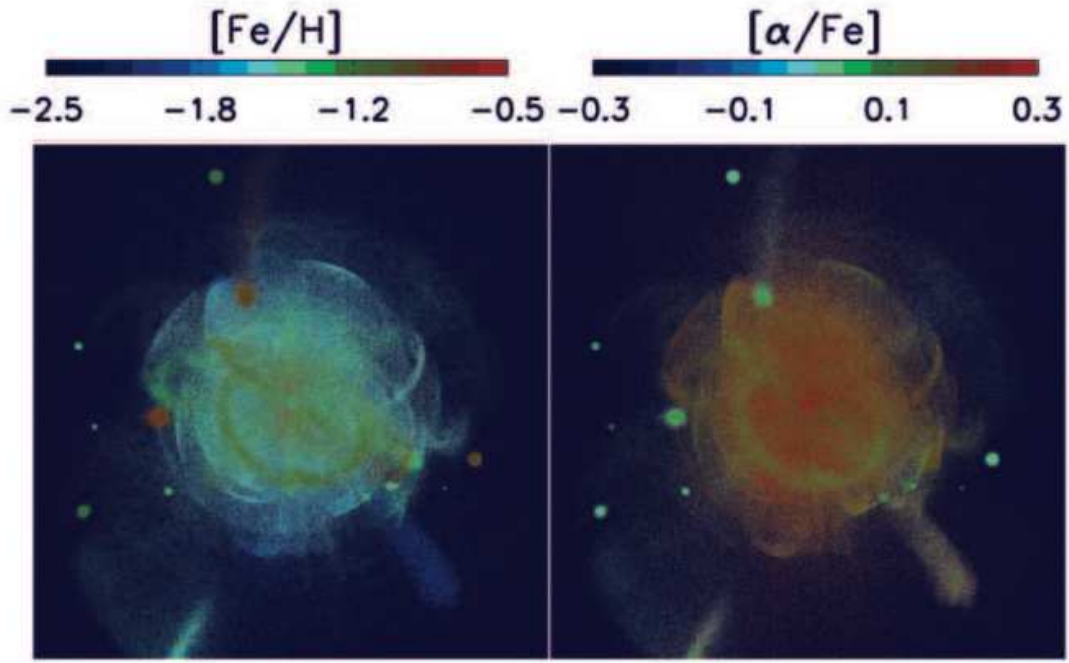} 
 \caption{Simulation of the  chemistry of a Milky Way-like environment. The maps span $300$ kpc on a side.  Note the enhanced levels of [$\alpha/\mathrm{Fe}$] in the Galactic halo while the ratio is nearly solar in the satellite galaxies, indicating a significant contribution of iron from thermonuclear (Type Ia) SNe. From \citet{font06}.}
   \label{alphahalo}
\end{center}
\end{figure}


\section{In search of the earliest chemical signatures}
\subsection{Sites of early star formation}\label{sites_of_early_sf}
Essentially all galaxies, including all known dwarf galaxies, appear to contain stars older than 10 Gyr. But only a fraction of these are likely to contain pristine signatures of the first and second stellar generations. So where do the most ancient stars reside today?
\textcite{scannapieco06} and \textcite{brook07} showed that metal-free stars, i.e., if they exist, should be spread over the entire Galaxy. This implies that slightly enriched, second generation stars, should be found more or less everywhere, and predominantly in the outer halo \citep{brook07,salvadori10}.  This is also where the most metal-poor stars have been found in past and present surveys. Together, the HK \citep{beers92} and the Hamburg/ESO \citep{wisotzki00} surveys of metal-poor stars have discovered roughly $2500$ stars below [Fe/H] $= -2$ and several hundred stars below [Fe/H] $=-3$ in the Galactic halo \citep{bc05}. The Sloan Digital Sky Survey (SDSS) and Sloan Extension for Galactic Understanding and Exploration (SEGUE) have now raised the numbers of known stars with $[\mathrm{Fe}/\mathrm{H}]<-2$ to over $25,000$ and the number of stars with $[\mathrm{Fe}/\mathrm{H}]<-3$ to roughly $1,000$ (Beers, private communication). Stars with different metallicities are distributed differently in the halo. The outer halo of the Milky Way appears to be comprised of a stellar population significantly more metal-poor than its inner halo (Carollo {\it et al.}, 2007, 2010\nocite{carollo07,carollo10}; but see Sch\"{o}nrich {\it et al.}, 2011\nocite{schonrich11}). According to these results, the metallicity distribution function (MDF) of the outer halo peaks around [Fe/H] $\simeq -2.2$, while the MDF of the inner halo peaks at \mbox{$[\mathrm{Fe}/\mathrm{H}]\simeq -1.6$}. The outer and inner halo may also differ in terms of oblateness and net rotation \citep{zinn93,kinman07,majewski92}. This suggests that the outer component was formed in a distinctly different way from the inner halo, probably  by dissipationless merging of small dwarf galaxies at a relatively late state of the assembly of the Milky Way (e.g., Carollo {\it et al.}, 2007\nocite{carollo07}, see also De Lucia and Helmi, 2008\nocite{delucia08}; Zolotov {\it et al.}, 2009\nocite{zolotov09}; Salvadori {\it et al.}, 2010\nocite{salvadori10}).  As a result of the inside-out construction of the Milky Way galaxy the {\it oldest}, most metal-poor stars are expected to reside in the innermost part of the Galactic halo \citep{bhp06,tumlinson10}, or the bulge \citep{ws00}. However, recent developments in galactic dynamics may force us to revise this expected picture, as discussed below (see Sec. \ref{churning}).
 
Extremely metal-poor stars are now also being found in dwarf galaxies in the Local Group. An increasing number of high-resolution spectrometric studies of the brightest stars in the faintest galaxies consistently find stars with [Fe/H] $\lesssim-3$ \citep{frebel10,simon10,norris10,feltzing09}.  Interestingly, the more luminous dwarf spheroidal (dSph) systems appeared, until recently, to be devoid of EMP stars all together \citep{helmi06}. In combination with observations of low [$\alpha$/Fe] ratios in the more metal-rich member stars \citep{shetrone01,venn04,tolstoy09}, this casted doubt on the present-day dSphs as being the formation sites of Galactic halo stars. Nor did they seem to be the surviving counterparts to the now dispersed systems in which the Galactic halo stars once were formed. However, new and refined measurements reveal the existence of EMP stars in these systems as well \citep{kirby09,frebel10b,starkenburg10}. The simulations by \citet{font06} depicted in Fig. \ref{alphahalo}, nicely illustrate the general understanding of the distribution of metals in the stellar component, in the Milky Way environment. 

At present, it is not clear in which environments the bulk of the EMP stars were formed originally.  Since they are believed to be second generation stars, born out of gas enriched only by the first SNe, the majority of these stars must have formed relatively early on, possibly in so-called `atomic cooling halos', which are small dark matter halos on the order of 10$^8$ $\mathcal{M_{\odot}}$ (see, e.g., Greif {\it et al.}, 2008\nocite{greif08}).  These halos were probably among the first objects able to form Pop II stars (e.g., Karlsson {\it et al.}, 2008\nocite{karlsson08}; Greif {\it et al.}, 2010\nocite{greif10}).  In fact, the majority of the stars in the newly discovered ultra-faint dSphs \citep{willman05,zucker06a,zucker06b,belokurov07,belokurov10}, suggested to be relic galaxies formed prior to reionization \citep{br09,salvadori09}, are very metal-poor with [Fe/H] $< -2$ \citep{kirby08}. If these galaxies should turn out to be surviving members of a population of `first galaxies' connected to the atomic cooling halos, it would give us a unique possibility to directly study the birth sites of second generation EMP stars \citep{fb10}. 

\subsection{Secular evolution, blurring and churning}\label{churning}
In the classical picture (e.g. Wielen, 1977\nocite{wielen77}), star clusters gradually dissolve and diffuse into the background
galaxy staying within a kiloparsec or so of the birth radius. But since \textcite{sellwood84}, we
have slowly come to realize that spiral perturbations are driven by accretion events and are likely to be
transient. In a seminal paper, \textcite{sellwood02} showed that in the presence of such perturbations, a star can migrate
over large radial distances. During interaction with a single {\it steady} spiral 
event of pattern speed $\Omega_P$, a star's energy ($E$) and angular
momentum ($J$) change while it conserves its Jacobi integral. In the $(E, J )$ plane, stars 
move along lines of constant $I_J  = E - \Omega_P J$. For transient spiral perturbations, 
the star undertakes a random walk in 
the $(E, J )$ plane, deflected by a series of presumably uncorrelated spiral arm events which occur on time intervals
of 500 Myr or so (the impulse acting on individual stars lasts for a fraction of this interval).
For example, the Solar family would have experienced about $\sim$10 of these events
in the vicinity of co-rotation if only a fraction of the stars are moved in each event \citep{hawthorn10b}.

Substantial variations in the angular momentum of a star are possible over its lifetime. 
A single spiral wave near co-rotation can perturb the angular momentum of a star over a broad
distribution in $J_z$ with a tail up to 50\%.
The star is simply moved from one circular orbit to another, inwards or outwards, by up to 
2 kpc or so.  Conceivably, the same
holds true for a star cluster except that most are dissolved in much less than an orbit period. 

Support for the Sellwood-Binney mechanism (churning) comes from recent N-body simulations that allow
for a steady accretion of gas from the outer halo (Ro\v{s}kar {\it et al.}, 2008\nocite{roskar08a}). These relatively ``cold'' simulations reveal 
some of the secular processes that may be acting over the lifetime of the disk. Averaged over the entire disk, 
these authors find that the rms radial excursion at $t=10$ Gyr is $\langle(\Delta R)^2\rangle^{0.5} \approx 2-3$ kpc.
For the outermost disk stars,
the simulations show that $\langle\Delta R\rangle \approx 3-4$ kpc, several times larger than their
epicyclic radial excursion. There are related mechanisms that may behave even more efficiently and
can even extract stars from the inner bulge and transport them across the disk \citep{quillen09,schonrich09,minchev09,minchev12}. The widespread population of supersolar metallicity 
stars throughout the disk lends further support to the idea that circumnuclear stars can travel far out
into the disk \citep{trevisan11}. This assumes that most supersolar stars are born in the inner regions which
finds support from high redshift observations \citep{hamann99,savaglio12}.
Thus, it is no longer clear that the most ancient stars are highly concentrated towards the centres of galaxies.

\section{Chemical signatures from the Galactic halo}\label{ece}

\subsection{General behaviour}\label{genb}
\begin{figure}[t]
\begin{center}
 \includegraphics[width=6.0in]{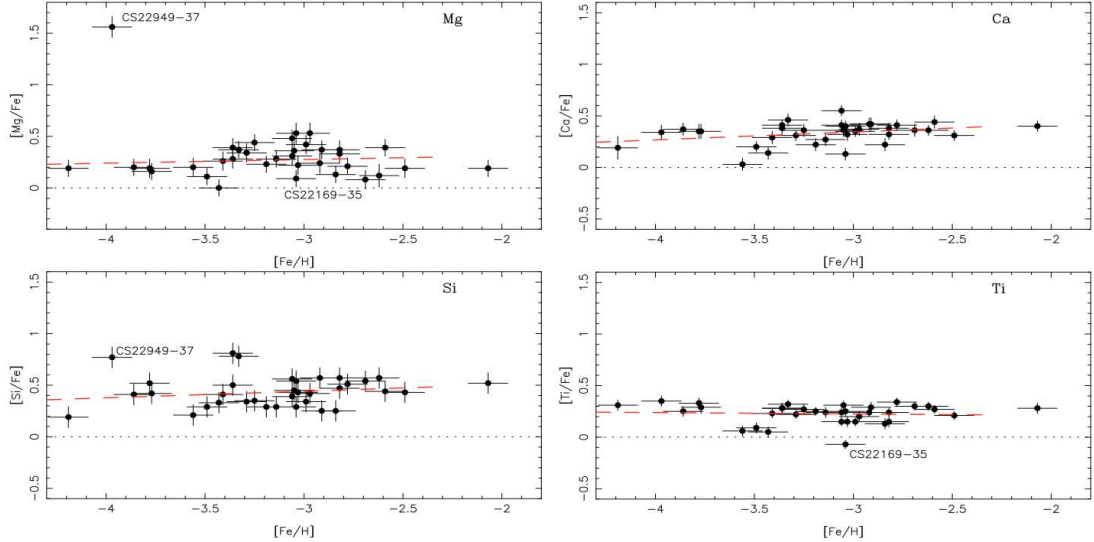} 
 \caption{The [$\alpha/$Fe] abundance ratios in very metal-poor Galactic halo stars. The bulk of the halo sample show a clear enhancement in the $\alpha$-group-to-Fe ratios, in particular for Mg, Ca, Si, and Ti, as compared to the Sun.  From \citet{cayrel04}.}
   \label{alphatofe}
\end{center}
\end{figure}

\begin{figure}[t]
\begin{center}
 \includegraphics[width=6.0in]{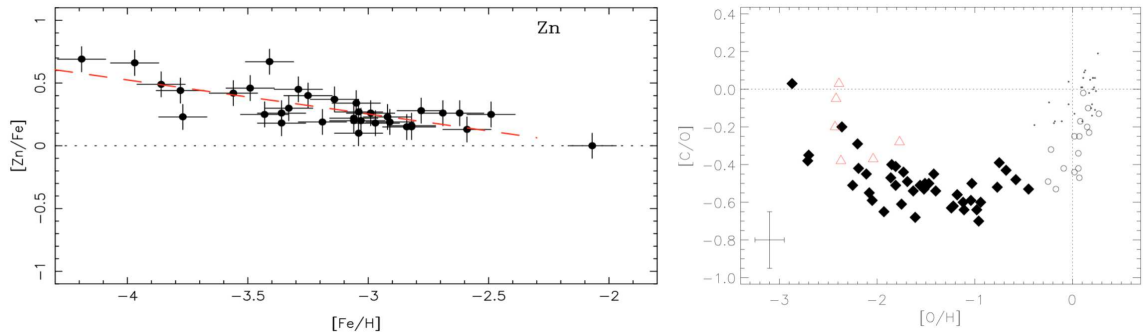} 
 \caption{Evolution of Zink and carbon in the metal-poor, Galactic halo. The data by  \textcite{cayrel04} show an increase of the [Zn/Fe] ratio with decreasing [Fe/H]  (left panel). This increase may be understood in terms of a contribution of Zn from HNe. The red, dashed line denotes the average slope of the data. Similarly, there is an upturn in the [C/O] ratio (right panel) at low metallicities (here measured by [O/H]). Pop III stars, or fast rotating Pop II stars, could be responsible for this upturn. Data by \textcite{fabbian09} are denoted by black diamonds while the red, open triangles are data of DLAs by \textcite{pettini08}. See \textcite{fabbian09} for further references.}
   \label{ZnCO}
\end{center}
\end{figure}

Detailed chemical abundances based on high-resolution ($R\gtrsim 30~000$) spectroscopy for many elements (e.g. C, N, O, Na, Al, Mg, Ca, Si, Sc,  Ti, Cr, Mn, Fe, Co, Ni, Zn, Sr, Y, Zr, Ba, Eu) are now available \citep{cayrel04,cohen04,spite05,barklem05,francois07,bonifacio09,lai08,zhang11}  for a large number of Galactic halo stars below [Fe/H] $=-2$. The general conclusion is clear: the large majority of the EMP stars were formed out of gas enriched by core collapse SNe in the mass range $10\lesssim m/\mathcal{M_{\odot}} \lesssim 40$ (see Sec. \ref{nucleosynthesis}) This is concluded primarily from their enhanced  [$\alpha$/Fe] ratios (see Fig. \ref{alphatofe}). There are, however, a number of exceptions to this observation. The behaviour of Zinc, or rather [Zn/Fe], at the lowest metallicities cannot easily be explained by normal core collapse SNe (Fig. \ref{ZnCO}). \textcite{un02,umeda02} showed that the production of Zn is favoured in their more energetic, hypernova models as a result of the extended Si-burning region (see also Umeda and Nomoto, 2005\nocite{un05}; Tominaga {\it et al.}, 2007\nocite{tominaga07}).  In contrast, \textcite{hw10} argue against HNe as the primary nucleosynthesis site for Zn and notice that this element has an uncertain contribution from the s-process, {\it inter alia}. Furthermore, since only a fraction of the massive stars is believed to explode as HNe, a large spread in [Zn/Fe] would be expected at the lowest metallicities, i.e., if the ISM was poorly mixed. Such a scatter is not observed, although a number of relatively low upper limits in [Zn/Fe], reported on by \textcite{bonifacio09}, may possibly be suggestive of a larger scatter. In addition, \textcite{honda11} found a strongly Zn-enhanced star at $[\mathrm{Fe}/\mathrm{H}]=-3.2$, that could be explained by a HN contribution.

Carbon and nitrogen are two other elements which behave rather curiously for `normal' stars in the metal-poor regime. The observed star-to-star scatter is significant for both elements (even if the population of C-enhanced stars are excluded), in particular for N (e.g. Spite {\it et al.}, 2005\nocite{spite05}). The evolution of [C/O] at low metallicities (Fig. \ref{ZnCO}) tends to increase with decreasing metallicity \citep{akerman04,fabbian09}. This has been interpreted as a unique signature of metal-free SNe \citep{akerman04}.  However, this signature may alternatively originate from massive Pop II stars in which significant amounts of C (and N) are dredged up and ejected in an extensive stellar wind induced by high rotational velocities \citep{meynet06,chiappini06,fabbian09}. If so, the high C/O ratios in the metal-poor stars are not uniquely a signature of Pop III.  

In general, odd-Z elements like K, Sc, V, and Co, but also Ti, appear to be difficult to synthesize in standard nucleosynthesis models of SNe \citep{kobayashi06}. Improvements can be achieved by including additional nucleosynthesis processes, like the $\nu$-process \citep{tyoshida08}, the $p\nu$-process (Fr\"{o}hlich {\it et al.}, 2006\nocite{frolich06}, could be important for the production of $^{64}$Zn and above), or modifications of parameters such as the electron fraction $Y_{\mathrm{e}}$. However, this discrepancy may not only be an effect of incomplete modeling of the nucleosynthesis \citep{hw10}, but could also result from systematic errors in the observations, such as non-LTE effects and/or the lack of a proper 3D abundance analysis \citep{andrievsky10,bonifacio09,lai08,aegp06,collet06,asplund05}. \textcite{bonifacio09} identify a number of discrepancies between the inferred chemical abundances in EMP giants and turn-off stars. They conclude that these discrepancies are predominantly, possibly except for C, due to shortcomings in the abundance analysis. Indeed, Andrievsky {\it et al.} (2010\nocite{andrievsky10}, see also Takeda {\it et al.}, 2009\nocite{takeda09}) find that a careful non-LTE analysis of Na, Al, Mg, and K tightens the abundance relations and some agreement with models of chemical evolution of these elements is attained. The same goes for silicon \citep{zhang11}. It is crucial that these systematic errors are identified and reduced.   

Curiously, the signature of PISNe is not detected in the Galactic halo EMP stars \citep{cayrel04,tumlinson04,ballero06}. This observation suggests that very massive stars were extremely rare in the early Universe, in contrast to some theoretical arguments \citep{abel02,bromm09}, but in agreement with other predictions \citep{salvadori07,tornatore07,karlsson08}. This issue will be discussed further in Sec. \ref{constrimf}.

\subsection{Scatters, trends and mixing}\label{mixissue}
\begin{figure}[t]
\begin{center}
 \includegraphics[width=5.0in]{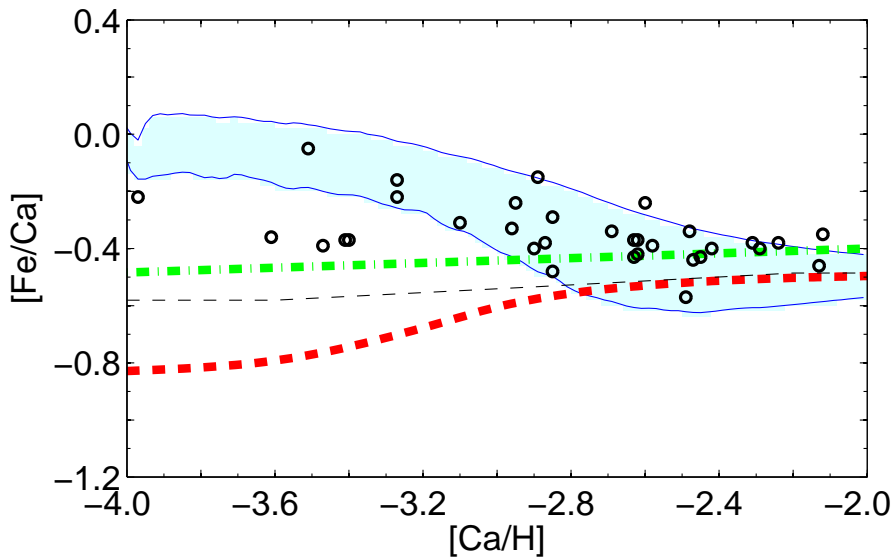} 
 \caption{\footnotesize{Predicted trends in the [Fe/Ca] -- [Ca/H] plane, assuming different mixing scenarios. The light blue shaded area within the blue solid lines denotes the 1-$\sigma$ dispersion in the inhomogeneous chemical evolution model by \textcite{karlsson08}, where SNe enrich their surroundings locally (although mixing will occur on a longer time scale). The negative slope is a result of the mass-dependence of the Ca yield where regions enriched by the least massive SNe have low [Ca/H] and high [Fe/Ca] while regions enriched by more massive SNe have high [Ca/H] and low [Fe/Ca]. The red thick dashed line denotes the trend predicted from a homogeneous chemical evolution model assuming instantaneous mixing. The positive slope originates, again, from the mass dependence of the Ca yield. In this case, however, the most massive stars are the first to explode and enrich the (entire) system, producing a low [Fe/Ca], while the least massive SNe enrich the system slightly later. The thick green dash-dotted line denotes a case where the cooling time scale of the gas is long as compared to the mixing time scale and the ejecta of SNe of all masses have time to mix before the next generation of stars is formed. Here, the positive slope is a result of a metallicity dependence of the SN yields such that the IMF averaged [Fe/Ca] ratio is slightly higher at higher metallicities. The thin black dashed line denotes a similar case where, instead, the IMF is allowed to change with metallicity, from a top-heavy to a Salpeter one (no metallicity dependent yields are assumed).  In all cases, the SN yields by \textcite{nomoto06} have been adopted. Data of Galactic halo giants \citep{cayrel04} are shown as black circles, for comparison.}}
   \label{trends}
\end{center}
\end{figure}

Recent observations by a number of authors (e.g., Cayrel {\it et al.}, 2004\nocite{cayrel04}; Arnone {\it et al.}, 2005\nocite{arnone05}; Barklem {\it et al.}, 2005\nocite{barklem05}; Andrievsky {\it et al.}, 2010\nocite{andrievsky10}; however, see Zhang {\it et al.}, 2010\nocite{zhang11}) have revealed surprisingly small star-to-star scatters in the $\alpha$-to-iron element ratios for very metal-poor and EMP stars, scatters which are consistent with observational uncertainties alone (Fig. \ref{alphatofe}).  This has been interpreted as evidence of a chemically well-mixed star-forming medium, even at very early epochs, in which the mixing timescale was extremely short. However, the star-to-star scatter in a given abundance ratio is only an indirect measure of the state of the early ISM/IGM. Indeed, a long star formation timescale produces about the same scatter as a short mixing timescale. \textcite{arnone05} speculated that the longer cooling times in metal-poor gas may prevent subsequent star formation until the ejecta from whole generations of SNe were efficiently mixed. As a result, a small star-to-star scatter can be achieved.  Very long cooling timescales would, however, wipe out any observed trend with metallicity, such as a decreasing $[\mathrm{Zn}/\mathrm{Fe}]$ with increasing $[\mathrm{Fe}/\mathrm{H}]$ observed for EMP giants \citep{cayrel04}, unless the SN yields, integrated over the IMF, show a significant metallicity dependence (see Fig. \ref{trends}).  At least in the case of $[\mathrm{Zn}/\mathrm{Fe}]$, theoretical yield calculations do not seem to predict such a dependence (e.g., Chieffi and Limongi, 2004\nocite{cl04}; Nomoto {\it et al.}, 2006\nocite{nomoto06}). An interesting example where metallicity dependent yields may be able to explain observations is the decreasing $[\mathrm{C}/\mathrm{O}]$ trend with increasing $[\mathrm{O}/\mathrm{H}]$ studied by Akerman {\it et al.}, (2004\nocite{akerman04}; but see the comment in Sec. \ref{genb}). In their analysis, they adopted the yields by \textcite{cl04}. 

A viable explanation to some observed trends is a change of the IMF itself, e.g., from a top-heavy IMF to an IMF which more resembles the present-day one (Fig. \ref{trends}). This route should be explored further. Interestingly, there are claims of IMF variations also in the local Universe which appear to arise in low surface brightness or low star-formation rate galaxies. There is now a very well established deficiency of H$\alpha$ luminosity relative to other star formation indicators (e.g. UV luminosity) that is not easily explained away without modifying the IMF \citep{meurer09}. One possibility is that ionizing photons are leaking out of low surface brightness galaxies, such that the H$\alpha$ luminosity does not reflect the true ionizing luminosity. But the effect could herald a non-uniform IMF at the present or any epoch as a consequence of galaxy evolution. Possible observational effects of an evolving early IMF are discussed further in Sec. \ref{constrimf}.

The presence of trends in the EMP Galactic halo may alternatively suggest that the ISM at this point was, in fact, fairly unmixed. In a poorly mixed, extremely metal-poor ISM, two low-mass stars can form out of gas enriched by two SNe of different masses producing different amounts of heavy elements.  In such a scenario, trends may simply result from the different mass-dependences of the SN yields \citep{kg05,nomoto06}. However, in contrast to what is a consequence in homogeneous chemical evolution models, where the most metal-poor stars are enriched by the stars with the shortest lifetimes, i.e., the highest-mass stars, the most metal-poor stars will instead, in most cases, be enriched by the least massive SNe as they eject the least amount of metals.

A number of elements show clear evidence of inhomogeneous chemical enrichment.  Large and real scatters are found in the n-capture elements (e.g., Fran\c{c}ois {\it et al.}, 2007\nocite{francois07}) as well as in C and N \citep{spite05}.  Although the scatter in the n-capture elements may result from the possibility that these elements are only produced in a small SN mass range as compared to light and intermediate-mass elements \citep{burris00,c08}, this would not explain the observed scatter in those other elements.  As regards nitrogen, \textcite{chiappini06} speculated that the large observed scatter may originate from variations in the initial rotational velocity of extremely metal-poor massive stars, which are shown to produce large variations in the N-yield \citep{h07}. In general, a large star-to-star scatter is realised in metal-poor and unmixed gas for elements with stellar yields that depend strongly on some parameter, like initial mass or rotation \citep{kg05,karlsson08,cescutti10}. The scatter dichotomy, where some elements show evidence of an unmixed ISM while others do not, may also need to be understood in the context of star cluster formation. If the majority of the EMP stars originally were formed in clusters, as are stars in the present-day Universe \citep{hawthorn10a}, extra mixing will initially occur on the scale of the $\mathrm{H}_{2}$ clouds out of which individual clusters were formed. Although this could provide enough mixing to explain the very small scatter e.g., in [$\alpha$/Fe] \citep{arnone05}, the large scatter observed for the neutron-capture elements must still be accounted for. This potential `cluster-mixing issue' requires further attention. In Sec. \ref{clustering}, we will come back to the importance of star clusters and the exciting prospects to use their chemical signatures as probes of the early star and galaxy formation.  

The issue of trends, as illustrated in Fig. \ref{trends}, and of the scatter dichotomy needs to be addressed in more detail. It is important to keep in mind that incomplete and/or biased samples of stars may not capture the true cosmic star-to-star scatter. Moreover, the effects of inhomogeneous chemical enrichment are only expected to be observed below $[\mathrm{Fe}/\mathrm{H}] \sim -3$ \citep{k05}, a metallicity regime which still is quite poorly sampled. Future stellar surveys need to go deeper, with a focus on minimizing the observational uncertainties. The signatures of chemical evolution in physically distinct stellar systems, such as the bulge, halo, and dwarf galaxies, can then be inter-compared and be used to disentangle progenitor mass-, age-, or metallicity dependences on abundance trends (Fig. \ref{trends}). In order to fully unveil the chemical signatures of Pop~III, it is of utmost importance to gain a deeper understanding on what drives the interstellar gas mixing, how efficient it is in various environments (e.g., Pan and Scalo, 2007\nocite{pan07}; Greif {\it et al.}, 2010\nocite{greif10}), and how it couples to star formation.

\subsection{Outlier objects and carbon-enhanced metal-poor stars}\label{umps}
The Hamburg/ESO survey of metal-poor stars has discovered the three most iron-poor stars known to date: HE~0107-5240 \citep{christlieb02}, HE~1327-2326 \citep{frebel05}, and HE~0557-4840 \citep{norris07}. All three stars are ultra-metal-poor, lying well below [Fe/H] $=-4$, while two of them, HE 0107 and HE 1327, are hyper-metal-poor ([Fe/H] $< -5$) according to the definition by \textcite{bc05}. Curiously, these stars, in particular the two most iron-poor ones, also exhibit several abundance anomalies as compared to the bulk of Galactic halo EMP stars, including strong enhancements of CNO elements relative to Fe (only an upper limit of $[\mathrm{N}/\mathrm{Fe}] < 1.0$ is reported for HE 0557, see Norris {\it et al.}, 2012\nocite{norris12}), Na (HE 0107 and HE 1327), Mg (HE 1327), Al (HE 1327), and Ca (HE 1327). Furthermore, despite HE1327-2326 being almost unevolved, it is strongly depleted in Li \citep{frebel08}.  Different scenarios have been put forward to explain the anomalous abundance patterns, in particular the high CNO abundances. These scenarios include pre-enrichment of the primordial cloud by a faint SN \citep{un03,iwamoto05}, enrichment by massive stellar winds \citep{meynet06,meynet10}, mass transfer from a binary companion (e.g., Suda {\it et al.}, 2004\nocite{suda04}; Tumlinson, 2007a\nocite{tumlinson07hmp}), and atmospheric dust-gas separation \citep{vl08}. So far, none of the stars show radial velocity variations indicative of binarity and none of them seem to be particularly enhanced in s-process elements, which otherwise often is the case for stars affected by mass transfer from an AGB companion. 

Recently, \textcite{spite11} argued that the abundance patterns of HE 0107 and HE 1327 presumably did not suffer from gas depletion, as suggested by \textcite{vl08}, since the somewhat more metal-rich star analogue, CS 22949-037, appears to show no enhancements in S and Zn. Enhancements in these elements would have been expected if the abundance pattern of CS 22949-037 had been locked to the dust condensation temperature, as it is in stars affected by dust depletion, like the $\lambda$ Bootis stars.  As an alternative explanation for the high C abundance, \textcite{k06} explored a scenario in which the C (and N, O) originated from massive rotating stars (see also Meynet {\it et al.}, 2010\nocite{meynet10}) with $m \gtrsim 40~\mathcal{M_{\odot}}$, while the Fe originated from somewhat less massive core collapse SNe in combination with a period of reduced star formation after the first generation of stars had formed (see below). For a similar scenario, see also \textcite{limongi03}. Despite the unusually high CNO and the enhancements of a few of the intermediate-mass elements, normal core collapse SNe in the range $15\lesssim m/\mathcal{M_{\odot}}\lesssim 30$ seem to be fairly successful in describing the abundance ratios of the heavier elements. 


It is not yet clear whether these stars tell us something fundamental about the first stars and the conditions of star formation in the early Universe, whether they are ``outlier objects'' formed under very special circumstances but not necessarily unique for Pop III star formation, or whether their chemical abundance patterns, at least some part of it, have been altered by a process subsequent to their formation, in which case the original signature would be hidden from us. It is important, however, not to disregard them in the context of chemical evolution purely based on the argument that they are ``chemically weird'' and lie too far away from the mean. Such a decision should be made based on more physical grounds and detailed observations (see Sec. \ref{clustering}). 


Carbon plays a special role in early star formation as it acts as an efficient cooling agent in metal-poor gas (\ref{pop2trans}). Indeed, a relatively large fraction of metal-poor stars below $[\mathrm{Fe}/\mathrm{H}] = -2$ have $[\mathrm{C}/\mathrm{Fe}] > 1$, as discovered by the HK and the Hamburg/ESO surveys. The fraction of these carbon-enhanced metal-poor (CEMP) stars is estimated to be $\sim10 - 20\%$ \citep{cohen05,lucatello06}, a fraction which tends to increase with decreasing metallicity.  About $80\%$ of the CEMP stars have been shown to be enhanced also in the s-process elements. This subgroup is accordingly named CEMP-s. There is now convincing evidence that the vast majority of the CEMP-s stars are members of binary systems and that they, like the more metal-rich CH and Ba stars, likely obtained their peculiar abundance pattern by mass transfer from an evolved binary companion that is now extinct (e.g., Fujimoto {\it et al.}, 2000\nocite{fujimoto00}; Lucatello {\it et al.}, 2005b\nocite{lucatello05b}).

\begin{figure}[t]
\begin{center}
 \includegraphics[width=5.0in]{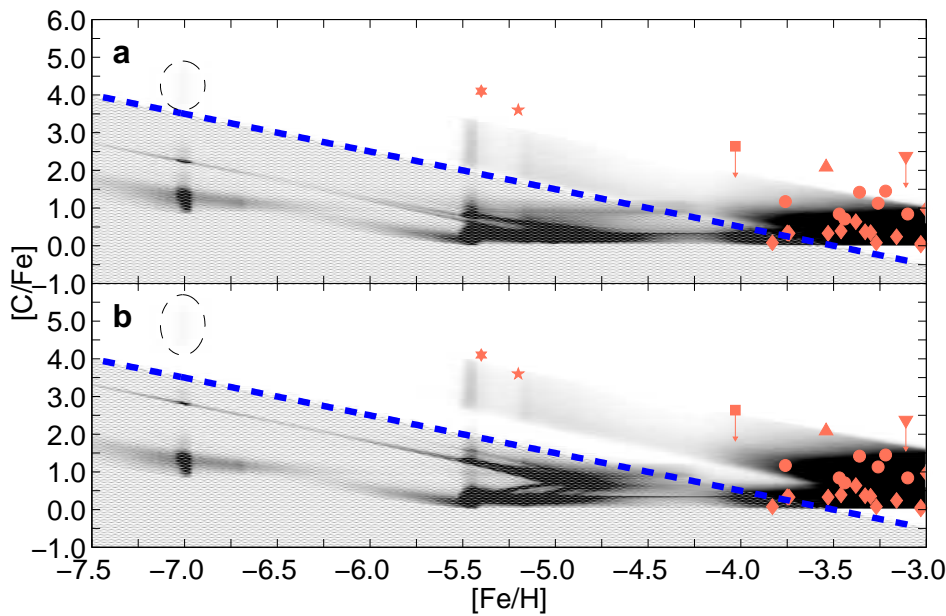}
 \caption{\footnotesize{The predicted distribution of stars in the $[\mathrm{C}/\mathrm{Fe}]-[\mathrm{Fe}/\mathrm{H}]$ plane. In both panels, the thick, dashed line indicates a carbon abundance of $[\mathrm{C}/\mathrm{H}]=-3.5$. The shaded area below this limit should, according to the `Frebel criterion' (see Sec. \ref{pop2trans}), contain very few low-mass stars and are therefore disregarded here. Symbols denote various observations of stars in the Milky way's halo (for details, see Karlsson, 2006).  The peak of the predicted group of mega metal-poor stars (i.e., $[\mathrm{Fe}/\mathrm{H}]<-6$) is encircled. In the lower panel, the fiducial carbon yield is increased by a factor of $4$ for stars in the mass range $30 \le m/\mathcal{M_{\odot}} \le 60$, in order to estimate the effect of stellar rotation \citep{mm02}. Observed abundance ratios are not corrected for 3D effects. From \textcite{k06}.}}
   \label{fig1}
\end{center}
\end{figure}

What about the remaining $20\%$ of the CEMP stars? The majority of these stars are either enriched in r-process elements, but no s-process, (CEMP-r, i.e. $\mathrm{CS}~22892-052$, Sneden {\it et al.}, 2003\nocite{sneden03}) or they show no overabundance in n-capture elements altogether (CEMP-no). Even though attempts have been made to explain the abundance patterns of all CEMP stars with the mass transfer scenario (e.g., Komiya \textit{et al.}, 2007; Masseron {\it et al.}, 2010\nocite{komiya07,masseron10}), it is not clear whether this is a necessary condition. Recently, \textcite{cooke11} reported on a metal-poor ($[\mathrm{Fe}/\mathrm{H}]\simeq -3$) damped Ly-$\alpha$ system in the spectrum of the quasar J0035--0918, showing a significant enhancement in carbon. This finding suggests that extremely metal-poor star-forming gas could indeed be C enhanced, in which case true CEMP stars may exist without having to originate from a mass-transfer event in a binary system.  Interestingly, three of the four ultra metal-poor stars \citep{christlieb02,frebel05,norris07} found below $[\mathrm{Fe}/\mathrm{H}] = -4.5$ are consistent with being CEMP-no stars. The only exception is the $[\mathrm{Fe}/\mathrm{H}]=-4.7$ star $\mathrm{SDSS}~\mathrm{J}102915+172927$ \citep{caffau11} which shows no evidence of CNO enhancements and is believed to have formed out of gas cooled by thermal dust emission \citep{klessen12,schneider12}.  Instead of treating these CEMP-no stars as individual, unique objects, it could, alternatively, be worthwhile to explore what one may learn if they were treated as `normal' stars in the general context of chemical evolution. If the gas out of which the EMP stars were formed experienced a period of low or delayed star formation, e.g., due to negative feedback effects from the first generations of stars, a small population of very C-enhanced, ultra metal-poor stars is to be expected \citep{k06}.  Figure \ref{fig1} shows the predicted distribution of EMP stars (grey and black shaded areas) in the $[\mathrm{C}/\mathrm{Fe}]-[\mathrm{Fe}/\mathrm{H}]$ plane.  Evidently, the predicted fraction of CEMP stars increases with decreasing $[\mathrm{Fe}/\mathrm{H}]$, which mainly is due to the fact that low-mass star formation is assumed to be inhibited in regions with low contents of C (and O, see Bromm and Loeb, 2003\nocite{bl03}; Frebel {\it et al.}, 2007\nocite{frebel07}).  In these models, the predicted fraction of CEMP-no stars below $[\mathrm{Fe}/\mathrm{H}] = -2$ is roughly $1-7\%$, depending on which set of stellar yields is used.  This is consistent with the observed fraction of $2-5\%$. Note that this prediction would be only marginally affected by the presence of ultra-metal-poor stars formed by the dust-cooling channel (see Sec. \ref{pop2trans}), unless this channel will be found to completely dominate over the fine structure line-cooling channel.     
As a result of the delayed star formation, the current model (Fig. \ref{fig1}) is also able to explain the apparent downturn of the number of stars in the metallicity range $-5\lesssim[\mathrm{Fe}/\mathrm{H}]\lesssim-4$.  The deficit of stars in this region should, in this scenario, rather be interpreted as an extension or a stretching of the Galactic halo MDF below $[\mathrm{Fe}/\mathrm{H}] \simeq -4$.  A small population of mega metal-poor stars ($[\mathrm{Fe}/\mathrm{H}]<-6$) is also predicted to exist as a result of an initial enrichment by electron capture SNe (Sec. \ref{ecapsne}). 

Chemical peculiarities which are a manifestation of mass transfer in binaries prove that it is difficult to match models of first star yields to the measured abundances of EMP stars \citep{mcwilliam09}. This poses a potentially serious problem, as a non-negligible fraction of stars formed at all epochs may be formed in binary or multiple systems \citep{stacy10,turk09}. Thus, the CEMP-s stars are not giving us a direct reading of the nucleosynthetic yields of the first stars. In particular, the light elements (e.g. CNO, n-capture elements, and possibly the $\alpha$-elements) may simply reflect the yields of nuclear fusion in a massive companion (e.g. AGB), dredged up and transferred to the surviving companion. If the massive companion was even higher mass, the heavy elements of the survivor may reflect the nucleosynthetic yields at the time of the explosion rather than in a slow wind or Roche lobe overflow. This is a more complex situation than the more direct case when a star's abundances simply reflect the ISM chemistry at the time of collapse of the parent molecular cloud.  In the latter case, the star's abundances are a true reading of the chemistry over some mass scale (i.e., the mixing mass or molecular cloud mass) in the early Universe. In present day star clusters, chemical homogeneity to a high degree is clearly observed \citep{desilva06,desilva07a,desilva07b}. We will explore this promising possibility of separating an original chemical signature from one that was imposed by a binary companion in more detail in Sec. \ref{clustering}. 

\begin{table}[t]
\caption{Redshift -- lookback time.}
{\begin{tabular}{cccc} \toprule
\hphantom{000}$z$\hphantom{000} & \hphantom{0}Age\footnote{Age of the Universe at redshift of $z$. Calculated from the combined data (Maximum Likelihood) of {\it WMAP}~$7$-year+BAO+$H_0$ (Komatsu {\it et al.}, 2011\nocite{komatsu11}).}\hphantom{0} & \hphantom{0}Lookback time\hphantom{0} & Upper stellar mass\footnote{Highest mass of still surviving stars that formed at redshift of $z$. These stars have a lifetime equal to the ``lookback time''. The first column gives the mass of stars with metallicity $Z=10^{-2.3}~\mathcal{Z_{\odot}}$ (which is taken as a proxy for $Z=0$) while the second column gives the mass of stars with solar metallicity. The hyphens indicate that no stars existed at the time of the Big Bang. Data are taken from {\sf http://stev.oapd.inaf.it/cgi-bin/cmd\_2.2} \citep{marigo08}.} \\ 
 & Gyr & Gyr & $\mathcal{M_{\odot}}$ \\ \colrule
$\infty$ & 0\hphantom{.00} & \hphantom{0}13.78\hphantom{$^c$}  & --\hphantom{0(..000}--\hphantom{.} \\
30 & 0.10 & \hphantom{0}13.68\hphantom{$^c$} & 0.799\hphantom{0()}0.974 \\
10 & 0.49 & \hphantom{0}13.30\hphantom{$^c$} & 0.806\hphantom{0()}0.981 \\
\hphantom{0}6 & 0.96 & \hphantom{0}12.83\hphantom{$^c$} & 0.814\hphantom{0()}0.991 \\
\hphantom{0}3 & 2.21 & \hphantom{0}11.58\hphantom{$^c$} & 0.837\hphantom{0()}1.018 \\
\hphantom{0}1 & 5.98 & \hphantom{10}7.80\hphantom{$^c$} & 0.934\hphantom{0()}1.130 \\ 
\hphantom{.0000}0.434 & 9.21 & \hphantom{10}4.57\footnote{Age of the Sun \citep{bonanno02}.} & 1.089\hphantom{0()}1.313 \\ \botrule
\end{tabular} \label{lookback}}
\end{table}

\subsection{Constraining the primordial IMF}\label{constrimf}
\subsubsection{The low-mass end of the IMF}
As the vast majority of the primordial stars presumably were massive objects with lifetimes $\lesssim 10$ Myr, we do not expect metal-free stars to exist today. This conclusion is well corroborated by observations \citep{rn91,beers92,carney96,christlieb02,cetal04,frebel05,schoerck09,yanny09}. Due to a number of observational constraints, the search for metal-free stars has concentrated on the solar neighborhood and the Galactic halo, leaving the presently unreachable Galactic bulge unexplored. However, the possibility that metal-free stars are exclusively hiding deep within the bulge seems slim (Brook {\it et al.}, 2007\nocite{brook07}, see also Sec. \ref{churning}).  Another concern has been raised that stars may sweep up enriched gas as they orbit around the Galaxy, thereby concealing their true nature, but this also appears unlikely (Frebel {\it et al.}, 2009\nocite{frebel09}). To date, not a single metal-free star in known to exist in the Milky Way, or in any other galaxy of the Local Group that is observable with current facilities.  This places direct constraints on the primordial IMF at the low-mass end \citep{k05,tumlinson06,salvadori07,oey03,bond81}, and indicates that metal-free stars with masses $\lesssim 0.8~\mathcal{M_{\odot}}$ never formed (see Table \ref{lookback}). 

CEMP stars may provide additional evidence for a higher mass scale of the IMF at early epochs.  As already discussed, the dominant fraction of the CEMP stars are believed to originate from binary mass transfer. A high CEMP fraction like the one observed can be understood if the formation of single stars with masses $\lesssim 0.8~\mathcal{M_{\odot}}$ was suppressed at metallicities below [Fe/H]$\simeq -2.5$, which results in an IMF with a higher characteristic mass, $m_c$, than that of the present-day one (e.g., Lucatello {\it et al.}, 2005a\nocite{lucatello05a}; Komiya {\it et al.}, 2007\nocite{komiya07}; Tumlinson, 2007b\nocite{tumlinson07imf}). For EMP stars, characteristic masses in the range $1\lesssim m_c/\mathcal{M_{\odot}}\lesssim 10$ are found in the literature. This result is broadly consistent with early star formation being dependent on metallicity. It appears, however, that metallicity may not be the sole parameter controlling the fraction of CEMP stars. Below [Fe/H]$=-2$, \textcite{frebel06} found an increasing CEMP fraction with increasing distance from the Galactic plane in the Hamburg/ESO data. This spatial variation led \textcite{tumlinson07imf} to argue for a time-dependent component of the IMF, coupled to the CMB (see Sec. \ref{pop2trans}). The observed spatial variation of the fraction of CEMP(-s) stars can be explained in terms of a time-dependent IMF, where the Jeans mass is set by the decreasing temperature of the CMB, if the stars closer to the Galactic plane (i.e., thick disk stars) were formed later than the stars further out in the halo. Although plausible, the origin of the spatial variation of the CEMP star fraction merits further investigation. The fraction of binaries in a single stellar population could have been quite different at early epochs and may have varied with the star formation efficiency. If so, that would affect the fraction of CEMP stars.  Nonetheless, irrespective of the precise explanation for the spatial variation of the CEMP fraction, the existence of such a variation can be utilized to obtain additional information on the origin of the Galactic halo and to help us discern the connection between the halo and the satellite dwarf galaxies. For example, a different CEMP fraction in dwarf galaxies as compared to that of, e.g., the inner halo, would be suggestive of different origins of the stars in the two type of systems.

\subsubsection{The high-mass end of the IMF}
As regards the high-mass end of the primordial IMF, theory and observations tend to disagree.  To put constraints on the high-mass end, we must resort to measurements of the nucleosynthetic signature of the first generations of stars, preserved in the second and higher-order generations.

As discussed in Sec. \ref{genb}, the emerging chemical signature of EMP stars in the Galactic halo indicates that the early stellar generations contained massive stars in a similar mass range and relative number fractions as the present-day IMF. There is no clear evidence of a significantly different shape of the IMF at the high-mass end. However, let us ask the question: assuming that primordial very massive stars were able to form and that at least some of them where able to explode as PISNe, in what metallicity regime should we expect to find their chemical signature?  Due to their very short life-times, PISNe must have been the first stars to enrich the ISM in metals.  In the classical picture of chemical evolution, an initial enrichment by primordial PISNe would generate a metallicity floor out of which the second generation, low-mass stars were able to form.  Naturally, these stars would be the most metal-poor stars to be found in the Galaxy and would show a unique PISN signature, characterized by, but not limited to, a pronounced odd-even effect, highly super-solar [Si/O] and [S/C] ratios, and a lack of r- and s-process elements.  Such a signature is, however, not observed, neither in the EMP stars, nor in the ultra metal-poor stars.  These stars all show chemical signatures more resembling those of normal core collapse SNe (see, however, Sec. \ref{umps} for a further discussion as regards the ultra metal-poor stars).  This has been taken as an indication that very massive stars were exceedingly few in the early Universe (e.g., Tumlinson {\it et al.}, 2004\nocite{tumlinson04}; Ballero {\it et al.}, 2006\nocite{ballero06}), if not altogether absent, in contrast to the predictions.

\begin{figure}[t]
\begin{center}
 \includegraphics[width=5.0in]{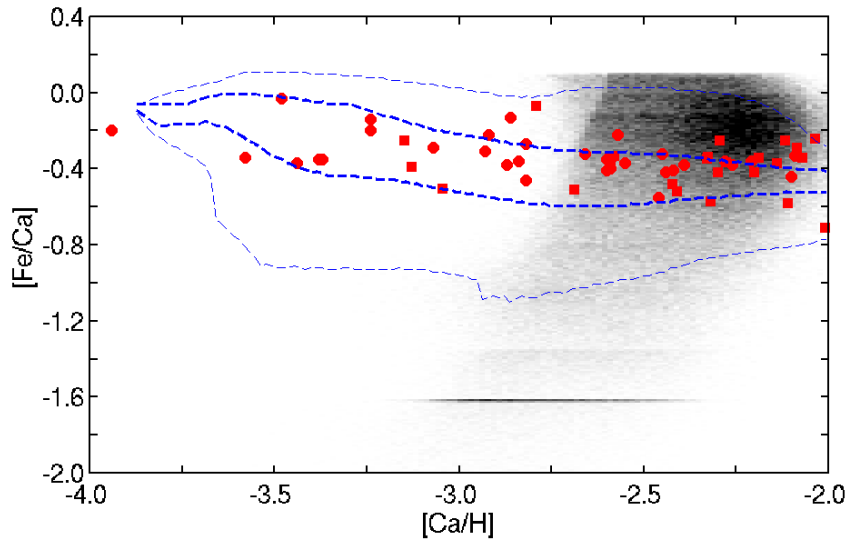} 
 \caption{Predicted distribution (shaded area) of low-mass stars predominantly ($>90\%$) enriched by PISNe in the $[\mathrm{Fe}/\mathrm{Ca}]-[\mathrm{Ca}/\mathrm{H}]$ plane. Note that this is only a very small subpopulation of the full sample of stars in the simulation. The dashed, thick and thin lines indicate, respectively, the $1\sigma$ (innermost $68.3\%$) and $3\sigma$ (innermost $99.7\%$) cosmic scatter of the full sample. Symbols denote various observations of Galactic halo stars. From \textcite{karlsson08}.}
   \label{fig2}
\end{center}
\end{figure}

A different conclusion may, however, be drawn if the instantaneous mixing approximation assumed in homogeneous chemical evolution modelling is relaxed and the enrichment by SNe instead is allowed to occur locally, where the ejecta are mixed relatively slowly with their surroundings by turbulent diffusion.  In order to simulate the stochastic enrichment by primordial PISNe in a cosmological context,  \textcite{karlsson08} followed the initial star formation and chemical enrichment of collapsing atomic cooling halos.  In the primordial star formation mode, only massive and very massive stars were able to form. As the gas became metal enriched, the primordial mode was suppressed and replaced by a normal Pop II star formation mode in which only low- and high mass stars were allowed to form.  The larger explosion energies of the PISNe was also taken into account as they sweep up larger amounts of gas before merging with the ISM. 

The result is shown in Figure \ref{fig2}. The predicted distribution of stars with a dominant contribution from primordial PISNe (gray-shaded area) is located around $[\mathrm{Ca}/\mathrm{H}]\sim -2.5$, which is significantly above the most metal-poor stars in the simulation (the distribution of the total population of simulated stars is indicated by the dashed lines). These stars have, in fact, such high Ca abundances that a fair fraction of them may risk to remain undetected in surveys of metal-poor stars, such as the HK and Hamburg/ESO surveys, in which the metal-poor star candidates are identified on the basis of the weakness (or absence) of the Ca \textsc{ii} K line.  A similar result was found by \textcite{greif10}, who used three-dimensional cosmological simulations to study the assembly process of the first galaxies (cf. Salvadori {\it et al.}, 2007\nocite{salvadori07}). For realistic estimates of the fraction of primordial very massive stars exploding as PISNe (say $\sim 10\%$ by mass in the primordial stellar population, Greif and Bromm, 2006\nocite{gb06}), the fraction of low-mass stars below $[\mathrm{Ca}/\mathrm{H}] =  -2$ with a dominant PISN signature is predicted to be very low, only $\sim 3\times10^{-4}$ \citep{karlsson08}. This may well explain the fact that such a star has not yet been found (see also Salvadori {\it et al.}, 2007\nocite{salvadori07}).  Current observational data are able to constrain the fraction of metal-free, very massive stars to $\lesssim 40\%$, by mass \citep{karlsson08}. 

It should be pointed out that the newly discovered ultra-faint dwarf galaxies may be a particularly favorable site for searching for the PISN signature. If these galaxies originated from the atomic cooling halos they may have been enriched by, or even formed Pop III stars (cf. Frebel and Bromm, 2010\nocite{fb10}). Since a relatively large fraction of the member stars have metallicities below [Fe/H]$=-2$, the fraction of stars with a PISN-dominated signature may be larger with a factor of about $3$, as compared to the corresponding fraction in the Galactic halo.   
 
We note in passing that Eta Carinae and the Pistol Star are two examples of very massive stars in the Galaxy. Recently, \textcite{crowther10} also claimed to have identified several massive stars in excess of $150~\mathcal{M_{\odot}}$ in the young super star cluster R136 in the Large Magellanic Cloud. Furthermore, \textcite{galyam09} reported on an extremely powerful SN explosion which appears to have resulted from the death of a very massive star of about $200~\mathcal{M_{\odot}}$ (see Smith {\it et al.}, 2007\nocite{smith07} for another example of a very energetic SN explosion).

\section{Chemical signatures from low mass galaxies}\label{chem_sign_from_lo_m_gal}

\subsection{Dwarf galaxies}

Low-mass galaxies in the Local Group have come increasingly into focus
as ideal laboratories to study the early enrichment history of the universe.
Here, we first introduce the relevant phenomenology, and proceed to discuss
what constitutes a bona fide `first galaxy'. These systems, in turn, could
serve as the host for the formation of the first low-mass, Pop~II stars,
thus allowing us to make the connection with the nucleosynthesis provided
by the first stars.

\subsubsection{Simplified chemical laboratories}

In recent years, we have come to realize that the Galactic halo is an amalgam
of many systems and fragments \citep{helmi08,tolstoy09}.  Rather than
concentrating on the general halo population, we anticipate that future studies will 
target metal-poor stars in halo subsystems. These include\footnote{We 
have not included globular clusters in this list. While they constitute a halo population, they have remarkably
high levels of enrichment for systems that are believed to be ancient \citep{gratton04}.} dwarf
spheroidals \citep{mateo98}, ultra-faint dwarfs \citep{sg07,kirby08}, stellar streams \citep{ibata95,chou10}, 
stellar associations \citep{walsh07}, and satellites to dwarf galaxies \citep{coleman04,belokurov09}. 
So to what extent can we learn about the first stars from these systems? 

Initially, things did not look good for fossil signatures in dwarf galaxies. Several groups pointed out that
stars enriched by the first stars are likely to be found at the centres of galaxies today \citep{bhp06,tumlinson10} 
although others have shown that first stars can be delayed to $z\sim 3$ such that present-day dwarfs may 
show the signatures of first stars \citep{scannapieco06,tornatore07,brook07}.
Observationally, the classical dwarfs Sculptor, Fornax, Carina and 
Sextans were compared to the halo stars, but no evidence for stars below [Fe/H]$=-3$ was found \citep{helmi06}. 
But a re-calibration of the Ca {\sc ii} triplet (CaT) lines has revealed numerous metal poor stars below [Fe/H]=$-3$ although
many of these have yet to be confirmed \citep{kirby08,norris08,starkenburg10}.
Seemingly like stars in the outer Galactic halo (e.g. Roederer, 2009\nocite{roederer09}), 
evidence is now emerging that star-to-star abundance variations exist in dwarf galaxies \citep{fulbright04,koch08,feltzing09}. 
We now explore the prospect  that dwarf galaxies are important sites to learn about the chemistry of the first stars after all.

The least massive dwarf galaxies, if these can be unambiguously identified, may provide the cleanest signatures
of the yields of the first stars \citep{fb10}. The very low rate of star formation
measured in the faintest dwarf galaxies presumably indicates that there have been relatively few enrichments by 
supernovae over the lifetime of the dwarf. This means that the mean metallicity is expected to be lower thereby giving
a higher fraction of metal poor stars. Furthermore, we may expect that the abundance scatter 
is large at low metallicity consistent with the small number of enrichments for the low baryon
mass fraction. Complicating factors are the effects of feedback and environment. The well known trend
of decreasing metallicity with declining galaxy luminosity has been argued to be a consequence 
of metals lost through galactic winds (see Dalcanton, 2007\nocite{dalcanton07}), but others have argued that this largely
reflects the lower star formation efficiencies in smaller galaxies (see Tassis {\it et al.}, 2008\nocite{tassis08}). Presently, 
it is unclear what impact this will have on the overall abundance scatter. Furthermore, the environmental 
influence of dynamical stripping is also uncertain. If faint dwarfs were the stripped-down
cores of larger dwarfs, would these look chemically distinct from present-day dwarfs of the same luminosity?
\textcite{kirby08} demonstrate that the luminosity-metallicity relation is present down to $\log(L/\mathcal{L_{\odot}})\simeq 4$
but it is unknown whether this reflects a simple mass-metallicity dependence or a more complex process
involving, e.g., the baryon fraction \citep{salvadori10}. 

\subsubsection{Minimum galaxy mass}

What then is the {\it minimum} mass a galaxy can have (Bromm and Yoshida, 2011\nocite{bromm11})? This question has been asked numerous times over the years in 
different contexts (see, e.g., Tolstoy {\it et al.}, 2009\nocite{tolstoy09}) and is particularly relevant today with the discovery of
ultra-faint dwarf galaxies \citep{willman05,sg07}.  In part, the answer depends on the epoch of galaxy 
formation. If dark matter is made up of weakly interacting massive particles ($m_{\rm W}\sim 100$\,GeV), it can 
fragment down to Jupiter mass scales \citep{diemand05,bertschinger06}.  But in order for the IGM to 
overcome thermal pressure and accrete onto dark matter, the mass of the minihalo must have exceeded 
$10^5~\mathcal{M_{\odot}}$ before reionization \citep{loeb06}. During the reionization epoch, in order for the local gas density 
to survive heating through photoionization (which suppresses star formation), the minimum halo mass exceeded
$10^8~\mathcal{M_{\odot}}$ \citep{rees86,ikeuchi86,haiman96}.  These simple calculations based
on ionization balance are supported by 3D hydrodynamical simulations \citep{quinn96,navarro97,gnedin00,okamoto08}. 
In contrast, others have argued that star formation 
was diminished but not fully suppressed, such that galaxies with masses below $10^8~\mathcal{M_{\odot}}$ form stars during 
and immediately after reionization \citep{gnedin06,br09,salvadori09}. 
\textcite{ricotti09} has argued for late-phase accretion in dwarfs as recently as $z=1-2$. Independent evidence of this 
may be the starbursts spaced by a few Gyr seen in the star formation histories of dwarf 
galaxies \citep{grebel04,tolstoy09};  these bursts were presumably triggered by accretion or re-accretion of 
cold gas, although a perigalactic passage is not ruled out.

Assuming that gas accretion can occur, we will take as the working definition of a ``minimum mass galaxy'' a dark 
matter halo whose baryons survive the impact of (at least) one supernova explosion. Afterwards, the enriched gas 
should be able to cool, collapse and form (at least) one subsequent generation of stars. Recently, \textcite{hawthorn10c} 
demonstrated that a clumpy medium is much less susceptible to SN sweeping 
(particularly if it is off-centred) because the coupling efficiency of the explosive energy is much lower 
than for a diffuse interstellar medium. With the aid of the 3D hydro code {\it Fyris} \citep{sutherland10}, they show 
that baryons are retained and stars are formed in dark matter haloes down to $3\times 10^6~\mathcal{M_{\odot}}$. These
systems are expected to have distinct chemical signatures which may be detectable in the near future,
either from direct observations of dwarf galaxies, or from quasi-stellar object (QSO) sight-line spectra due to
the presence of dwarfs around distant galaxies.

This limiting case opens up a profitable line of enquiry, particularly when we consider the chemical signatures 
produced by the discrete supernova events \citep{karlsson10}.  While the minimum mass galaxies may not contribute 
a large fraction of baryons and dark matter to the formation of galaxies, they are likely to carry important 
information on the stars that were formed at the earliest cosmic times. The possible connection between the minimum mass 
galaxies and the faintest dwarfs in the Local Group must be further investigated.  In fact, given that stars below 
[Fe/H]$=-3$ are detected in the dwarf galaxies \citep{frebel10,norris10,kirby09,starkenburg10}, 
we could, without fully realising it, already be looking at the very formation sites of the first stars (cf.
White and Springel, 2000\nocite{ws00}).

\subsection{Damped Lyman-$\alpha$ systems}\label{dampeda} 
An interesting development is the recent discovery of very
metal-poor DLA systems along QSO sight 
lines \citep{pettini08,penprase10}. If these are protogalactic
structures that have recently formed from the IGM, there is the
prospect of identifying the chemical imprint of early generations of
stars (e.g. Pettini {\it et al.}, 2002\nocite{pettini02}). Interestingly, two 
of these systems with [Fe/H] $\sim -3$ may bear the hallmarks of early 
stellar enrichment \citep{erni06,cooke11}.

First, \citet{erni06} identify a DLA toward the QSO Q0913+072 ($z=2.785$)
with an iron abundance characteristic of the IGM at that redshift. The C, N, O,
Al, Si abundances show an odd-even effect reminiscent of the most
metal-poor stars in the Galactic halo. This pattern is created in models where
the neutron flux is low (e.g. Heger and Woosley, 2002\nocite{hw02}), presumably due to the 
low overall metal abundance. A more striking signature is the strong [N/H] depletion
which \citet{pettini02} has argued is further evidence for a system 
that has recently formed from the IGM. \citet{erni06} argue that the abundances 
are in good agreement with 10$-$50 M$_\odot$ zero metallicity Pop III models.

Secondly, \citet{cooke11} identified a DLA towards the QSO J0035-0918 ($z=2.340$)
with a remarkably strong [C/Fe] enhancement and with the same odd-even effect in
light elements. This result is particularly striking because it is reminiscent of the recent
discovery of CEMP stars in the Galactic halo \citep{bc05,lucatello06}. They infer the total mass of
neutral gas to be $\lesssim 3\times 10^6$ M$_\odot$ within a linear scale
of $\lesssim 100$ pc. The inferred amount of carbon is consistent with one or two SN 
enrichment events from primordial, or very metal-poor high mass stars. At face value, 
these observations suggest that at least some fraction of the CEMP stars, in particular those
which are not s-process enhanced  (see Sec. \ref{umps}), may have formed out of 
carbon-enhanced gas, and are therefore, in a sense, truly C-enhanced stars \citep{k06}.

The importance of such observations is two-fold: (i) they provide a chemical signature on
a mass scale that is orders of magnitude larger than the mass of a stellar
envelope, and is therefore expected to be more representative of early star formation;
(ii) they allow us to probe {\it in situ} enrichment at early cosmic epochs. The fact that
we recover chemical signatures that are strikingly similar to those observed in ancient halo
stars is a striking example of the powerful synergies between near-field and far-field
cosmology \citep{freeman02}. In the next section, we look at another example of a 
chemical signature on comparable mass scales.

\section{The impact of star clusters}\label{clustering}
As we have seen, the chemical information arising from the most metal-poor stars is 
very difficult to unravel \citep{nomoto05,kirby08}. One possibility is that not all of the stars are providing us with an 
unambiguous `reading' of the early enrichment of the primordial interstellar gas. 
As mentioned in Sec. \ref{umps}, a significant 
fraction of extremely metal poor stars appear to have undergone mass transfer with a companion 
\citep{ryan05,lucatello05b} which complicates any attempt at inferring the progenitor yields for
elements such as CNO and light $\alpha$ elements. Binarity may explain why the elemental abundances of
the most metal poor stars defy a clear explanation at the present time \citep{joggerst10a,mcwilliam09}. It is 
therefore imperative that we can distinguish the effects of binarity from the effects of inhomogeneous 
mixing, where we may stand a better chance of learning about the yields of the first stars.

Recently, \textcite{hawthorn10a} demonstrate how this issue may be resolved by incorporating a missing ingredient into existing models of stochastic chemical evolution -- the formation of star clusters. If the surface abundances of a star are altered, e.g. by a mass transfer from an evolved binary companion, they no longer constitute a {\it bona fide} imprint of the chemical state of the gas cloud out of which the star once was formed. However, if one could have access to a large number of stars formed out of the same gas cloud and with identical surface abundances to begin with, one could afford a fraction of these stars to be `polluted', without suffer the loss of information on the parent gas cloud. This is where the star clusters come in. If clusters were able to form in the early Universe, they should be incorporated in the models to allow for a correct interpretation of the observational data.

\subsection{The distribution of cluster masses and chemical homogeneity}
In the present-day Universe, most stars are born in a single burst within compact clusters and stellar fragments,
rather than in isolation. This fact is well established in the local Universe \citep{lada03} 
and a growing body of evidence suggests that it is likely to be true also at the time of the first galaxies \citep{clark11a,stacy10,turk09,clark08,karlsson12}.

In order to derive the impact of star clusters on the abundance plane, we must consider the
progenitor mass distribution of star clusters. It is now well established (e.g. Larsen, 2009\nocite{larsen09}) 
that star clusters have a range of masses extending from a minimum mass ($M_{\rm min}$) to a 
maximum mass ($M_{\rm max}$). Although the particular situation for dwarf galaxies is not clear, we 
know from the cloud mass distribution that the range of masses covers many decades 
(e.g. Escala and Larson, 2008\nocite{el08}). The birth distribution of stellar clusters is known as 
the initial cluster mass function (ICMF) and is assumed to have the form
\begin{equation}
\mathrm{d}N/\mathrm{d}M_{*} = \chi(M_{*}) = \chi_0 M_{*}^{-\gamma},
\label{icmf}
\end{equation}
where $M_{*}$ denotes the cluster mass. At least in the nearby Universe, the observations may support a universal 
slope of $\gamma \approx 2$ in most
environments, i.e. equal mass per logarithmic bin \citep{fall05,fall09,lada03,elmegreen10}, although a flatter 
slope may be more applicable in the early Universe \citep{hawthorn10a}. However, the slope cannot be too flat. Since the 
star-to-star scatters, particularly for the r-process elements, decrease with decreasing $\gamma$, the slope must 
be steep enough to ensure consistency with the observed scatters (e.g., Fran\c{c}ois {\it et al.}, 2007\nocite{francois07}).

\textcite{desilva06,desilva07a,desilva07b} have shown that both old ($\sim$ 10 Gyr) and intermediate-age ($\sim 1$ Gyr) open clusters are chemically homogeneous \footnote{Apart from a few light elements, the same
holds true for globular clusters \citep{gratton04}, although few systems (e.g., $\omega$Cen and NGC 1851) recently
have been found to show evidence for more than one burst of star formation (Lee {\it et al.}, 2009\nocite{lee09}).} to a high degree ($\Delta$[Fe/H] $\lesssim 0.03$ dex). \textcite{hawthorn10b} provide a condition that explains why chemical 
homogeneity is expected in most star-forming clusters. For a star cluster to be chemically uniform, it must form
from a gas cloud that has collapsed and finished forming stars before the first supernova goes off within the cloud.
A `homogeneity condition' is given that depends only no the column density through the cloud for a given cloud 
mass. A typical open cluster is expected to be chemically uniform up to about 10$^{4-5}$ M$_\odot$, and a 
globular cluster is expected to be uniform up to 10$^{6-7}$ M$_\odot$.
The homogeneity condition has an important consequence for the 
distribution of stars in the abundance plane, and the effects should be observable even in the limit of only a 
few data points.

As a result of stars being formed in chemically homogeneous clusters, ``clumping'' of stars in abundance space, say, in the [Eu/Fe] -- [Fe/H] plane, is to be expected. It is important to understand that the majority of the clusters considered here are supposed to have been dissolved into the diffuse field. However, while the stellar kinematical information is lost, the chemical information is still preserved and the prospects of reconstructing clusters in abundance space are high, as long as the elemental abundance determinations are accurate enough (i.e., $\Delta [\mathrm{Fe}/\mathrm{H}]\lesssim 0.1$ dex). This is the key point. If detected, the most metal poor abundance groupings are likely to reflect the conditions in the gas at the onset of star formation. The spectra of these stars can be added without loss of information to produce a more accurate measurement of the progenitor conditions. This abundance measurement is averaged over a substantial amount of gas and is therefore not subject to mixing anomalies \citep{kg01} or mass transfer in binaries (e.g. Lucatello {\it et al.}, 2005b\nocite{lucatello05b}). 

A clean signature of clustering, particularly at low metallicity, is important because it indicates the presence of massive star clusters in the early Universe and conceivably provides a constraint on the mass of the first systems. Particularly strong clumping in abundance space, for a given number of stars in the sample, would indicate a highly flattened ICMF, or high mass cut-offs ($M_{\mathrm{min}}$,$M_{\mathrm{max}}$).  This could herald the onset of the formation of massive star clusters in dwarf galaxies (e.g. Bromm and Clarke, 2002\nocite{bc02}). If the star formation efficiencies were low at that time, this may require supermassive gas clouds ($\gtrsim 10^7~\mathcal{M_{\odot}}$) to have formed even at the earliest times \citep{abel00}, possibly consistent with the regular occurrence of massive star-forming clumps at high redshift \citep{genzel06,forster06,elmegreen05}. Conversely, if such clustering was not observed, then we would infer that the slope of the early ICMF is steep, or the maximum cluster size is relatively small compared to the present day. But the observed scatter in the abundance plane would need to be consistent with the non-detection of clustering.

\subsection{The Galactic halo}\label{haloclstr}
Clumping in abundance space can also be used as an independent probe of galaxy formation. The detection of clumping in the Galactic halo would give us unique insight into how the different components of the Galactic halo were formed and whether the building blocks of the halo possibly differed, in terms of how they formed stars, from the dwarf galaxies that presently orbit around the Milky Way. However, since the stellar mass of the halo is large as compared to the mass of a typical cluster, the number of clusters building up the halo is relatively big (cf. Sec. \ref{dwarf_clumping}) and clumping may be difficult to detect.

\begin{figure}[t]
\includegraphics[angle=270,width=6.5in]{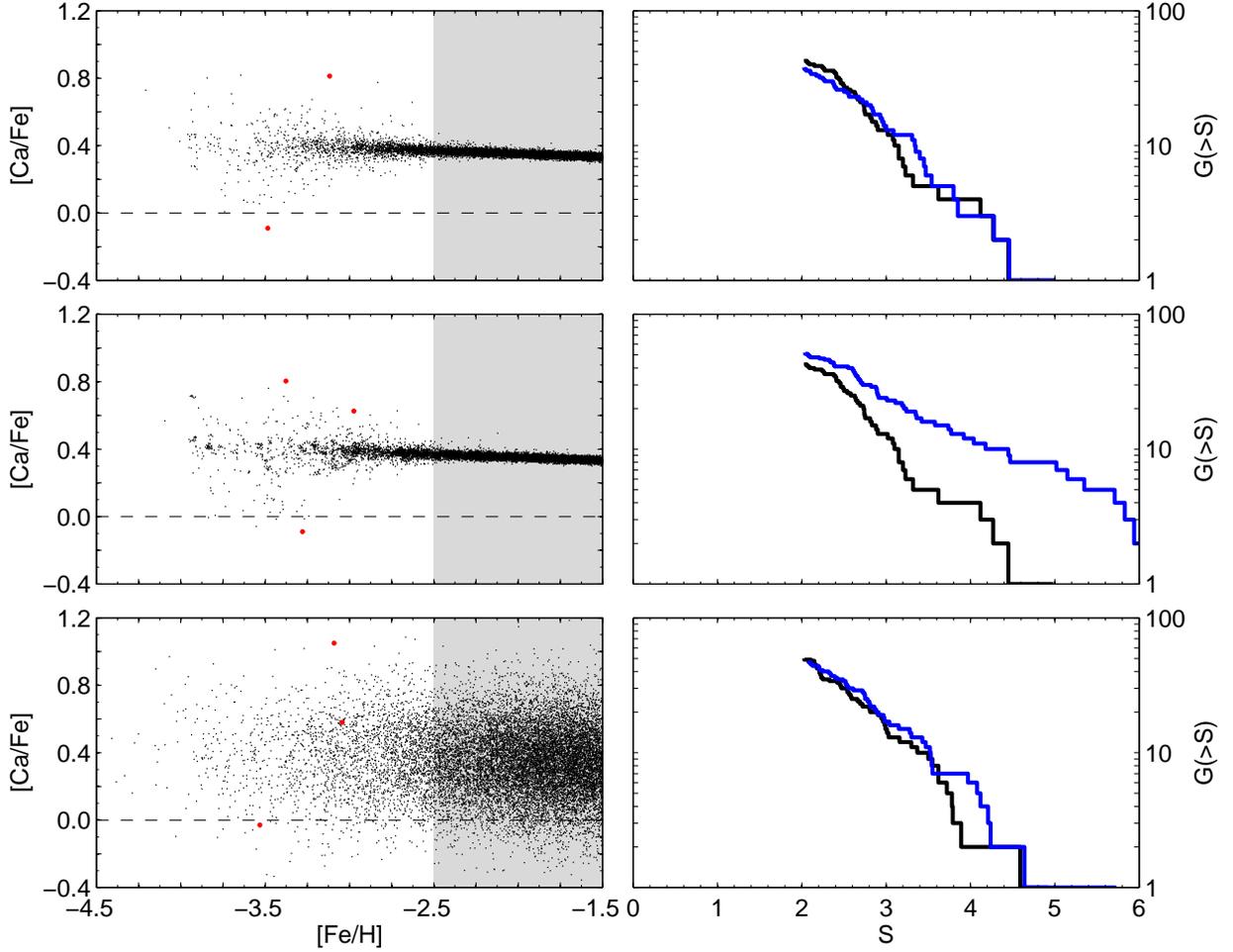}
\caption{
The possibility of detecting star clusters in the Milky Way halo. The left panels show the predicted distribution of $21,000$ stars in the [Ca/Fe] -- [Fe/H] plane. The intrinsic ``observational'' uncertainties are set to $0.01$ dex in both axes. The right panels indicate the corresponding degree of excess clumping (blue lines) as compared to spurious clumping in an ``infinitely massive'' model where not two stars originate from the same cluster (black lines). The clumping statistics is calculated for stars below $[\mathrm{Fe}/\mathrm{H}]=-2.5$ (white areas) using EnLink \citep{sharma09}. The parameter $G(>S)$ is defined as the number of clumps with $>S$ stars relative to the background (S is defined as a relative overdensity). The stars in the top panel probe the entire initial stellar mass of the Galactic halo ($M_{\mathrm{halo,form}}=1\times 10^{9}~\mathcal{M_{\odot}}$) while the stars in the middle panel only probe $10\%$ of the halo mass (i.e., $10\%$ of the number of star clusters), which is comparable to the mass probed by the halo G dwarfs in the SEGUE survey. In contrast to the top panels, clumping is clearly observed. In the bottom panel, the stars from the middle panel are additionally impaired by an observational uncertainty of $0.2$ dex in both abundance ratios. As a consequence, the clumping is almost entirely washed out.  The red dots in the left panels denote stars strictly enriched by material from a single SN. These stars may not necessarily be the most metal-poor stars in the sample.}
\label{clstrs_in_halo}
\end{figure}

Let us illustrate with a specific example. The large SDSS/SEGUE surveys (see Sec. \ref{sites_of_early_sf}) contain data on hundreds of thousands of stars, most of which belong to the Galactic thick disk and halo. Although the spectral information is limited, these surveys have unveiled a new multi-component Galaxy rich in substructure and whose formation and subsequent evolution generally is consistent with a hierarchical build-up from smaller systems \citep{carollo10,bell08,ivezic08,dejong10}. Among the SEGUE stars, $\sim60,000$ are G dwarfs \citep{yanny09}. Only a minor fraction of these stars likely belong to the halo. Assuming a three-component, smooth model of the Galaxy (see, e.g., de Jong {\it et al.}, 2010\nocite{dejong10}), we estimate that about $35\%$ of the SEGUE G dwarfs are halo stars, which corresponds roughly to $21,000$ stars. 

Given that they can be identified unambiguously, how much of the total (stellar) halo mass is probed by these G dwarfs? If the stars are completely phase mixed, they will probe the entire initial formation mass $M_{\mathrm{halo,form}}\simeq1\times 10^9~\mathcal{M_{\odot}}$ of the Galactic halo. An extreme lower limit would be to assume that the stars only probe the instant volume of the survey. We can, however, do slightly better by assuming that all stars follow circular orbits around the Galaxy and integrate the mass over the spherical shell whose boundaries are defined by by the inner- and outermost G dwarf orbits. Again, making use of a smooth model of the stellar halo, and assuming that the G dwarfs are found within $6-14$ kpc from the Galactic center, the mass fraction probed by the SEGUE halo stars is estimated to $\sim 0.1$. Since halo stars usually have non-zero eccentricities, this estimate should be a lower limit. 

The two scenarios are displayed in Fig. \ref{clstrs_in_halo} (top and middle panels). The left-hand-side panels show the distribution of simulated Galactic halo stars in the [Ca/Fe] -- [Fe/H] plane. These are generated by the stochastic chemical evolution model presented in \textcite{hawthorn10a}, assuming a present-day-like ICMF with $\gamma=2$ and cluster masses between $50$ and $2\times 10^5~\mathcal{M_{\odot}}$. The corresponding right-hand-side panels show the excess amount of detected clumping (blue lines) as compared to spurious, random clumping (black lines). The group-finding algorithm EnLink \citep{sharma09} is used to produce the clumping statistics. EnLink calculates the number of clumps having a given number of stars within them, both for the wanted distribution and for a ``smooth'' background distribution. A comparison between these two numbers indicate whether excess clumping is present in the wanted distribution. The ``observational'' uncertainties are set to $0.01$ dex both in [Ca/Fe] and [Fe/H].  Although beyond reach for present-day abundance analyses, we choose such a small uncertainty in order to disentangle the observational ``smearing'' from the ``smearing'' due to under-sampling. Clearly, clumping is intrinsically difficult to detect for the larger halo mass $M_{\mathrm{halo,form}}\simeq1\times 10^9~\mathcal{M_{\odot}}$, and for the given number of stars in the survey (top panels).  In a $10\times$ smaller halo mass, however, excess clumping is present (middle panels). Note that the clumping of stars is visible for the naked eye in the middle left panel. The clumping statistics displayed in the middle right panel clearly shows the presence of excess clumping. The bottom panels show the resulting amount of excess clumping taking into account the observational uncertainty of the SEGUE survey. A fiducial uncertainty of $0.2$ dex in both axes will almost entirely wash out any excess clumping. This suggests that evidence of clustering will be difficult to detect in the SEGUE sample, even if stars did form in clusters also at the epoch of formation of the Galactic halo stars.

We note in passing that stars strictly enriched by a single SN (red dots in Fig. \ref{clstrs_in_halo}) constitute a fraction of $1.5\times 10^{-4}$ of the total number of halo stars in this particular model. The corresponding fraction below [Fe/H]$=-2.5$ is $1.3\times 10^{-3}$. These fractions depend on several factors such as the IMF, the ICMF and the amount of turbulent mixing of the ISM.  These stars generally appear in a range of metallicities from $-4\lesssim [\mathrm{Fe}/\mathrm{H}] \lesssim -2.5$ and may not necessarily be the most metal-poor stars in the sample. Note that these stars are enriched by normal core collapse SNe and are distinct from the `second generation stars' predominantly enriched by the PISNe, which may show up at somewhat higher metallicities, as discussed in Sec. \ref{constrimf}.  The identification of  `single-SN-star-clusters' will give us direct insight into primordial stellar yields. 

\begin{figure}[t]
\includegraphics[angle=270,width=6.5in]{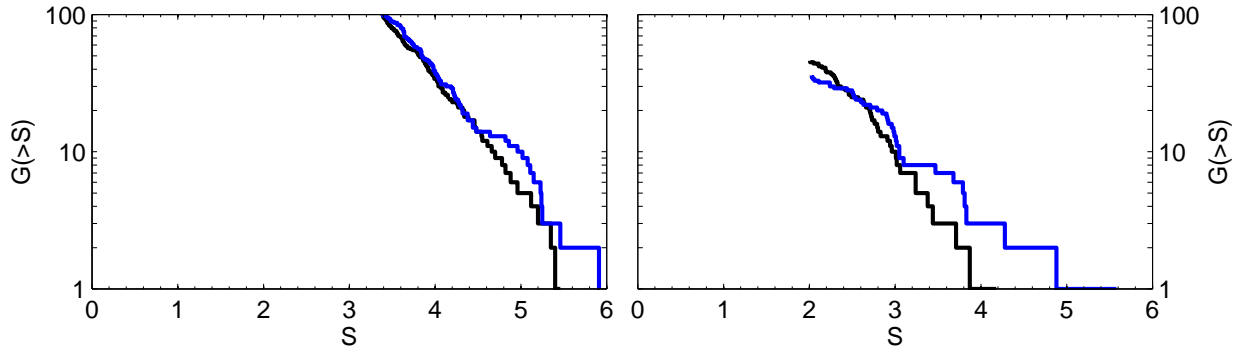}
\caption{ \label{singlesn}
The expected amount of excess clumping (blue lines) in two different model halo samples. The left panel display the amount of clumping in a sample $10$ times the size of the SEGUE halo G dwarf sample ($\sim210,000$ stars, 0.2 dex uncertainty in both axes) while the right panel display the amount of clumping in a SEGUE-sized sample ($\sim 21,000$ stars) for which the observational uncertainty is reduced to 0.05 dex in both axes. Excess clumping is detected in this sample, while the sample displayed in the left panel only show very little evidence of excess clumping. See caption of Fig. \ref{clstrs_in_halo} for a definition of $S$ and $G(>S)$.}
\label{detect_clstrs}
\end{figure}

Figure \ref{detect_clstrs} shows the amount of excess clumping predicted for a sample $10$ times larger than the SEGUE halo G dwarf sample (left) and the amount predicted for a SEGUE-sized sample for which the observational uncertainty instead is reduced to $0.05$ dex in both axes (right). Enhanced excess clumping is detected in the latter case. This shows the necessity of acquiring as small observational errors as possible, sometimes in favor of a larger number of stars, since crucial information may be buried in the details. If the abundance errors are worse than about 0.1 dex, to improve our chances of detection, we should look to use as many distinct elements as possible \citep{hawthorn04}. The probability of detecting clumping will significantly increase with increasing number of elements, in particular for n-capture elements like Eu and Ba, which are expected to show a large scatter in their abundance ratios (Fig. \ref{noisy1}, see also discussion in Bland-Hawthorn {\it et al.}, 2010a\nocite{hawthorn10a}). The number of stars must, however, be large enough to ensure an adequate sampling rate per cluster. 

In the (worst) case of complete phase mixing of the Galactic halo stars (i.e., a, by the survey, sampled mass of $M_{\mathrm{survey}}=M_{\mathrm{halo,form}}=1\times 10^9~\mathcal{M_{\odot}}$), we need roughly $2.4\times 10^6$ stars to obtain an average sampling rate of one star per cluster (note that the most massive clusters will be sampled by $\gg 1$ stars while the least massive ones will generally fall below a sampling rate of one star per cluster), assuming a present-day-like ICMF with cluster masses in the range $50-2\times 10^5~\mathcal{M_{\odot}}$ and a slope of $\gamma = 2$.  This is $10$ times more stars than the total number of stars in the SEGUE survey.  Hence, depending on parameters like the ICMF and the amount of phase mixing, there is an optimal survey size for identifying evidence of clustering. However, in order to probe, e.g.,  `single-SN-star-clusters',  the low-mass end of the ICMF, or the enrichment by rare types of SNe, bigger samples may be required.

\subsection{Dwarf galaxies}\label{dwarf_clumping}
So far, we have mainly focused on the Galactic halo when discussing the possibility of finding evidence of ancient star clusters by 
searching for clumps or aggregates in the chemical abundance diagrams. What about the stars in dwarf galaxies? As suggested 
from the results in the last section (see Fig. \ref{clstrs_in_halo}), the amount of excess clumping is expected to be larger for less massive systems, 
i.e., for a fixed number of stars in the sample, and the crowding should be less severe. Unfortunately, the larger distances to the 
dwarf galaxies make it a challenge to observe many stars.

\begin{figure}[t]
\includegraphics{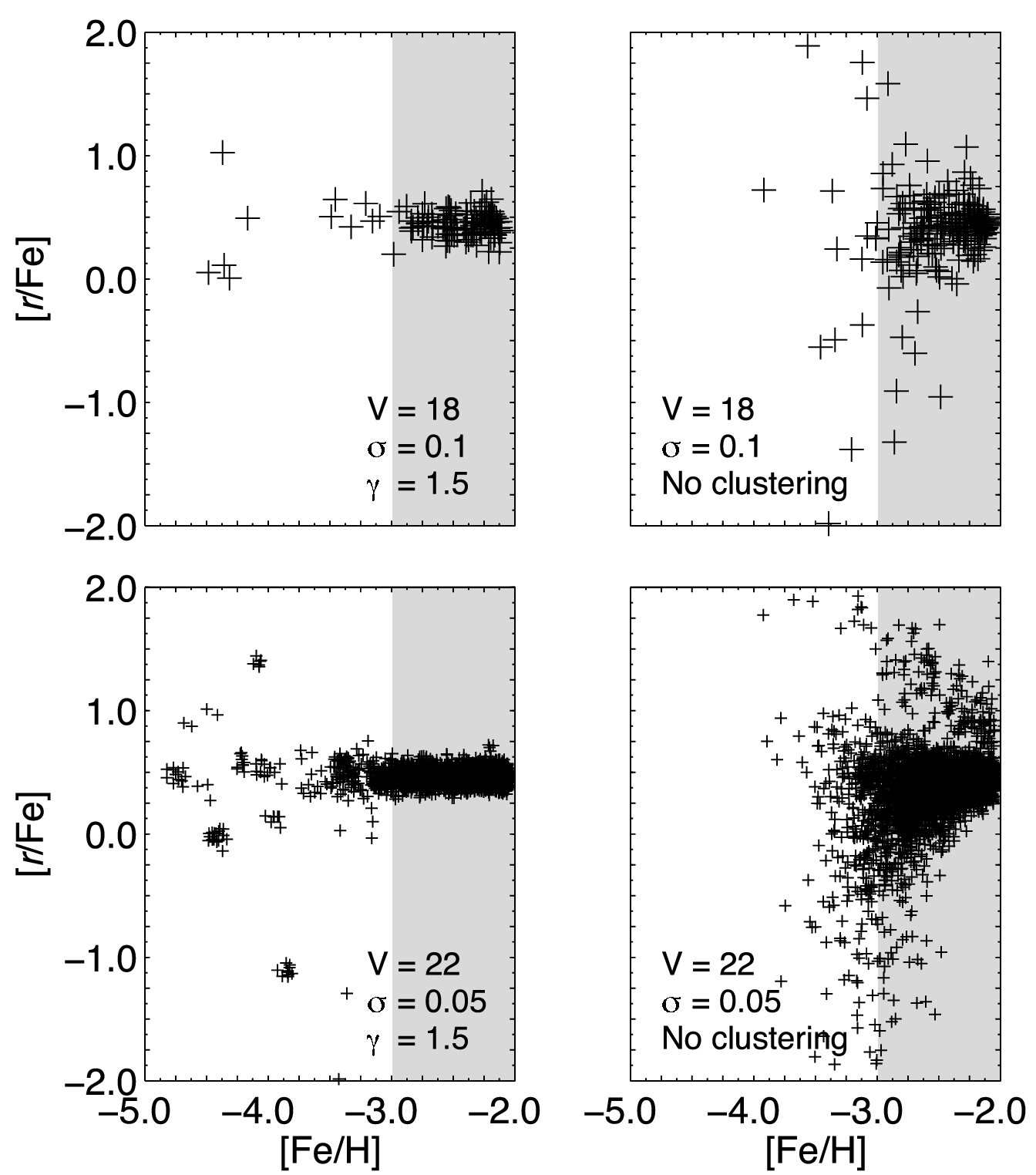}
\caption{ \label{noisy1}
A simulation of a targeted study of a nearby ($D=30$ kpc) dwarf galaxy 
($M_{\star}=3\times 10^5~\mathcal{M_{\odot}}$) on 
an 8m (top) and 30m (bottom) telescope, assuming a limiting magnitude of $V=18$ and $V=22$, respectively. 
Stars are assumed to form in clusters according to an ICMF with a slope of  $\gamma=1.5$ (left). The panels 
on the right show a corresponding galaxy
with no clustering, for comparison. Simulated errors are $0.1$ dex in both [{\it r}/Fe], {\it r} standing for any
 {\it r}-process element, and [Fe/H] in the top figures and $0.05$ dex in the bottom figures. There is evidence of 
 clustering at [Fe/H] $<-3.0$ from a sample of $\sim10$ stars on an 8m class telescope; the clustering is easily 
 detected in the 30m telescope experiment. Above [Fe/H]$=-3$ (gray shaded area), crowding becomes increasingly 
 severe which makes clumping increasingly more difficult to detect. From \textcite{hawthorn10a}.}
\end{figure}

To emphasize the impact of clustering with low number statistics, i.e., we simulate the abundance 
measurements for a dwarf galaxy at a distance of 30 kpc (see Fig.~\ref{noisy1}),
as observed on 8m class and 30m class telescopes, respectively.
We consider a galaxy with a stellar mass of $3\times 10^5~\mathcal{M_{\odot}}$, typical of
a faint dwarf galaxy. This object has about $10^6$ stars, a luminosity of $10^5~\mathcal{L_{\odot}}$ 
and an absolute V mag of $M_V = -7.6$ assuming a Salpeter IMF.
For echelle observations on an 8m telescope (Fig.~\ref{noisy1}),
we assume measurement errors of 0.1 dex in both [{\it r}/Fe] and [Fe/H]. 
Since the mass distribution of star-forming 
clouds in dwarf galaxies is not well established, we treat the slope of the ICMF in 
Eq. (\ref{icmf}) as a free parameter. 
For $\gamma \lesssim 2$, the effects of clustering start to become evident.
For the same experiment on a 30m telescope (Fig.~\ref{noisy1}),
we assume a general improvement in the atmospheric models and
the experimental errors, and therefore adopt errors of 0.05 dex. The effects of
clustering, which are easily seen, remain clearly visible even after a twofold
increase in the measurement errors in both axes. 

In summary, the simulation in Fig.~\ref{noisy1} is a powerful statement of the importance of 
multi-object echelle spectrographs on ELTs. It should be relatively straightforward to detect 
the decreasing abundance scatter and the imprint of early star clusters for individual dwarf 
galaxies. And, by specifically targeting subgiants, for which accurate ages can be 
determined, we will have an additional, independent measure of cluster membership. The 
reconstructed cluster must be a coeval population by definition.

\subsubsection{Signatures of ancient star clusters: the case of the MDF}
Signatures of relic clusters does not have to be identified in multi-dimensional abundance space. Already in one dimension, ``bumps'' and irregularities in the MDF, particularly in the metal-poor tail, could be a indication of disrupted clusters. \textcite{karlsson12} report on a tentative detection of a relic star cluster in the Sextans dSph. They re-analysed the excellent abundance data of six very metal-poor stars below [Fe/H] $=-2.5$ by \textcite{aoki09} and found that at least three stars clumped together in chemical abundance space, particularly in Mg, Ti, Cr, Ba, and Fe. While two of the ``clump'' stars have similar [Na/Fe], the third star have a significantly lower [Na/Fe]. This could possibly be a manifestation of the Na-O anti-correlation observed in Galactic globular clusters, and may suggest that these stars were once members of a very metal-poor globular cluster. The weighted iron abundance of the ``clump'' stars is $[\mathrm{Fe}/\mathrm{H}]=-2.7$, which would make the parent cluster the most metal-poor cluster to date. Interestingly, in the re-calibrated CaT data of Sextans by \textcite{starkenburg10}, there is a clear sign of a ``bump'' in the MDF at $[\mathrm{Ca}/\mathrm{H}]\simeq -2.65$, very close to the mean Ca abundance of the three stars observed by \textcite{aoki09}. \textcite{karlsson12} made use of this ``bump'' to estimate the initial mass of the cluster to $M_{*,\mathrm{init}}=1.9^{+1.5}_{-0.9} (1.6^{+1.2}_{-0.8}) \times 10^5~\mathcal{M_{\odot}}$, assuming a Salpeter (or Kroupa, Kroupa, 2001\nocite{kroupa01}) IMF.  

There is also evidence for a kinematical substructure, possibly a remnant star cluster, in the central region of Sextans (Kleyna {\it et al.}, 2004;\nocite{kleyna04} Battaglia {\it et al.}, 2011;\nocite{battaglia11} see, however, Walker {\it et al.}, 2006\nocite{walker06}). This substructure has a metallicity of $[\mathrm{Fe}/\mathrm{H}]\sim -2.6$ \citep{battaglia11}, which is close to the metallicity of the ``clump'' stars and a common origin would not be implausible. This lends further support to the hypothesis of the existence of a now disrupted star cluster in Sextans, whose signature can be detected both in velocity space, multi-dimensional abundance space, {\it and} in the MDF.   

Metal-poor environment of dwarf galaxies like dSphs provides an ideal base for searching for ancient clusters.  In the data of \textcite{kirby11}, several MDFs exhibit ``bumpy'' tails, such as Leo II, Draco and Ursa Minor. Interestingly, the Ursa Minor dSph is known to possess a second concentration of stars roughly 14' from the center. \textcite{kleyna03} claimed that these stars belong to a cold substructure, consistent with the remnants of a disrupted star cluster. 

It would also be interesting to compare ensembles of MDFs for different types of galaxies. In a first attempt, \textcite{karlsson12} compared the MDFs of Sextans, Sculptor, Fornax, and Carina dSph \citep{starkenburg10} with the MDFs of eight ultra-faint dwarf galaxies observed by \textcite{kirby08,kirby11}. They found an excess scatter between the MFDs of the dSphs, consistent with the presence of clusters distributed in mass according to the present-day ICMF. No such excess scatter was found between the MDFs of the eight ultra-faints. Together with an observed offset between the average MDF of the ultra-faints and that of the dSphs, the distinction in the degree of ``MDF bumpiness'' may suggest that the two groups of galaxies were formed in different environments. Again, this shows that the identification and reconstruction of disrupted star clusters in chemical abundance space is important, not only to gain a deeper understanding of star formation and stellar yields, but also of galaxy formation and the build-up of galactic stellar haloes. Any ``bump'' in the MDF of a dwarf galaxy need, however, to be confirmed/refuted as a cluster signature, e.g. in multi-dimensional abundance space.

\section{Peering into the future}
It is an extraordinary fact that we can probe back to the earliest generations of stars from observations of the 
{\it local} Universe $-$ this is what we mean by `near field cosmology.' The first stars were unique to their time: 
they were responsible for the first
chemical elements \citep{ryanweber06} and for reionizing the neutral fog of hydrogen that permeated the early
Universe \citep{fan02}. Our goal in describing pre-galactic enrichment is to understand the formation of the first stellar generations and their yields. Did the first stars form in isolation or in groups? Were relatively few stars responsible for reionization or was it triggered by the collective effect of massive star clusters? Just what are the processes that govern star formation at extremely low metallicity? Is this exclusively the domain of the most massive stars, or can substantial intermediate and low mass stars form?  In other words, did stellar populations observable today exist before reionization? We have few if any answers at the present time. While this is a field that is very much in its infancy, 
we believe there has been sufficient progress to warrant our review. 

The chemical signatures imprinted in old, metal-poor stellar populations make up a vast repository of information that takes us back to the first few hundred million years of the Universe. It is an invaluable tool that allows us to witness the very process of primordial and early star formation in unparalleled detail. But not only that, chemical signatures on galactic scales help us to understand the connection between local dwarf galaxies and those systems which once built up the Milky Way. A more coherent and complete picture of the galaxy formation process is now emerging from observations of nearby stars.

We anticipate a great deal of progress over the next decade, particularly in an era of
extremely large, ground-based telescopes (25-40m) and a new generation of space telescopes. We anticipate
that many of the great advances will come from more extensive observations of the near field, in particular, 
metal poor stars in the vicinity
of the Galaxy. But we should not rule out the extraordinary. For example, at least one gamma-ray burst has been
visible to the naked eye in recent years \citep{bloom09} allowing for high resolution spectroscopy
which can in turn provide important constraints on the local stellar populations (e.g. Castro-Tirado {\it et al.}, 2010\nocite{castro10}). High redshift quasars beyond $z\sim 3$, even with the added boost from gravitational 
lensing, are likely to be too faint to provide
useful absorption line spectra of intervening galaxies \citep{glikman08}. We may obtain future constraints
from rest-frame UV ionization diagnostics of high-redshift sources. Several authors have identified a population of
CNO-rich galaxies being internally ionized by star clusters with a top-heavy IMF \citep{fosbury03,glikman07,raiter10}.

The study of the high redshift Universe is complementary to studies of the local Universe \citep{bhf00}. For example, measurements of the integrated spectral signature of Pop III stars is an exciting prospect although separating the signal from the infrared foreground is difficult (see, e.g., Raue {\it et al.}, 2009\nocite{raue09}).  Even though tantalizing progress has been made \citep{kashlinsky07}, we must await the advent of next generation telescopes, such as the {\it JWST} and ALMA to be able to resolve the first galaxies, and detect individual sources and their birth sites \citep{wc08}. It should be stressed that the level of detail on the first stars and galaxies, buried in the chemical abundance patterns of old stellar populations, can never be reached by studying the direct light from these objects themselves, nor by studying the high-$z$ IGM.  The information needs to be extracted from {\it resolved stars}. The modeling of the build-up of chemical elements over cosmic time, in particularly at early stages, is absolutely essential to the interpretation of this information.

But our expectation is that the most important insights on pre-galactic enrichment will come from the near field. We may be in for some surprises. \textcite{karlsson08} has described how a significant fraction of second generation stars that formed out gas enriched by PISNe are, contrary to what may be expected, predicted to be found at relatively high metallicities around [Fe/H] $\simeq  -2.5$. For this reason, they may not have been picked up in existing surveys of metal-poor stars, like the HK and Hamburg/ESO surveys. These same surveys have tended to target the diffuse Galactic halo, but dwarf galaxies, the inner halo, the Galactic Centre, or another site, may prove to be a more prosperous hunting ground in future years (see Sec. \ref{sites_of_early_sf} and Sec. \ref{chem_sign_from_lo_m_gal}).

In the closing statements of many reviews in astronomy, there are the obligatory requests for more theoretical
and numerical work in support of a rich harvest of observations. But in the field of pre-galactic enrichment, the 
observations appear more limited, in part because we have not yet learnt to properly interpret the data. There is a
pressing need for more computational work on many fronts. These include:

\begin{itemize}
\item better stellar atmospheric models to improve abundance calibrations for metal poor stars; \\
\item a new generation of stellar evolution models (especially giants, sub-giants) with updated nuclear reaction networks, diffusion and mixing; \\
\item self-consistent stellar models that take into account the possible binary nature of the first stars; \\
\item self-consistent supernova models that properly treat the explosion mechanism, fallback, rotation and mixing; \\
\item better fine-scale and coarse-scale mixing theories and algorithms for metal transport in turbulent media; \\
\item binary star mass transfer (zero metallicity) with a proper treatment of Roche-lobe overflow, slow and fast winds; \\
\item higher resolution hydrodynamical models of the formation of the first stars and star clusters.
\end{itemize}

Just where to look for the most ancient and/or second generation stars (and possibly metal free stars) is a hot topic 
at the present time with some groups favouring the inner parts of galaxies (e.g. Tumlinson, 2010\nocite{tumlinson10}; 
White \& Springel (2000)), and others finding a 
wide spread in radius \citep{scannapieco06,brook07,salvadori10}. More advanced numerical
simulations may well provide better guidelines on where to look and even the specifics of the orbit parameters of the
oldest stars (e.g. Tumlinson, 2010\nocite{tumlinson10}). In this respect, we look forward to the {\it Gaia} astrometric 
satellite which is due to launch in the next few years. This will provide exceptional kinematic information for up to a 
billion stars in the Galaxy. We recommend that future simulations attempt to predict the energy-angular momentum 
space (e.g. Helmi, 2008\nocite{helmi08}) expected for these ancient stars.

Regardless of the simulations, it is important to keep in mind that metal-rich gas is observed in quasar/AGN 
spectra to the highest redshifts \citep{hamann99} consistent with detections of dust and molecular gas at these
early times \citep{yun00,cox02,klamer04}. This reflects the fact that the dynamical timescales in the cores of
 galaxies are very short ($\lesssim 10^7$yr). Thus ancient stars with solar abundances may well exist in the 
 cores of galaxies, assuming these have not been flung out into the disk by the formation of a central bar 
 (e.g. Minchev and Famaey, 2009\nocite{minchev09}; see Sec. \ref{churning}).

Precisely when the first star clusters formed is unclear. Interestingly, the difficulty in identifying metal-free stars
at the present epoch may already provide a constraint on the primordial initial mass function \citep{scannapieco06} and 
the possible existence of star clusters. As shown in Table~\ref{lookback}, a star must be less massive
than the Sun to have survived a Hubble time. If the star formed at a redshift as late as $z\sim 1$, its mass must be
less than $0.9-1.1~\mathcal{M_{\odot}}$, depending on the metallicity, in order to have reached the giant branch today. If these 
stars did not exist at those early epochs due to a truncated IMF, then we must infer the progenitor yields from later 
generations of stars. Interestingly,  because `clustered' abundance signals will be unaffected by mass transfer in 
binaries, these may give us a truer reading of the IGM metallicity at early times.

A new approach to exploring the evolution of the early initial {\it cluster} mass function has been 
proposed by \textcite{hawthorn10a}. Stars born in clusters are extremely uniform in their chemical
elements \citep{hawthorn10b,desilva07a}. This should lead to clumping in the abundance
plane, e.g. [Fe/H] vs. $[\alpha$/Fe]. While the effect can be detected in 8m class data, a much richer return is
expected in an era of extremely large telescopes. To this end, it will be necessary to equip these telescopes
with wide-field multi-object spectrographs that operate at high spectroscopic resolution ($R\gtrsim$20~000).
Such an instrument can be profitably targetted at dwarf galaxies, the Galactic bulge or populations identified
by the {\it Gaia} satellite.

\section*{Acknowledgments}
TK and JBH gratefully acknowledge the hospitality and the inspiring environment of the Beecroft Institute of Particle Astrophysics and Cosmology, where this review was finalized. We are grateful to Philipp Podsiadlowski for insightful comments on the effects of stellar binarity, to Mark Krumholz for discussions of the IMF, and to Sanjib Sharma for enlightening discussions on the formation of the Galactic halo and for running the group-finding code EnLink for us. We acknowledge useful conversations with Ryan Cooke and Bengt Gustafsson. TK is supported by ARC grant FF0776384 through a Federation Fellowship held by JBH. JBH is indebted to the Leverhulme Foundation and to Merton College, Oxford for fellowships in support of this work.  VB acknowledges support from NSF grants AST-0708795 and AST-1009928, as well as NASA ATFP grants NNX08AL43G and NNX09AJ33G.


\bibliographystyle{apsrmp}
\bibliography{rmp_refs}

\end{document}